\documentclass[twocolumn]{aastex701}

\usepackage{enumitem}
\usepackage[symbol,flushmargin,hang]{footmisc}
\usepackage{makecell}
\usepackage{textcomp}
\usepackage{fix-cm}
\usepackage{amsmath}
\usepackage{listings}
\newsavebox{\promptbox}
\newcommand{\gap}{\vspace{-3.5pt}}
\lstdefinestyle{prompt}{
  basicstyle=\fontsize{7.5}{7.5}\selectfont\ttfamily,
  columns=flexible,
  showstringspaces=false,
  escapeinside={(*}{*)},
  literate={~}{{\raisebox{-0.6ex}{\textasciitilde}}}1
           {×}{{$\times$}}1 {↑}{{$\uparrow$}}1 {↓}{{$\downarrow$}}1
           {−}{{$-$}}1 {≤}{{$\leq$}}1 {≥}{{$\geq$}}1
           {→}{{$\rightarrow$}}1 {—}{{\textrm{---}}}1,
  breaklines=true,
  breakatwhitespace=true,
  breakautoindent=false,
  breakindent=0pt,
  resetmargins=true,
  frame=none,
  xleftmargin=0pt,
  xrightmargin=0pt,
  aboveskip=0pt,
  belowskip=0pt
}
\defcitealias{gemmateam2026}{Gemma Team et al. 2026}

\graphicspath{{figures/}}

\shorttitle{Agentic Active Learning for Anomaly Detection in ASAS-SN}
\shortauthors{Pe\v{s}ta \& Ting}

\begin{document}

\bibliographystyle{aasjournalv7}

\title{Agentic Active Learning Meets Visual Embeddings: \\ Finding Anomalies among 370\,000 Variable Stars from ASAS-SN}

\author[0000-0003-3649-9811]{Milan Pe\v{s}ta}
\email{pesta.22@osu.edu}
\affiliation{
Institute of Theoretical Physics, Faculty of Mathematics and Physics, Charles University, V Hole\v{s}ovi\v{c}kách 2, 180 00 Prague, Czech Republic}
\affiliation{Department of Astronomy, The Ohio State University, 140 West 18th Avenue, Columbus, OH, 43210, USA}
\affiliation{Center for Cosmology and AstroParticle Physics, The Ohio State University, 191 West Woodruff Avenue, Columbus, OH 43210, USA}
\correspondingauthor{Milan Pe\v{s}ta}
\email{pesta.22@osu.edu}

\author[0000-0001-5082-9536]{Yuan-Sen Ting}
\email{ting.74@osu.edu}
\affiliation{Department of Astronomy, The Ohio State University, 140 West 18th Avenue, Columbus, OH, 43210, USA}
\affiliation{Center for Cosmology and AstroParticle Physics, The Ohio State University, 191 West Woodruff Avenue, Columbus, OH 43210, USA}
\affiliation{Max-Planck-Institut f\"{u}r Astronomie, K\"{o}nigstuhl 17, D-69117 Heidelberg, Germany}

\begin{abstract}
Unusual light-curve morphologies can point to rare physical configurations or new phenomena, but automatic searches for anomalies are often dominated by artifacts. Separating genuine anomalies from false positives has traditionally required manual vetting, which does not scale to modern surveys. We present an active learning framework for detecting anomalies in samples of periodic variable stars, with the vetting delegated to multimodal large language model agents. The initial ranking comes from isolation forests trained on DINOv2 ViT-g/14 embeddings of phase-folded light curves. The agents iteratively review the light-curve images of the top-ranked candidates and assign relevance scores, which are propagated through the embedding space to prioritize the next targets. A logistic regression step then extends the search beyond label propagation, and a multi-agent consensus review filters false positives. Using Gemini~3 Flash agents, we applied the pipeline to $373\,646$ periodic variables from ASAS-SN Sky Patrol V2.0. Across $51$ iterations, the agents labeled ${\sim}1\%$ of the sample in roughly $3$ hours at a cost of ${\approx}\$40$, yielding our final catalog of $24$ anomalies ($18$ newly reported) and $153$ potentially interesting objects. The anomalies include eclipsing binaries with extremely deep primary eclipses, high-amplitude contact binaries and pulsators, and a symbiotic nova in outburst for over two decades. At the same labeling budget, the initial ranking would have recovered only $30\%$ of our catalog, while reaching its lowest-ranked anomaly would have required ${\sim}70$ times the budget. These results demonstrate the feasibility of agentic active learning for anomaly detection in existing and upcoming photometric surveys.
\end{abstract}

\keywords{\uat{Astronomy data analysis}{1858} --- \uat{Light curves}{918} --- \uat{Outlier detection}{1934} --- \uat{Periodic variable stars}{1213} --- \uat{Sky surveys}{1464} --- \uat{Time series analysis}{1916}}

\section{Introduction}\label{sec:intro}

The current era of time-domain astronomy is defined by a deluge of photometric data. Wide-field ground-based surveys such as the All-Sky Automated Survey for SuperNovae \citep[ASAS-SN;][]{shappee2014,kochanek2017} and the Zwicky Transient Facility \citep[ZTF;][]{bellm2019} provide multi-epoch observations for hundreds of millions of sources. The upcoming Vera C. Rubin Observatory Legacy Survey of Space and Time \citep[LSST;][]{ivezic2019} will expand this further, delivering light curves for billions of objects over its projected lifetime. Given the scale of these data sets and the limited resources available for follow-up observations, efficient and scalable methods are needed to determine which sources merit detailed examination.

Anomalies are a natural target for such prioritization. These objects deviate from typical source types and behaviors, which can point to extreme physical conditions or new phenomena. Historically, many discoveries have emerged from examining sources that defied initial classification or showed unusual behavior \citep[e.g.,][]{hewish1968,schmidt1963,leavitt1912}. Identifying anomalies in large surveys is therefore essential for testing theoretical models and uncovering exotic objects.

Several unsupervised methods for anomaly detection are commonly used in the literature. Techniques such as isolation forests \citep{liu2008,villar2021}, autoencoders \citep{hinton2006,nicolaou2026}, and normalizing flows \citep{rezende2015,liang2023} can identify outliers without requiring predefined labels. However, their effectiveness depends strongly on how the data is represented. In the case of light curves, manually curated features such as Fourier coefficients, color indices, and light-curve statistics are effective for population studies and classification \citep[e.g.,][]{simon1981,debosscher2007,kim2016}, but they may not capture the subtle morphological nuances that distinguish unusual objects from standard sources. If the representation is not sensitive to the relevant form of anomalous behavior, such sources can remain hidden within the bulk distribution.

An alternative is to represent each light curve by its visual appearance rather than a fixed set of hand-crafted features. Vision foundation models trained on large collections of natural images, such as CLIP \citep{radford2021}, MAE \citep{he2021}, and DINOv2 \citep{oquab2023}, learn general-purpose representations that transfer to new domains with little or no fine-tuning. Applied to light-curve images, they capture both the global morphology and local irregularities, such as asymmetries, scatter, and phase-dependent features. Visual embeddings have proven effective for light-curve classification tasks \citep{moreno_cartagena2025}, but remain largely unexplored for anomaly detection.

In practice, even a representation that preserves the anomalous nature of the sources cannot prevent unsupervised methods from finding artifacts alongside genuine astrophysical anomalies, since these contaminants also appear as outliers in the data. The scale of this contamination can be severe. For instance, \citet{etsebeth2024} applied isolation forests to ${\sim}4$ million galaxy images from the Dark Energy Camera Legacy Survey (DECaLS) and found that $1\,763$ of the top $2\,000$ candidates were instrumental artifacts or masked sources, with only a single scientifically interesting anomaly present in the initial ranking.

Removing such contaminants traditionally requires labor-intensive human vetting, which does not scale to modern survey sizes. Active learning mitigates this by directing expert review only to the most informative candidates \citep{lochner2021,lochner2025}, sharply reducing the labeling effort needed to separate artifacts from genuine anomalies. However, when expert time is scarce or the labeling effort must be regularly repeated, even active learning becomes prohibitive. The labor bottleneck could be relieved by large language models (LLMs), which have been shown to perform well in a variety of contexts \citep[e.g.,][]{stoppa2025,jia2026,ferber2024,xu2025}, yet active learning applications in astronomy have so far relied on human experts \citep[e.g.,][]{leoni2022,ishida2019,richards2012,lochner2021,ishida2021,pruzhinskaya2023}.

In this work, we introduce an active learning framework for anomaly detection in which LLM agents\footnote[2]{By agent, we mean an independently prompted instance of a multimodal LLM acting as an automated annotator. Our agents are deliberately minimal, with no memory across evaluations, access to external tools, or ability to plan their own actions.} take over the expert role, and demonstrate its feasibility on a sample of over $370\,000$ periodic variable stars from the ASAS-SN survey. The active learning aspect is central to our approach, as evaluating every source with an LLM agent would be too costly at survey scale. Our pipeline derives an initial anomaly ranking from isolation forests trained on visual embeddings of the light curves, which we obtain using an off-the-shelf vision model without any fine-tuning or survey-specific adaptation. LLM agents then iteratively refine this ranking through active learning, and the resulting labels feed a classifier search and a final multi-agent consensus review of all candidates. Together, these steps form an automated workflow for systematic anomaly detection that is no longer limited by expert availability.

The paper is organized as follows. In Section~\ref{sec:data}, we describe the ASAS-SN data and our sample selection. We detail the pipeline in Section~\ref{sec:methods} and present the discovery statistics and the anomaly catalog in Section~\ref{sec:results}. In Section~\ref{sec:discussion}, we discuss the implications of our findings and conclude in Section~\ref{sec:conclusions}.

\section{Data}\label{sec:data}

The All-Sky Automated Survey for SuperNovae (ASAS-SN) is a ground-based photometric survey originally designed to discover bright supernovae and other transients \citep{shappee2014}. Currently, it comprises twenty $14$-cm telescopes across five stations, monitoring the entire sky nightly in the $g$ band to a depth of $g$~$\lesssim$~$18.5$\,mag. Including the archival $V$-band data collected in 2014--2018, ASAS-SN has accumulated nearly $13$ years of photometry for the longest-observed sources. To make the data accessible to the community, the ASAS-SN team launched Sky Patrol V1.0 \citep{kochanek2017}, allowing users to retrieve light curves for arbitrary sky coordinates. They also released pre-computed light curves for millions of sources in both $V$ and $g$ bands as part of their search for variable stars \citep{jayasinghe2021,christy2023}. With the launch of Sky Patrol V2.0 \citep{hart2023}, they provided access to continuously updated light curves for ${\sim}111$ million targets derived from numerous external catalogs. The primary stellar source table, which serves as the basis for this work, comprises $98\,602\,587$ sources from the ATLAS Reference Catalog \citep[REFCAT2,][]{tonry2018} with $g<18.5$\,mag for which the target contributes $>50\%$ of the flux within a $20$\,arcsec radius.

To assemble our sample of variable stars, we queried the Sky Patrol V2.0 database through the \texttt{pyasassn} Python client \citep{hart2023} and performed an inner join of the \texttt{stellar\_main} table with \texttt{aavsovsx}, corresponding to the International Variable Star Index (VSX)\footnote[3]{\url{https://vsx.aavso.org/}}. The join returned $502\,747$ ASAS-SN sources listed in VSX. Because our morphology-based analysis relies on phase-folded light curves, we retained only sources with reported periods, reducing the sample to $375\,410$ variable stars. For each source, we downloaded all available $V$- and $g$-band photometry (as of November 2025) and catalog metadata.

\section{Methods}\label{sec:methods}

\begin{figure*}
  \centering
  \includegraphics[width=0.99\textwidth]{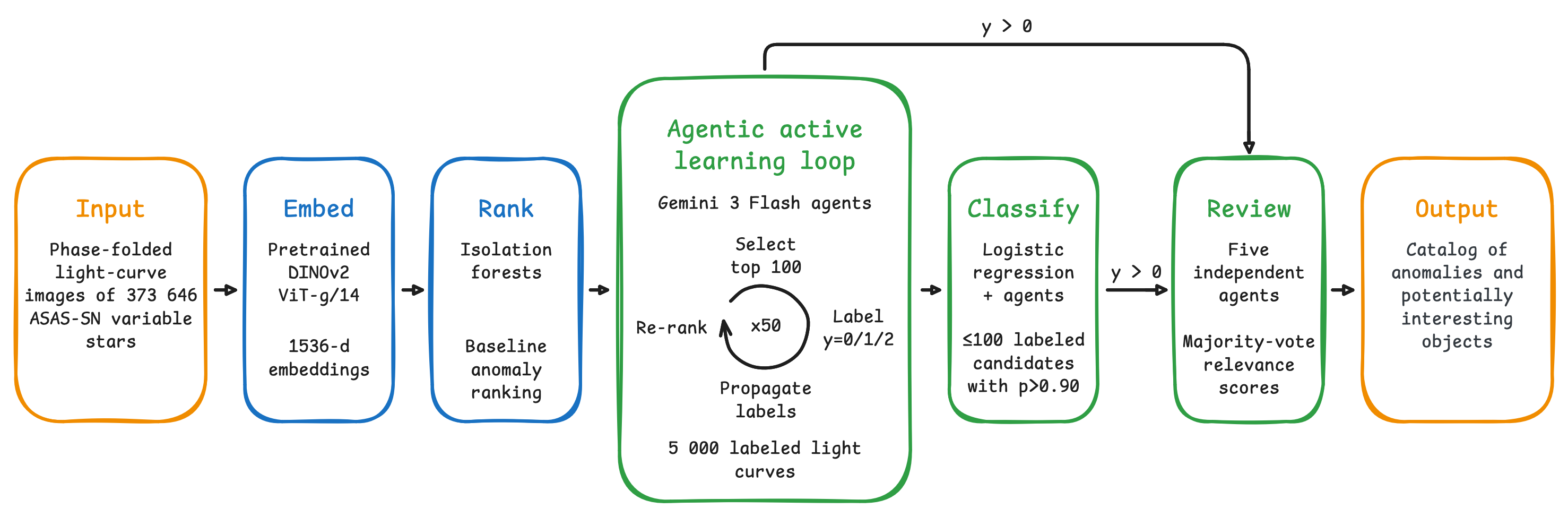}
  \caption{Schematic of our agentic active learning pipeline. Phase-folded light-curve images of $373\,646$ variable stars from ASAS-SN Sky Patrol V2.0 are embedded with DINOv2 ViT-g/14 and ranked with class-conditional isolation forests. Gemini~3 Flash agents then repeatedly review the $100$ highest-ranked candidates over $50$ iterations of an active learning loop and assign relevance scores of $y=0$ (not interesting), $y=1$ (potentially interesting), or $y=2$ (anomaly). After each iteration, the scores are propagated to the neighboring sources in the embedding space, prioritizing promising candidates for the next batch. A logistic regression step (iteration~$51$) then trains a classifier on the accumulated labels, rescores the full pool, and selects up to $100$ sources with anomaly probability $p>0.90$ for another round of agent labeling. Finally, a multi-agent consensus review of sources with $y\geq1$ yields a catalog of anomalies and potentially interesting objects.}
  \label{fig:pipeline}
\end{figure*}

In Figure~\ref{fig:pipeline}, we show a schematic of our pipeline. First, we render raw ASAS-SN time series as phase-folded light curve images and embed them with a pre-trained vision model (Section~\ref{subsec:embedding}), then we establish a baseline anomaly ranking with class-conditional isolation forests trained on these embeddings (Section~\ref{subsec:isolation}). Next, we iteratively refine the initial ranking through an active learning loop in which multimodal LLM agents repeatedly label the top-ranked anomaly candidates, and their labels are propagated to neighboring sources in the embedding space (Section~\ref{subsec:agentic}). After the main loop, we train a logistic regression classifier on the accumulated labels, score the full unlabeled pool, and submit the top-ranked candidates to the agents for an additional round of labeling (Section~\ref{subsec:logreg}). Finally, we subject all anomaly candidates flagged by the agents in the previous steps to a multi-agent consensus review (Section~\ref{subsec:review}), producing our final catalog of anomalies and potentially interesting objects.

\subsection{Visual Embedding with DINOv2}\label{subsec:embedding}

We obtained visual embeddings of the ASAS-SN light curves using the DINOv2 ViT-g/14 model \citep{oquab2023}, which is a vision transformer model pre-trained on a diverse data set of approximately $142$ million natural images. The self-supervised training of DINOv2 combined image-level and patch-level objectives, encouraging the model to learn semantically meaningful representations that captured both global structure and local details without requiring human-annotated labels. We deliberately chose a general-purpose vision model rather than one trained or fine-tuned on astronomical data, allowing us to test how far off-the-shelf visual representations transfer to light-curve morphology without any domain-specific adaptation.

To prepare inputs for DINOv2, we rendered each light curve as a $518\times518$ pixel image and let the standard preprocessing rescale it to $224\times224$ pixels, matching the input resolution used during the main training stage of the model. This supersampling reduced the blending of neighboring data points in densely sampled regions and preserved finer detail than rendering directly at the input resolution. Before rendering, we removed observations with bad quality flags, negative fluxes, or $\sigma_{\text{mag}} > 99$\,mag (corresponding to non-detections). We then phase-folded the remaining observations on the VSX period and plotted magnitude against phase over $[-0.5, +0.5]$, with the $V$ band in blue and the $g$ band in green. We excluded all axes, labels, and annotations so that the embeddings would capture morphology rather than spurious visual cues, but retained the magnitude range $\Delta m$ of each plot and the period $P$ of each light curve as metadata for the subsequent analysis.

We applied quality cuts to remove sparsely observed and noise-dominated sources with fewer than $50$ observations or magnitude ranges $\Delta m > 15$\,mag, yielding our final sample of $373\,646$ variable stars. We passed the light-curve images from this sample through the DINOv2 backbone and extracted their \texttt{[CLS]} tokens from the last layer, representing each light curve as a $1536$-dimensional embedding vector. After extracting all embeddings, we centered them by subtracting the sample mean. This removed the common component dominating the representation, which made the embeddings more sensitive to morphological variations between sources.

\begin{figure*}
  \centering
  \includegraphics[width=\textwidth]{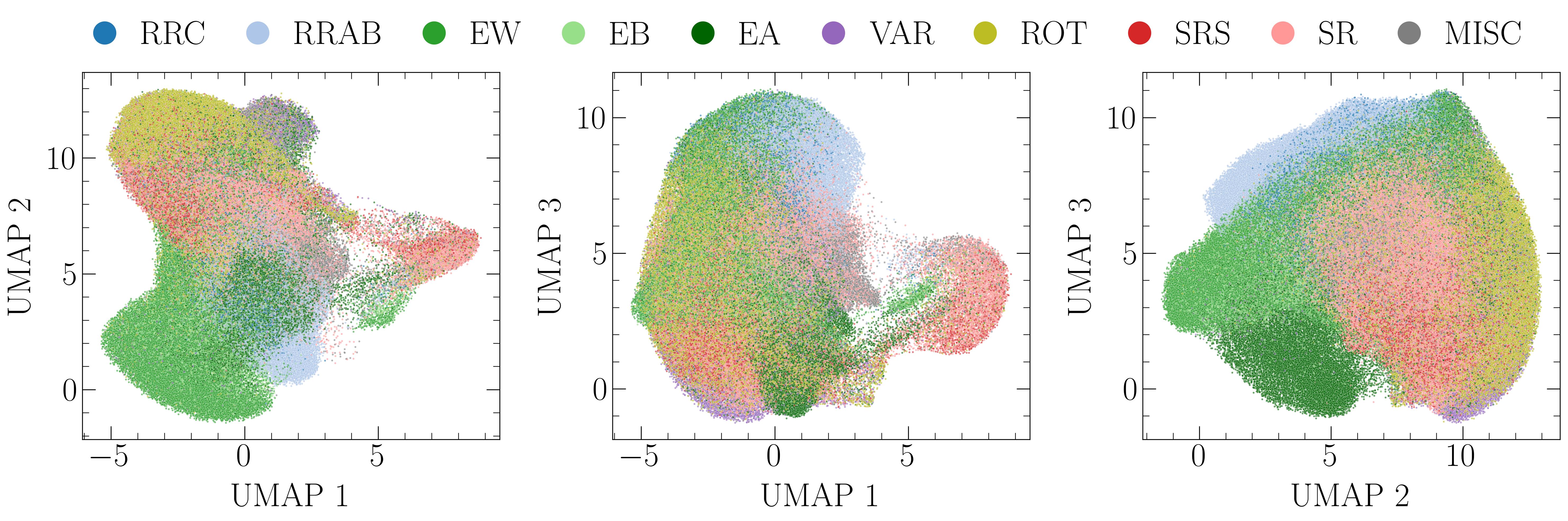}
  \caption{Three-dimensional UMAP projection of the DINOv2 embeddings for the ten most frequent VSX variability classes in our sample, spanning RR~Lyrae pulsators (RRAB, RRC), eclipsing binaries (EA, EB), contact binaries (EW), semiregular variables (SR, SRS), and rotating variables (ROT), along with the unspecified (VAR) and miscellaneous (MISC) categories. Distinct variability types occupy well-separated regions and morphologically similar classes lie adjacent, even though the embeddings were not trained on light-curve images and the labels were not used in the projection.}
  \label{fig:umap}
\end{figure*}

Before attempting anomaly detection, we confirmed that the embeddings encode the morphological structure of the light curves by visualizing them in three dimensions with UMAP \citep{mcinnes2018}. In Figure~\ref{fig:umap}, we show the projection for the ten most common VSX variability types in our sample, which together account for over $80\%$ of the data. Although the embeddings were not trained on light-curve images and the class labels were not used when constructing the projection, distinct variability types occupy well-separated regions while morphologically similar classes lie adjacent.

\begin{figure*}
  \centering
  \includegraphics[width=\textwidth]{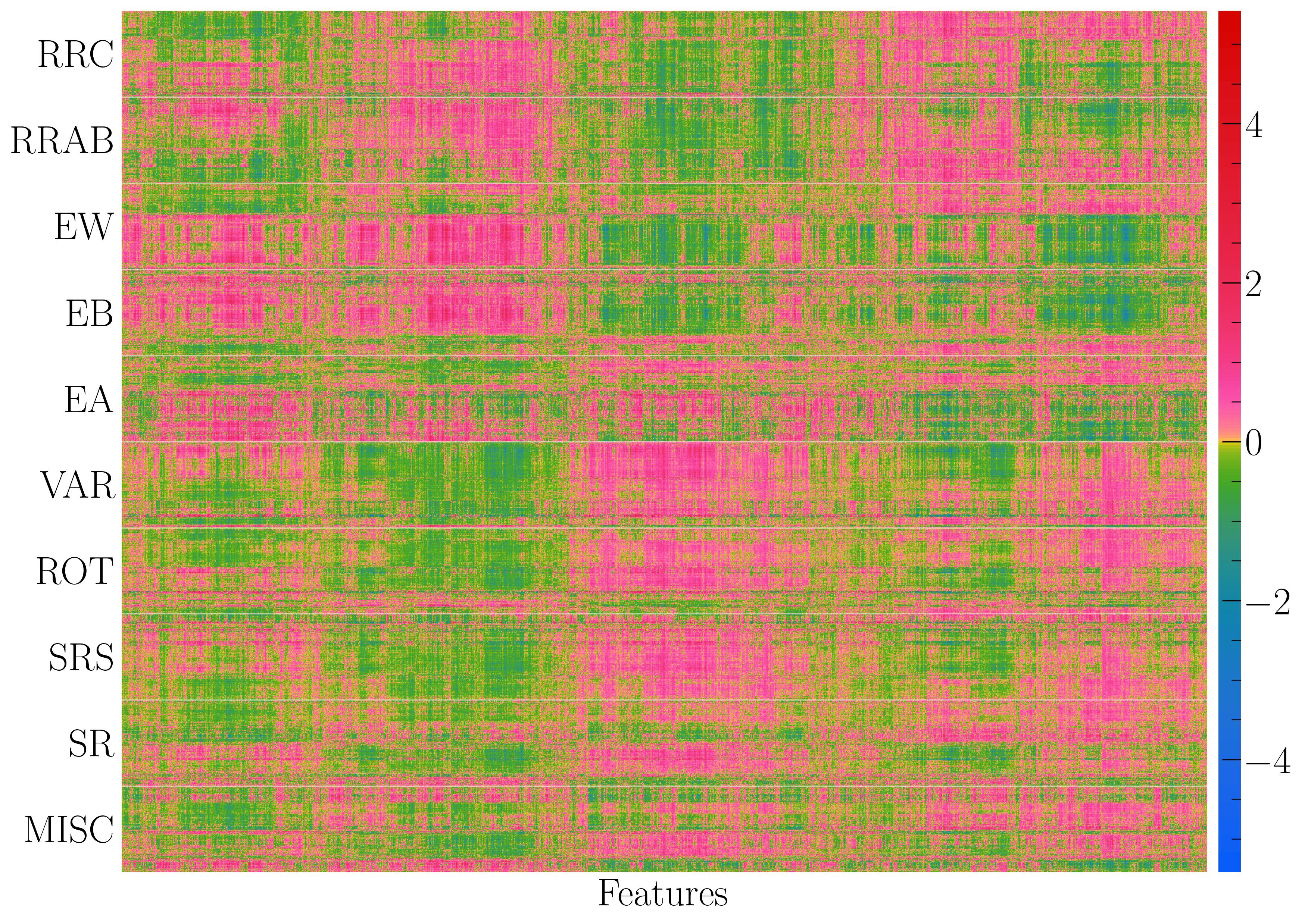}
  \caption{Heatmap of DINOv2 embeddings for $10\,000$ sources randomly selected from the ten most frequent VSX variability classes in our sample, with $1\,000$ sources per class. The color encodes feature values after centering the embeddings on the full-sample mean. The rows are grouped by class and ordered within each block by hierarchical clustering, and the feature columns are reordered to bring correlated features together. Morphologically related classes form contiguous blocks with smooth transitions, while the sharp EA--VAR boundary separates regular from irregular variability.}
  \label{fig:heatmap}
\end{figure*}

The same pattern appears in an embedding heatmap of $10\,000$ randomly sampled sources from the ten most frequent classes ($1\,000$ per class; Figure~\ref{fig:heatmap}). The rows are grouped by class and ordered within each block by hierarchical clustering. The columns are reordered to bring correlated features together. Morphologically related classes form contiguous blocks with continuous transitions, while the sharp discontinuity between the EA and VAR blocks marks the morphological divide between regular and irregular variability. The smooth within-block gradients demonstrate that the embeddings are expressive enough to resolve the intrinsic variations within each class.

\subsection{Class-conditional Isolation Forests}\label{subsec:isolation}

We used isolation forests \citep{liu2008} to establish a baseline anomaly ranking from the DINOv2 embeddings. An isolation forest is an ensemble of randomized decision trees, each of which recursively partitions the data by selecting a random feature and a random split value until every point is isolated or a maximum depth is reached. The method is based on the assumption that anomalies, being rare and occupying sparse regions of the feature space, require fewer random splits to be separated from the rest of the sample than objects residing in densely populated regions. For a single tree, the path length $h(x)$ is the number of splits along the branch connecting the root to the leaf at which source $x$ becomes isolated. Averaging $h(x)$ for a given source over all trees in the ensemble gives the expected path length $\mathrm{E}[h(x)]$, which we converted into the normalized anomaly score
\begin{equation}\label{eq:iforest_score}
  s(x,n) = 2^{-\mathrm{E}[h(x)] / c(n)},
\end{equation}
where $n$ is the subsample size and $c(n)$ is the average path length in a tree of size $n$. This transformation bounds the score between $0$ and $1$, with values near unity indicating anomalies and values $\lesssim0.5$ typical objects.

Applied directly to our full sample of variable stars, a single isolation forest would tend to flag all members of rare variability classes as anomalies, even when their light curves are typical for their type, simply because those classes occupy sparsely populated regions of the embedding space by construction. To avoid this, we trained a separate isolation forest for each VSX class containing at least $1\,000$ members, yielding $26$ class-specific models, and an additional forest for all remaining sources. Each forest consisted of $500$ trees grown on random subsamples of $256$ sources. Since the embeddings encode only the morphology of the light curves, we included $P$ and $\Delta m$ as additional input features.

We scored each source $x$ in our sample against all forests, resulting in $27$ separate anomaly scores for each source. We then converted these scores into percentile ranks within each forest and took the minimum across all ranks, yielding the baseline anomaly score $s^{(0)}(x)$. This cross-class minimum strategy ensured that a source was flagged as anomalous only if it appeared unusual relative to every variability type, filtering out common objects that were assigned to the wrong class.

\subsection{Agentic Active Learning Loop}\label{subsec:agentic}

Although isolation forests are effective at identifying outliers in high-dimensional data, they have no way of distinguishing between genuine astrophysical anomalies and data artifacts. This task has traditionally required human insight and domain knowledge, often applied in an active learning workflow for improved cost-efficiency of the labeling process \citep{lochner2021,etsebeth2024}. Here, we followed a similar approach but replaced human experts with agents powered by Gemini~3 Flash, a multimodal LLM capable of interpreting both text and images. Our motivation for doing so was twofold. First, multiple agents deployed in parallel can review anomaly candidates much faster than human experts while maintaining consistent evaluation criteria and avoiding fatigue. Second, anomaly detection provides an ideal testbed for evaluating how well the reasoning and visual capabilities of off-the-shelf models generalize to astronomy-specific tasks.

\begin{figure*}
  \centering
  \includegraphics[width=0.99\textwidth]{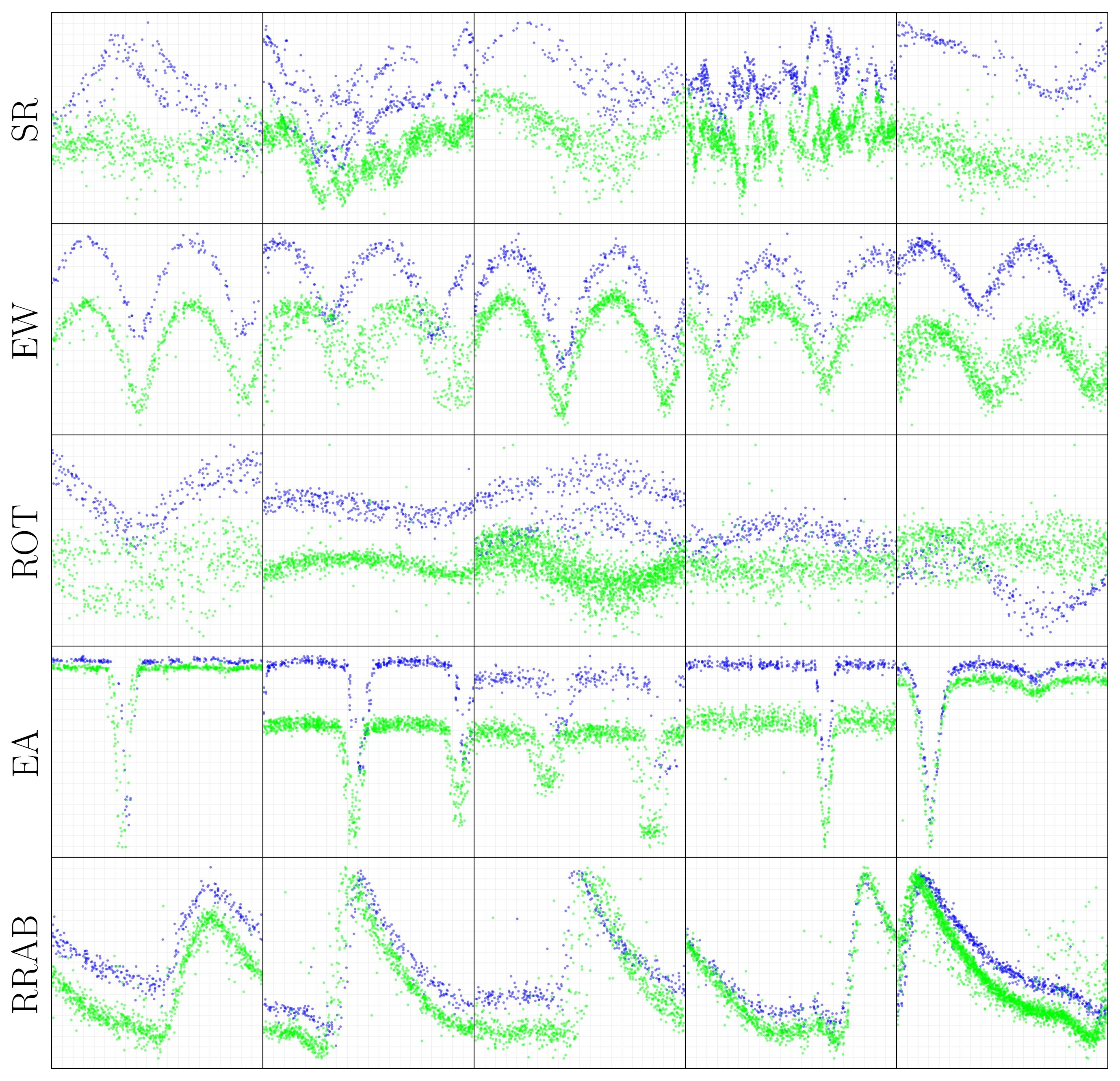}
  \caption{Examples of light-curve images submitted to Gemini~3 Flash agents for review. From top to bottom, the rows correspond to the five most frequent VSX classes in our sample: semiregular variables (SR), contact binaries (EW), rotating variables (ROT), eclipsing binaries (EA), and fundamental-mode RR~Lyrae variables (RRAB). The $V$ band is shown in blue, the $g$ band in green. The gray lines indicate a $20\times20$ reference grid used to aid visual inspection.}
  \label{fig:example_input}
\end{figure*}

\begin{figure*}
  \centering
  \includegraphics[width=0.85\textwidth]{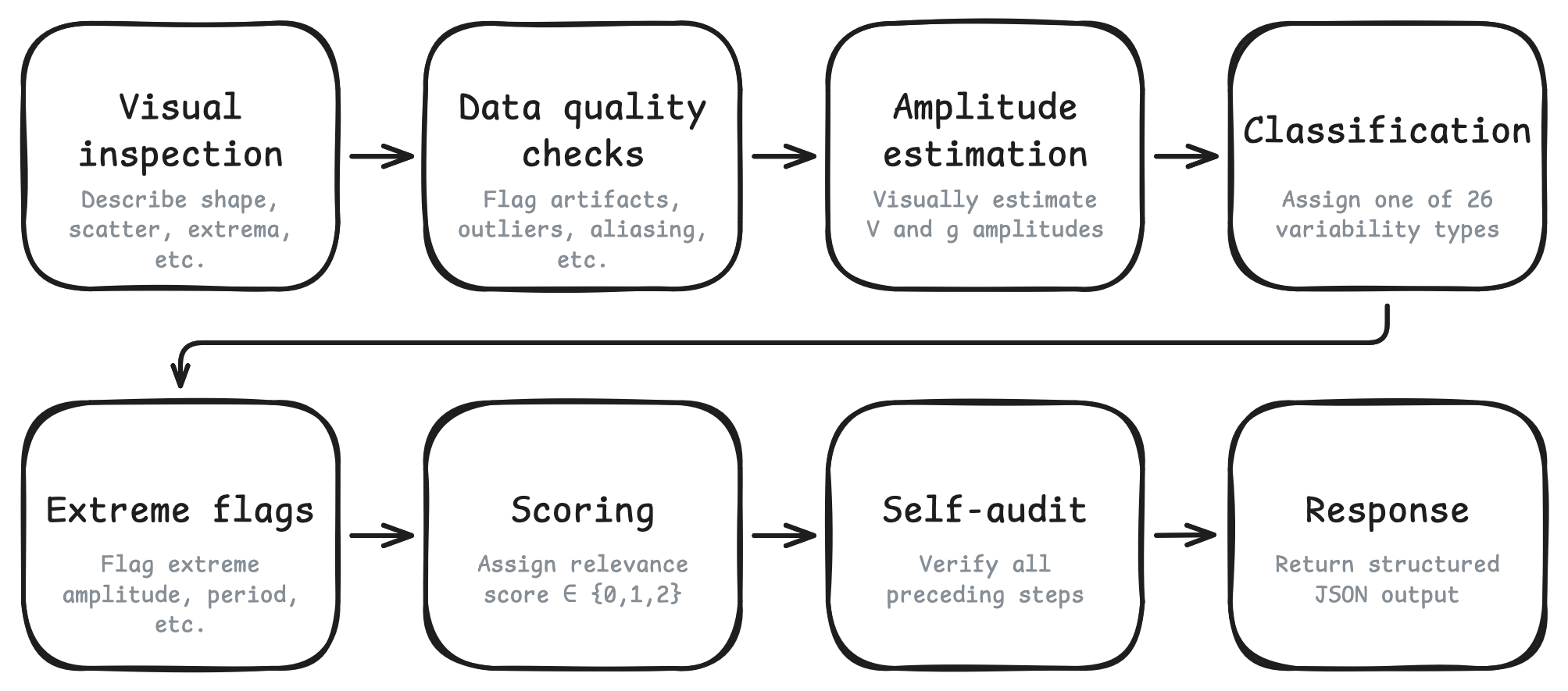}
  \caption{Schematic of the eight sequential steps that Gemini~3 Flash agents followed when evaluating anomaly candidates. The first five steps accumulate the evidence supporting the relevance score assigned in the sixth step. The self-audit then re-checks the verdict before the last step produces a structured JSON response. A condensed version of the prompt is given in Appendix~\ref{app:prompt}.}
  \label{fig:steps}
\end{figure*}

Because evaluating all sources in our sample with LLM agents would be too costly, each iteration in our active learning loop considered only the $100$ unlabeled sources with the highest anomaly scores. Each light curve in the batch was evaluated independently by a single Gemini~3 Flash agent with dynamic thinking enabled, allowing the model to allocate its own reasoning effort. The agent received a phase-folded light-curve image (same as the input to DINOv2 but overlaid with a gray reference grid; Figure~\ref{fig:example_input}), a concise metadata block (ASAS-SN ID, $P$, $\Delta m$, and VSX variability type), and a prompt instructing it to assess the light curve following the eight-step procedure shown in Figure~\ref{fig:steps}.

We designed the steps to mimic how human experts might approach similar tasks while mitigating the main failure modes of large language models: hallucinations and inconsistent evaluation criteria. The analysis begins with a visual inspection of the image, followed by a series of checks assessing the data quality and correctness of the period folding. The agents then estimate the amplitude of the light curve, classify the source as one of predefined variability types, flag extreme properties, and assign a relevance score of $0$ (not interesting), $1$ (potentially interesting), or $2$ (anomaly). Finally, they audit their reasoning before producing a structured JSON output.

A few of the design choices behind this procedure are worth highlighting. We treated amplitude estimation as a separate step because our experiments showed that incorporating it into other steps reduced the accuracy of the estimates. Instead of relying on VSX labels, we had the agents classify each source and override the VSX label in case of disagreement. The assignment of the relevance score $y\in\{0,1,2\}$ followed a strict priority order in which the score could be elevated above zero only after the data quality and folding checks had been cleared, ensuring that artifacts could never receive a positive score. Finally, we capped long-period variables at $y\leq1$, since large amplitudes and long periods are intrinsic to this class. We provide a condensed version of the prompt in Appendix~\ref{app:prompt}.

Once the agents had scored all sources in iteration $t$, we updated the anomaly score of each unlabeled source $u$ for iteration $t+1$ as
\begin{equation}
  s^{(t+1)}(u) = \left\{
  \begin{array}{ll}
      1.0, & \mbox{if } y_l = 2 \mbox{ for any } l \in \mathcal{N}_u \neq \emptyset, \\
      \langle y_l / 2 \rangle_{l \in \mathcal{N}_u}, & \mbox{if } y_l \leq 1 \mbox{ for all } l \in \mathcal{N}_u \neq \emptyset, \\
      s^{(0)}(u), & \mbox{if } \mathcal{N}_u = \emptyset,
  \end{array} \right.
\end{equation}
where the angle brackets denote an average and $\mathcal{N}_u = \{\,l \in \mathcal{L}^{(t)} : \cos\theta_{ul} > 0.90\,\}$ is the neighborhood of $u$ comprising all members of the labeled set $\mathcal{L}^{(t)}$ accumulated up to iteration $t$ whose cosine similarity to $u$ exceeds $0.90$. This propagation mechanism promoted sources with anomaly neighbors to the top of the ranking and demoted those surrounded by low-scoring neighbors to the bottom, while sources with no labeled neighbors retained their baseline anomaly scores.

To balance exploration of new regions with exploitation of known anomaly clusters, we capped the number of elevated anomaly-neighbor sources reviewed in each iteration at half the batch size. We ran the active learning loop for a total of $50$ iterations, yielding $5\,000$ labeled sources, which then served as the training set for a logistic regression classifier.

\subsection{Logistic Regression Classification}\label{subsec:logreg}
The neighbor-based propagation of Section~\ref{subsec:agentic} can only elevate sources whose similarity to an already-identified anomaly exceeds the conservative $0.90$ limit, leaving genuine anomalies that fall just below this threshold out of reach. To recover them, we trained a logistic regression classifier on the relevance labels accumulated across all iterations and used it to score the entire unlabeled pool directly, without imposing a hard similarity limit.

Before training, we re-centered the embeddings on the labeled subset alone to avoid leakage from the unlabeled pool, then augmented each vector with $P$ and $\Delta m$, and standardized the features. We collapsed the three-level scale into a binary label ($y=2$ positive, $y\in\{0,1\}$ negative) and trained with balanced class weights to offset the class imbalance. We used the classifier to score the remaining unlabeled pool and passed up to $100$ sources with anomaly probability above $0.90$ to the agents for labeling, utilizing the same prompt and scoring protocol as in the main loop (Figure~\ref{fig:steps}).

\subsection{Multi-Agent Consensus Review}\label{subsec:review}
Even when provided with detailed instructions, LLM agents can still make mistakes. For example, they may misjudge a morphological feature, misapply a class-specific threshold, or overlook a data quality issue that should disqualify a candidate. While active learning is effective at prioritizing sources for review, it does not reduce errors in individual agent assessments, allowing those errors to propagate into the anomaly candidate list. To mitigate this, we subjected all sources that received $y\in\{1,2\}$ in the main iterative loop or the logistic regression step to a multi-agent consensus review, concentrating resources on the objects most likely to be genuine anomalies.

We submitted each candidate to five independent Gemini~3 Flash agents, each following the same eight-step prompt as in the main iterative loop (Figure~\ref{fig:steps}). The final label was assigned by majority vote, breaking ties in favor of the lower relevance class to be conservative. Since the evaluations were fully independent, the mean score across the five agents provides a continuous measure of relevance, while the variance of the scores yields a measure of confidence in the final assignment. We preserved the full reasoning chains from all five agents, ensuring that each labeling decision remains fully auditable.

\section{Results}\label{sec:results}

We applied the pipeline described in Section~\ref{sec:methods} to our sample of over $370\,000$ periodic variable stars from ASAS-SN. In Section~\ref{subsec:stats}, we report the discovery statistics and the results of the multi-agent consensus review. We present our final catalog of anomalies and potentially interesting objects in Section~\ref{subsec:anomalies}, before assessing the impact of adaptive re-ranking in Section~\ref{subsec:adaptive_reranking}.

\subsection{Discovery Statistics}\label{subsec:stats}

The main agentic iterative loop followed by the logistic regression step, which we collectively refer to as the single-agent stages, yielded a total of $5\,100$ labeled light curves. Out of these, $233$ sources were assigned $y=1$ (potentially interesting) and $43$ were assigned $y=2$ (anomaly). We subjected all $276$ sources to a multi-agent consensus review, where each candidate was evaluated by five independent Gemini~3 Flash agents. Of the $43$ sources initially flagged as anomalies, only $20$ retained the classification, while $9$ were downgraded to potentially interesting and $14$ were reclassified as not interesting. Conversely, $4$ sources originally rated as potentially interesting were promoted to anomaly status after the consensus review, while $85$ were demoted to not interesting. We show the full label-transition matrix in Table~\ref{tab:transition}. Overall, the consensus review demoted $99$ of the $276$ flagged sources to not interesting, resulting in our final catalog of $24$ anomalies and $153$ potentially interesting objects. 

\begin{deluxetable}{lcccc}
\setlength{\tabcolsep}{10pt}
\tablecaption{Label transitions between the single-agent stages (rows) and the multi-agent consensus review (columns). The review demoted $99$ of the $276$ positive single-agent labels to not interesting, yielding our final catalog of $24$ anomalies and $153$ potentially interesting objects.\label{tab:transition}}
\tablewidth{0pt}
\tablehead{
  \colhead{} & \multicolumn{3}{c}{Consensus review} & \colhead{} \\
  \cline{2-4}
  \colhead{Single-agent} & \colhead{Anomaly} & \colhead{Potentially interesting} & \colhead{Not interesting} & \colhead{Total}
}
\startdata
Anomaly                 & $20$ & $9$   & $14$ & $43$ \\
Potentially interesting & $4$  & $144$ & $85$ & $233$ \\
\hline
Total                   & $24$ & $153$ & $99$ & $276$ \\
\enddata
\end{deluxetable}

\begin{figure}
  \centering
  \includegraphics[width=\columnwidth]{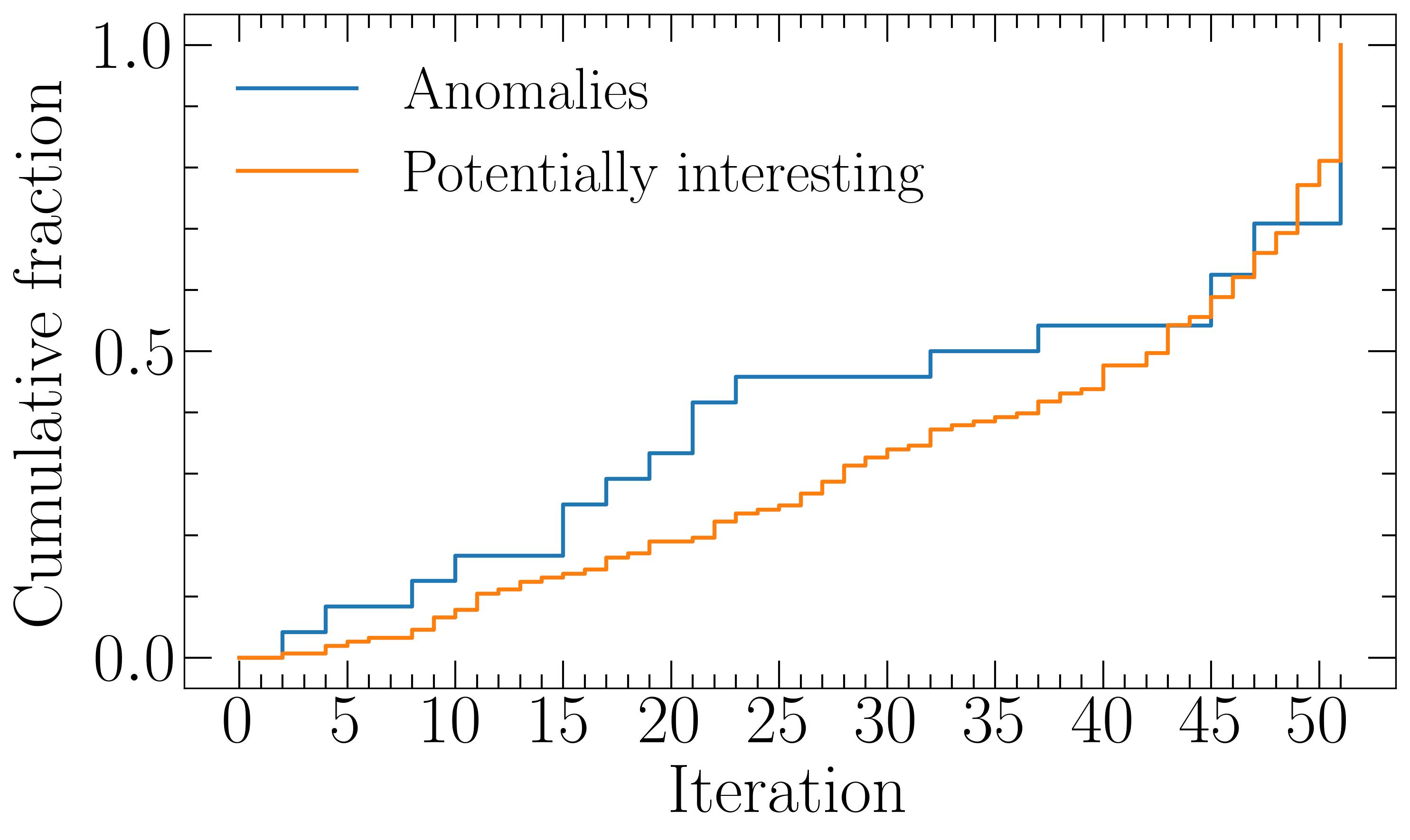}
  \caption{Cumulative fractions of the anomalies and the potentially interesting objects from our final catalog as a function of the number of iterations needed to identify them. The jump at iteration~$51$ corresponds to the logistic regression step, which contributed $29\%$ of the anomalies and $19\%$ of the potentially interesting objects to the catalog.}
  \label{fig:cumulative}
\end{figure}

In Figure~\ref{fig:cumulative}, we show the cumulative fractions of the anomalies and the potentially interesting objects from our final catalog as a function of the number of iterations it took to identify them. After $50$ iterations, approximately $71\%$ of the anomalies and $81\%$ of the potentially interesting objects had been identified. The remaining sources were surfaced in the logistic regression step (iteration $51$), which contributed $7$ anomalies and $29$ potentially interesting objects, extending the search beyond the reach of similarity-based label propagation.

\begin{figure}
  \centering
  \includegraphics[width=\columnwidth]{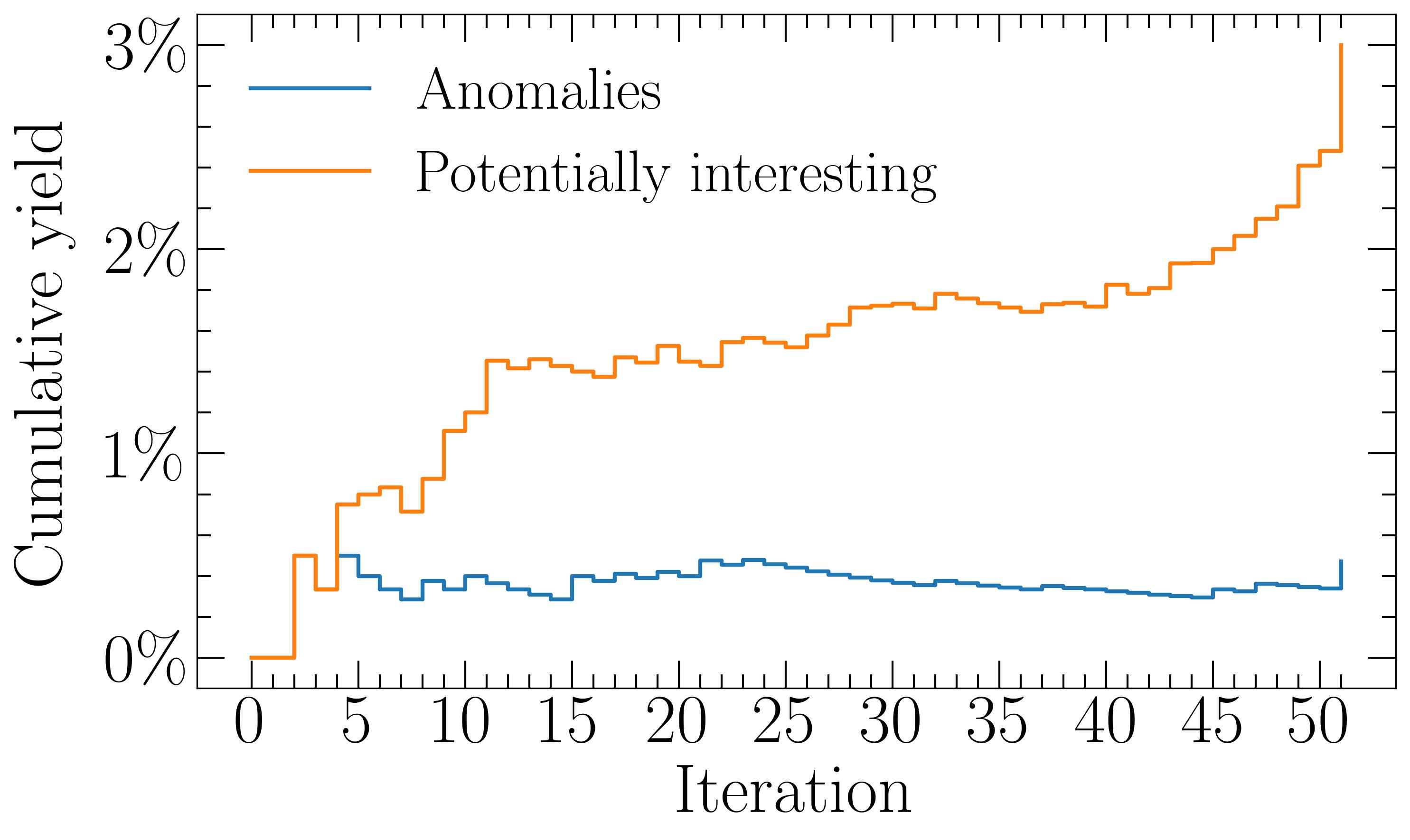}
  \caption{Cumulative yield, computed as the fraction of all sources labeled up to a given iteration that ended up in our final catalog, shown separately for anomalies and potentially interesting objects. The anomaly curve stays nearly flat at approximately $0.4\%$ throughout iterations $2$--$50$, while the cumulative yield of the potentially interesting objects rises steadily from about $1\%$ to roughly $2.5\%$ between iterations $10$ and $50$, reflecting the different efficiency with which the active learning loop enriches successive batches in the two cases. The jump in both curves at iteration~$51$ corresponds to the logistic regression step.}
  \label{fig:yield}
\end{figure}

The cumulative yield curves in Figure~\ref{fig:yield} show the fractions of all sources labeled up to a given iteration that ended up in our final catalog as anomalies and potentially interesting objects. The anomaly curve remains nearly flat at roughly $0.4\%$ throughout iterations $2$--$50$, while the yield of the potentially interesting objects rises steadily from approximately $1\%$ to $2.5\%$ between iterations $10$ and $50$. The reason is that in the latter case both similarity-based elevation and demotion of uninteresting objects contribute comparably to the enrichment of each successive batch, whereas for anomalies, which are naturally scarce and isolated, elevation is much less effective. Despite this, both curves show a jump at iteration~$51$, reflecting the increased yield of the logistic regression step.

Table~\ref{tab:gemini_class} breaks the $276$ flagged sources down by majority-vote variability class. EA-type eclipsing binaries, contact binaries, and RRab variables (fundamental-mode RR~Lyrae) together account for $87.5\%$ of the anomalies, with Type~II Cepheids and a single dwarf nova making up the rest. The potentially interesting objects follow a similar distribution, with the same top three classes comprising $86.9\%$ of them. The remaining sources span many classes, including EB-type eclipsing binaries, rotating variables, and semiregular variables. This long tail most likely reflects the larger sample size of potentially interesting objects rather than any systematic difference between the two relevance categories.

The per-class demotion rates (DRs) in Table~\ref{tab:gemini_class} show the fraction of each variability class reclassified as not interesting during the consensus review. Among the three most frequent classes, the DRs range from $0.24$ to $0.38$, with EA-type eclipsing binaries exhibiting the lowest rate. Their sharp V-shaped eclipses against flat out-of-eclipse baselines likely made the light curves straightforward to interpret, resulting in the low DR. In contrast, EB-type eclipsing binaries, rotating variables, and semiregular variables were all demoted at rates exceeding $0.6$. Most EB-type variables were misclassified as contact binaries, which led to their periods and amplitudes being evaluated against the tighter ranges of that class, making otherwise typical values appear anomalous. Meanwhile, the noisy light curves of rotating and semiregular variables made it difficult for the agents to reliably estimate amplitudes and identify data quality issues, causing them to flag ordinary variables as anomalies. These failure modes can also be seen in an $n$-gram analysis of the rationales from the consensus review, where the vocabulary for the demoted sources splits between misflagged class-standard objects and overlooked data quality issues (Appendix~\ref{app:ngram}).

\begin{deluxetable}{lccccc}
\tablecaption{Consensus relevance labels and demotion rates by majority-vote variability class.\label{tab:gemini_class}}
\tablewidth{0pt}
\tablehead{
\colhead{Majority-Vote Class} &
\colhead{Anomalies} &
\colhead{Potentially Interesting} &
\colhead{Not Interesting} &
\colhead{Total} &
\colhead{DR}
}
\startdata
Eclipsing binary (EA) & $11$ & $60$ & $22$ & $93$ & $0.24$ \\
Contact binary & $7$ & $49$ & $35$ & $91$ & $0.38$ \\
RR Lyrae (RRab) & $3$ & $24$ & $10$ & $37$ & $0.27$ \\
Eclipsing binary (EB) & $0$ & $9$ & $16$ & $25$ & $0.64$ \\
Rotating variable & $0$ & $2$ & $6$ & $8$ & $0.75$ \\
Semiregular variable & $0$ & $2$ & $4$ & $6$ & $0.67$ \\
RR Lyrae (RRc) & $0$ & $3$ & $2$ & $5$ & $0.40$ \\
Type II Cepheid & $2$ & $2$ & $0$ & $4$ & $0.00$ \\
$\delta$~Scuti & $0$ & $1$ & $1$ & $2$ & $0.50$ \\
Classical Cepheid & $0$ & $1$ & $0$ & $1$ & $0.00$ \\
RV~Tauri & $0$ & $0$ & $1$ & $1$ & $1.00$ \\
Contact binary/Eclipsing binary (EB) & $0$ & $0$ & $1$ & $1$ & $1.00$ \\
Dwarf nova & $1$ & $0$ & $0$ & $1$ & $0.00$ \\
Unknown & $0$ & $0$ & $1$ & $1$ & $1.00$ \\
\hline
Total & $24$ & $153$ & $99$ & $276$ & $0.36$ \\
\enddata
\tablecomments{Sample comprises $276$ candidates initially flagged as ``Anomaly'' or ``Potentially Interesting'' in the single-agent stages. \textit{DR} (demotion rate) is the fraction of each class demoted to ``Not Interesting'' in consensus review. \textit{Majority-Vote Class} is assigned by majority vote of five agents in the consensus review. Slash (/) denotes a tie in the majority vote. Parenthetical labels denote a subtype, if any. Some agent-defined classes group multiple VSX labels (e.g., Type~II Cepheid groups CW, CWA, CWB). See our \href{https://github.com/milanpesta/aal_anomalies_asassn}{GitHub} repository for the full agent-to-VSX class mapping.}
\end{deluxetable}

\subsection{Anomalies}\label{subsec:anomalies}
\begin{figure*}
  \centering
  \gridline{
    \fig{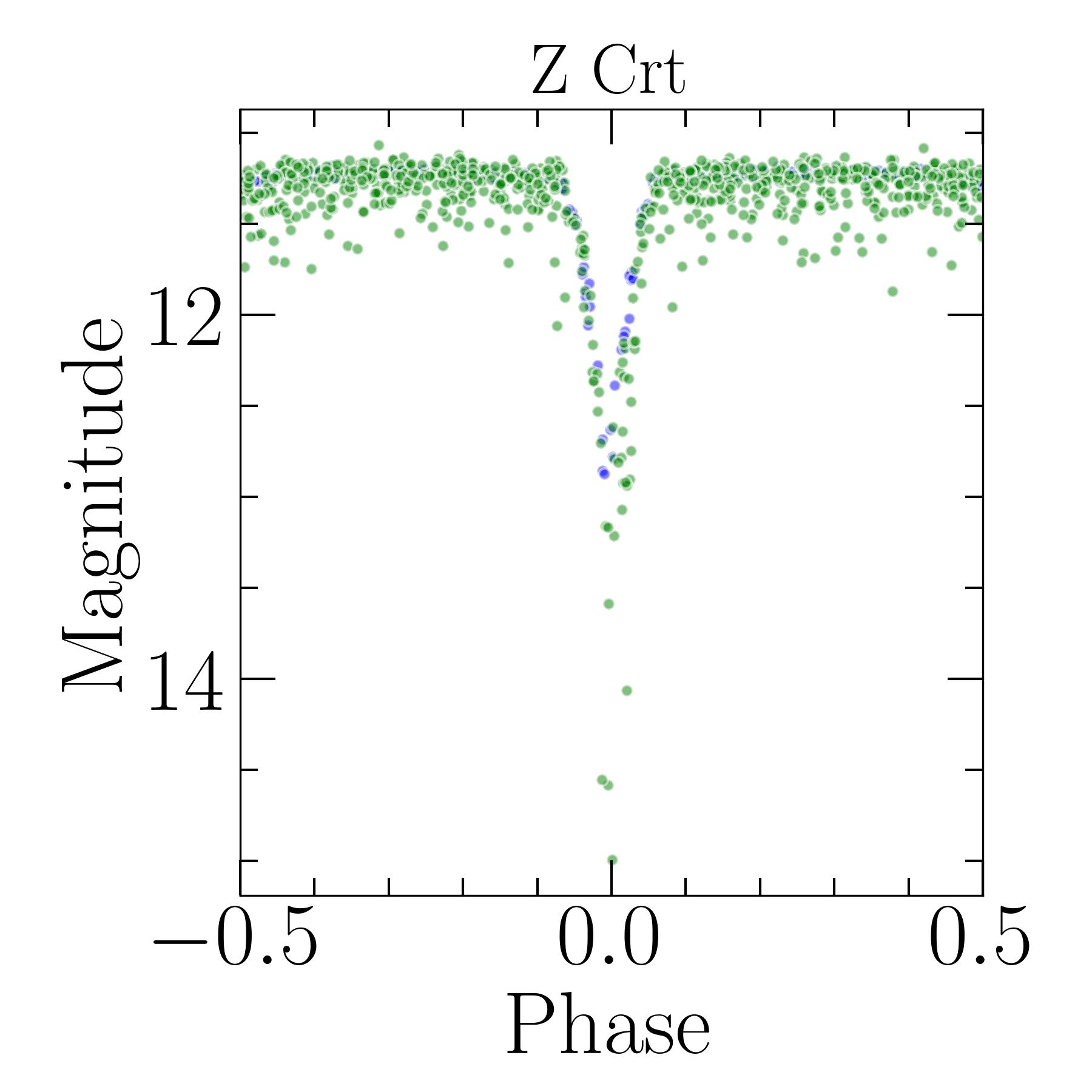}{0.245\textwidth}{}
    \fig{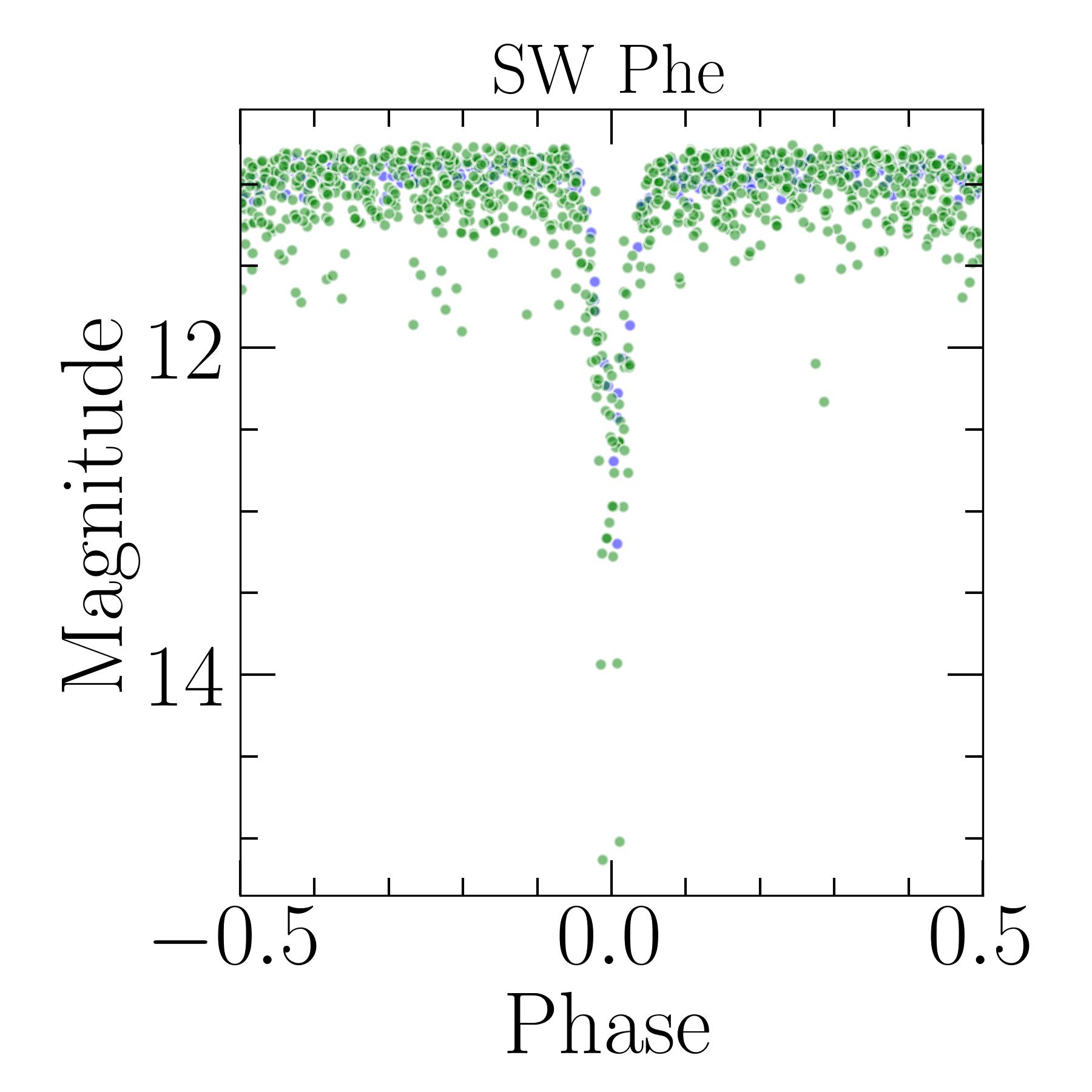}{0.245\textwidth}{}
    \fig{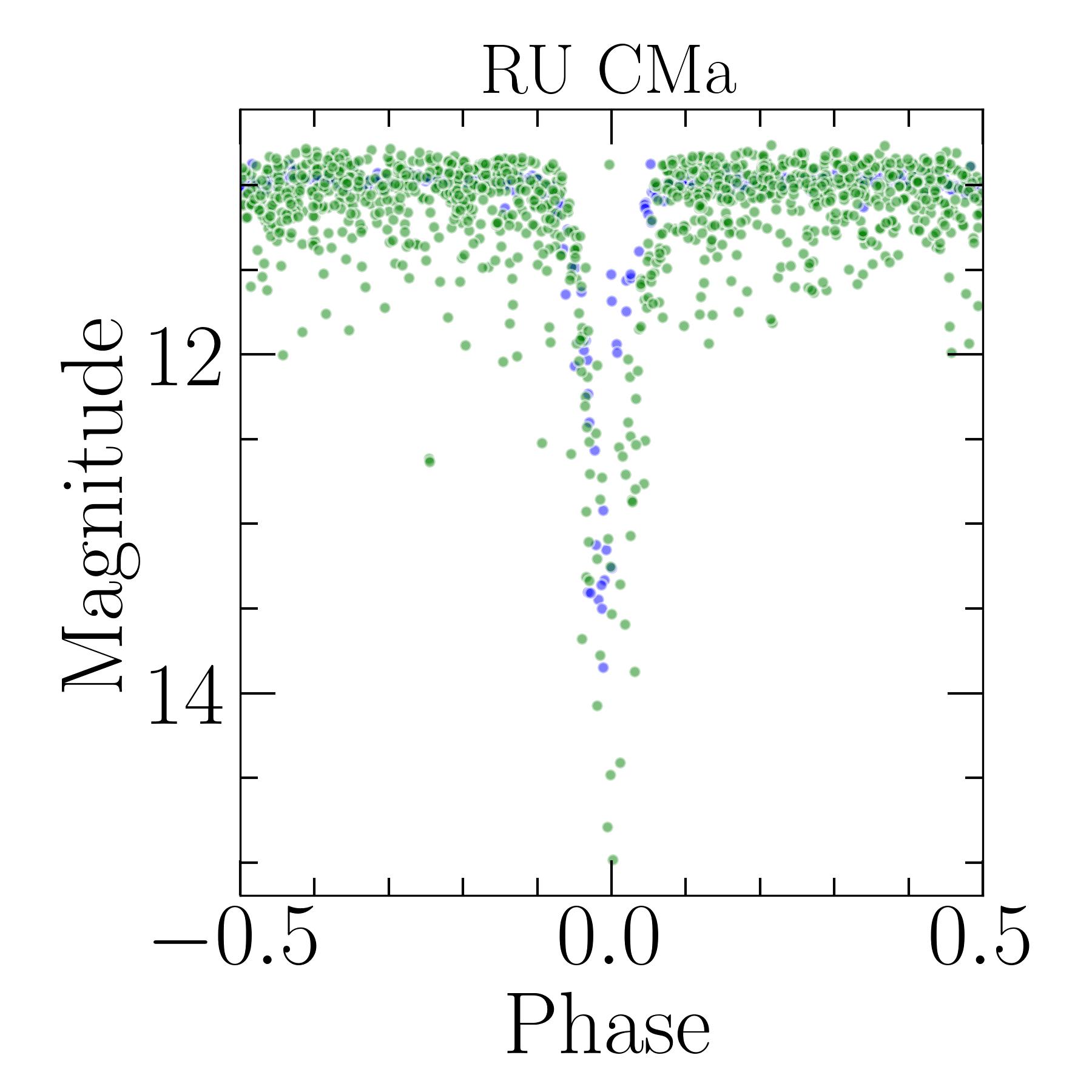}{0.245\textwidth}{}
    \fig{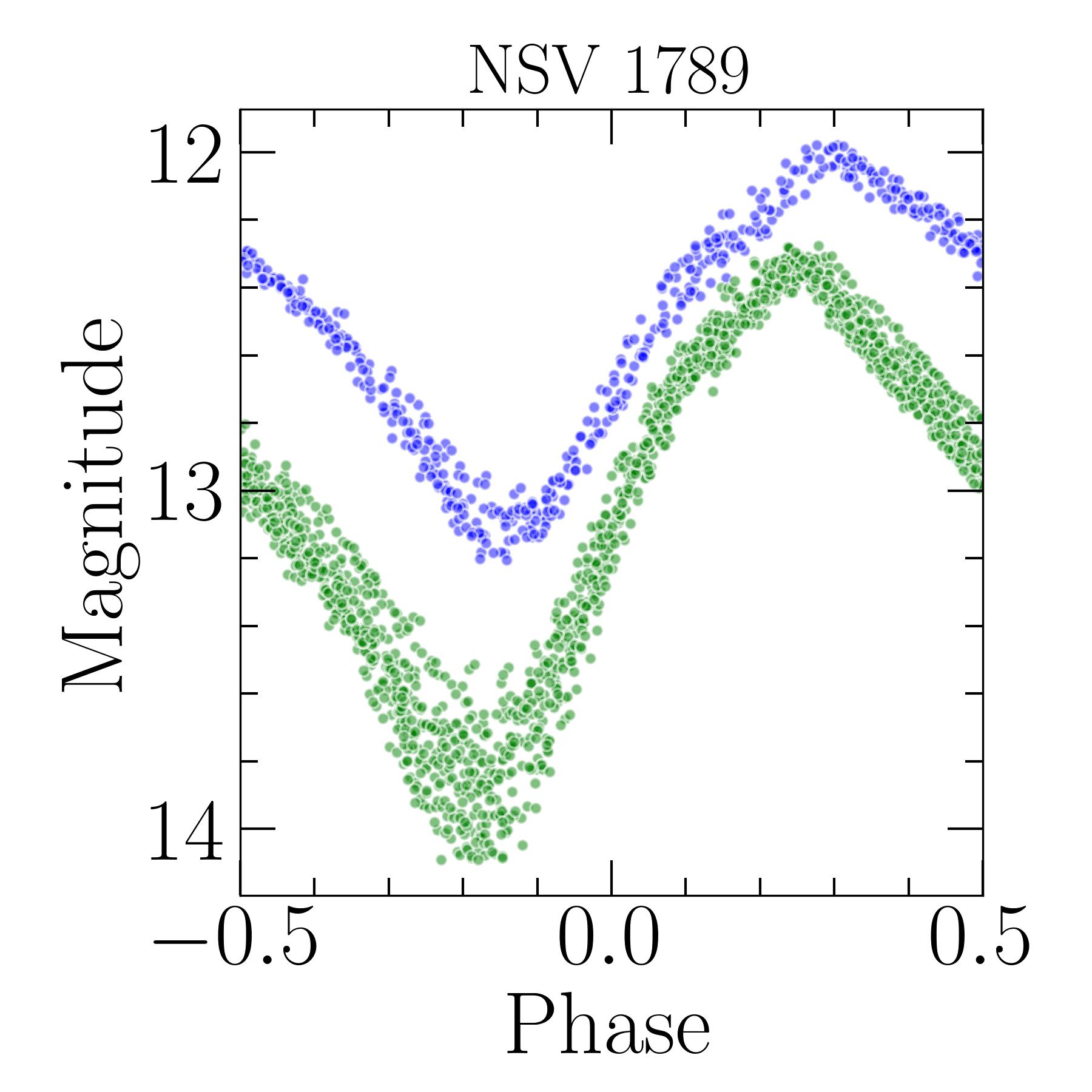}{0.245\textwidth}{}
  }
  \vspace{-2em}
  \gridline{
    \fig{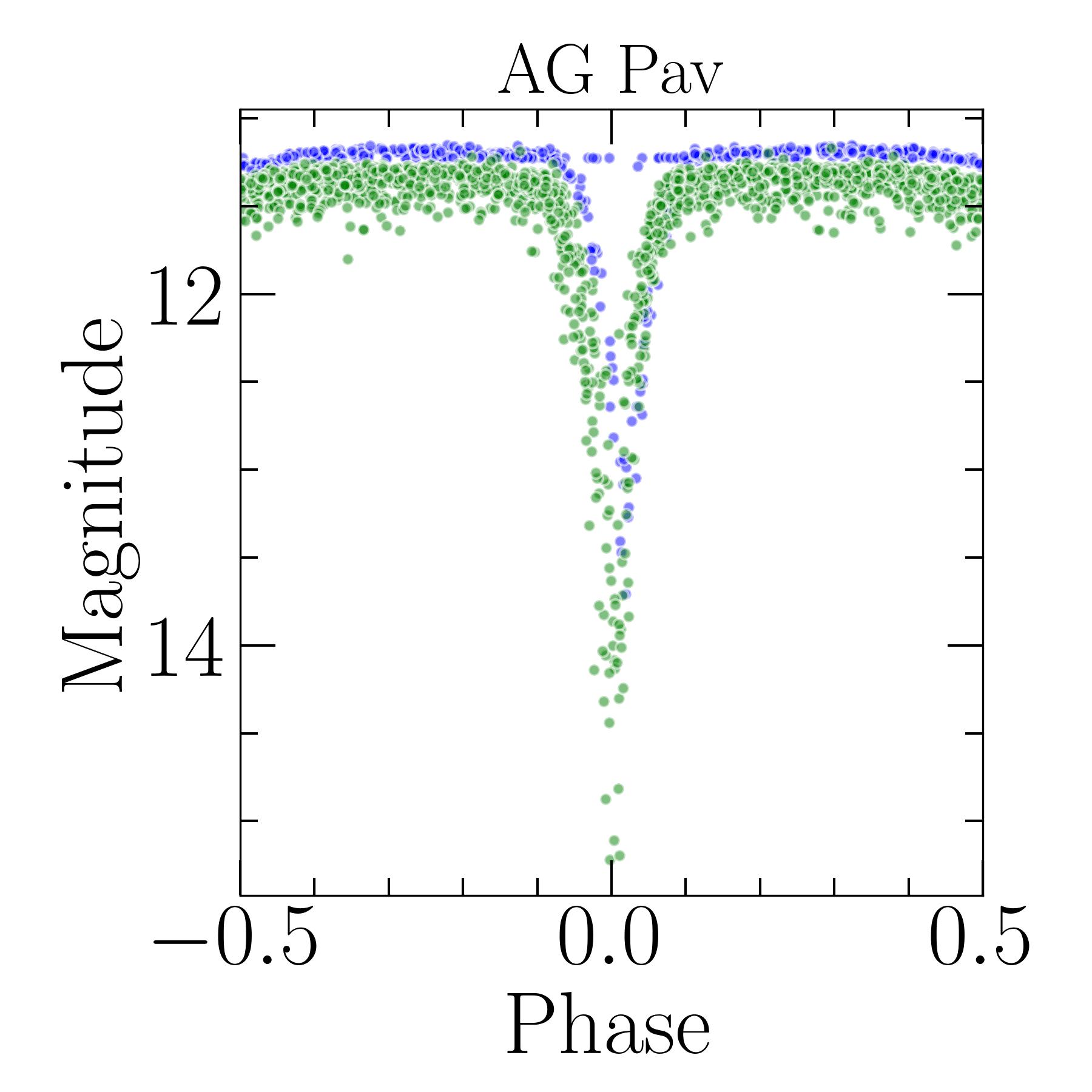}{0.245\textwidth}{}
    \fig{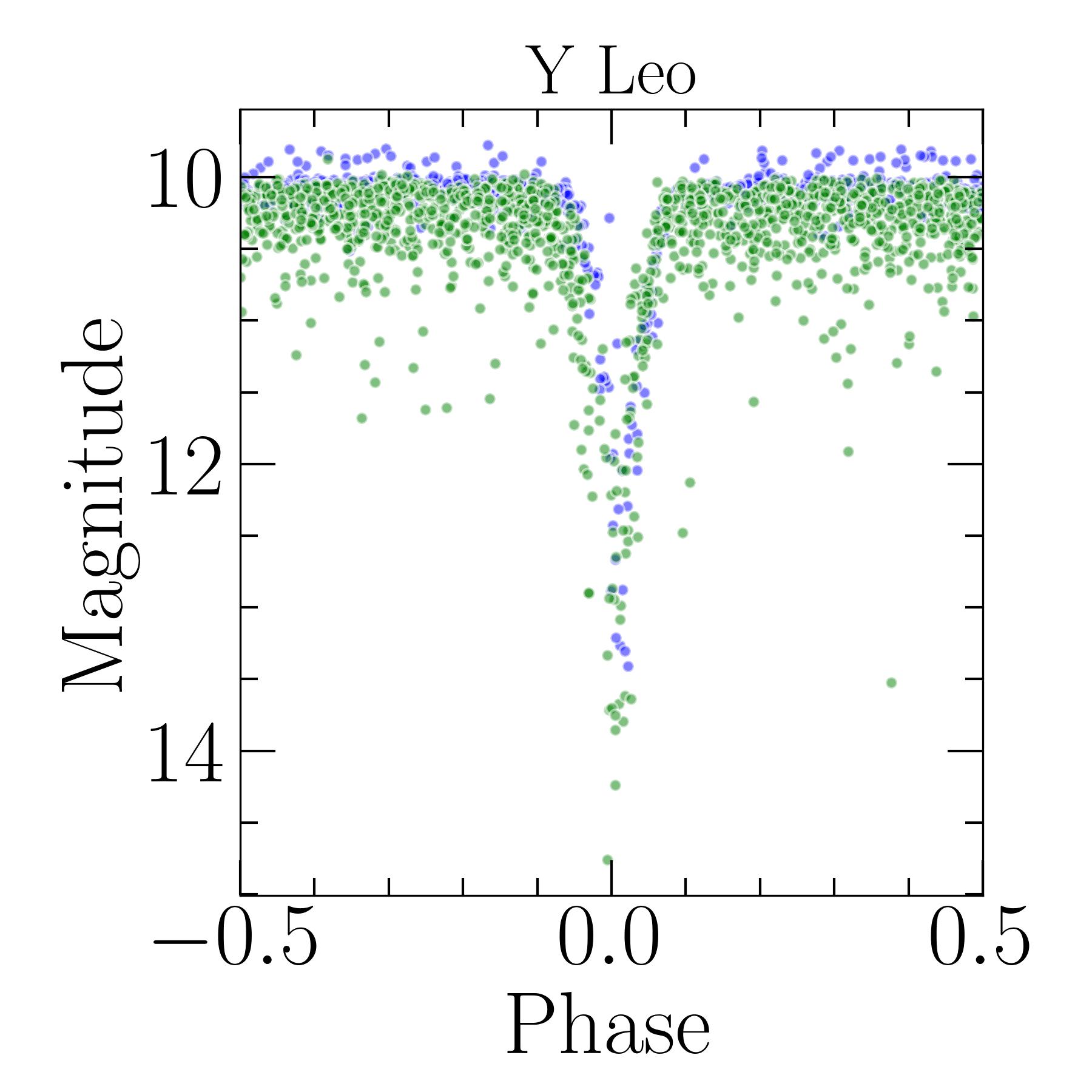}{0.245\textwidth}{}
    \fig{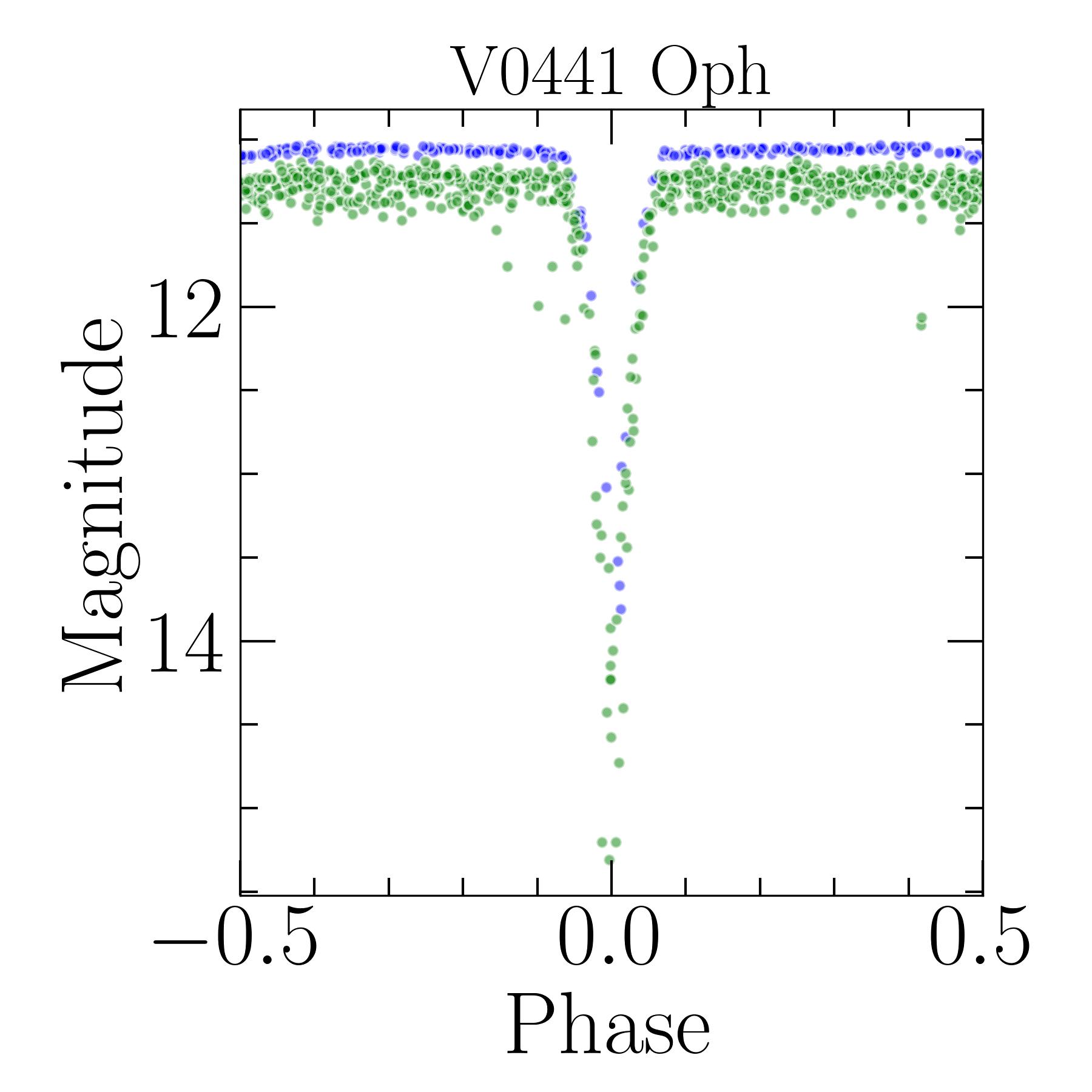}{0.245\textwidth}{}
    \fig{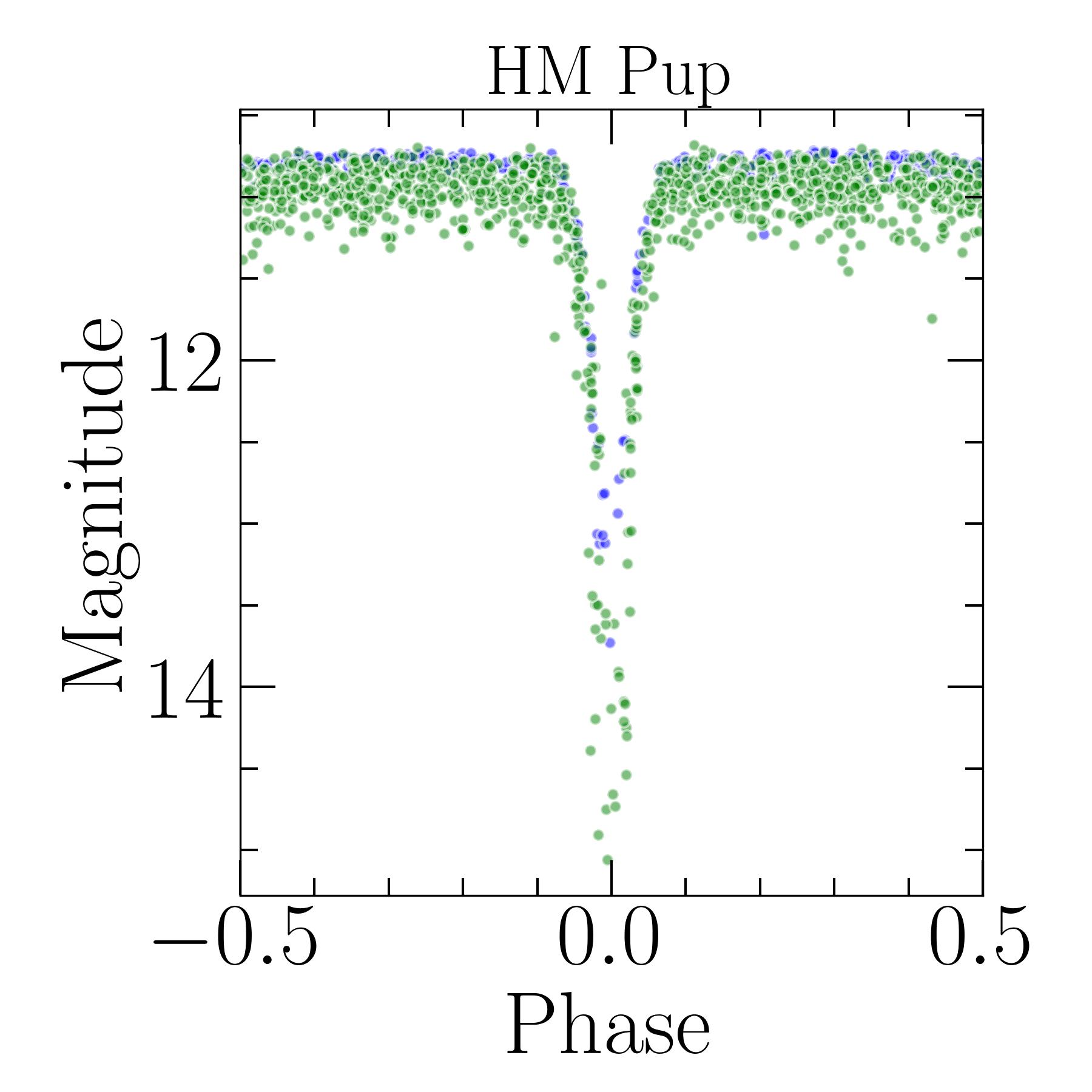}{0.245\textwidth}{}
  }
  \vspace{-2em}
  \gridline{
    \fig{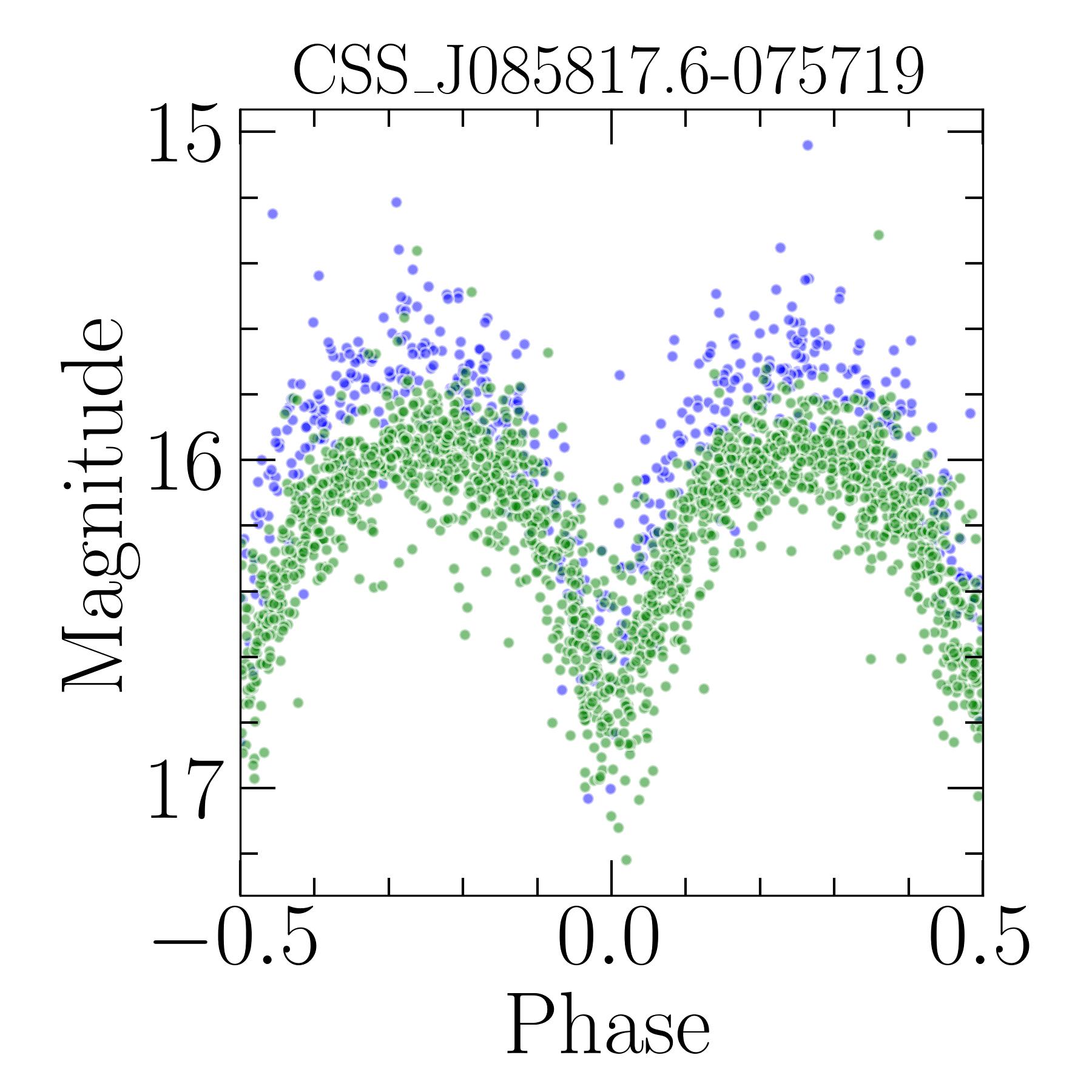}{0.245\textwidth}{}
    \fig{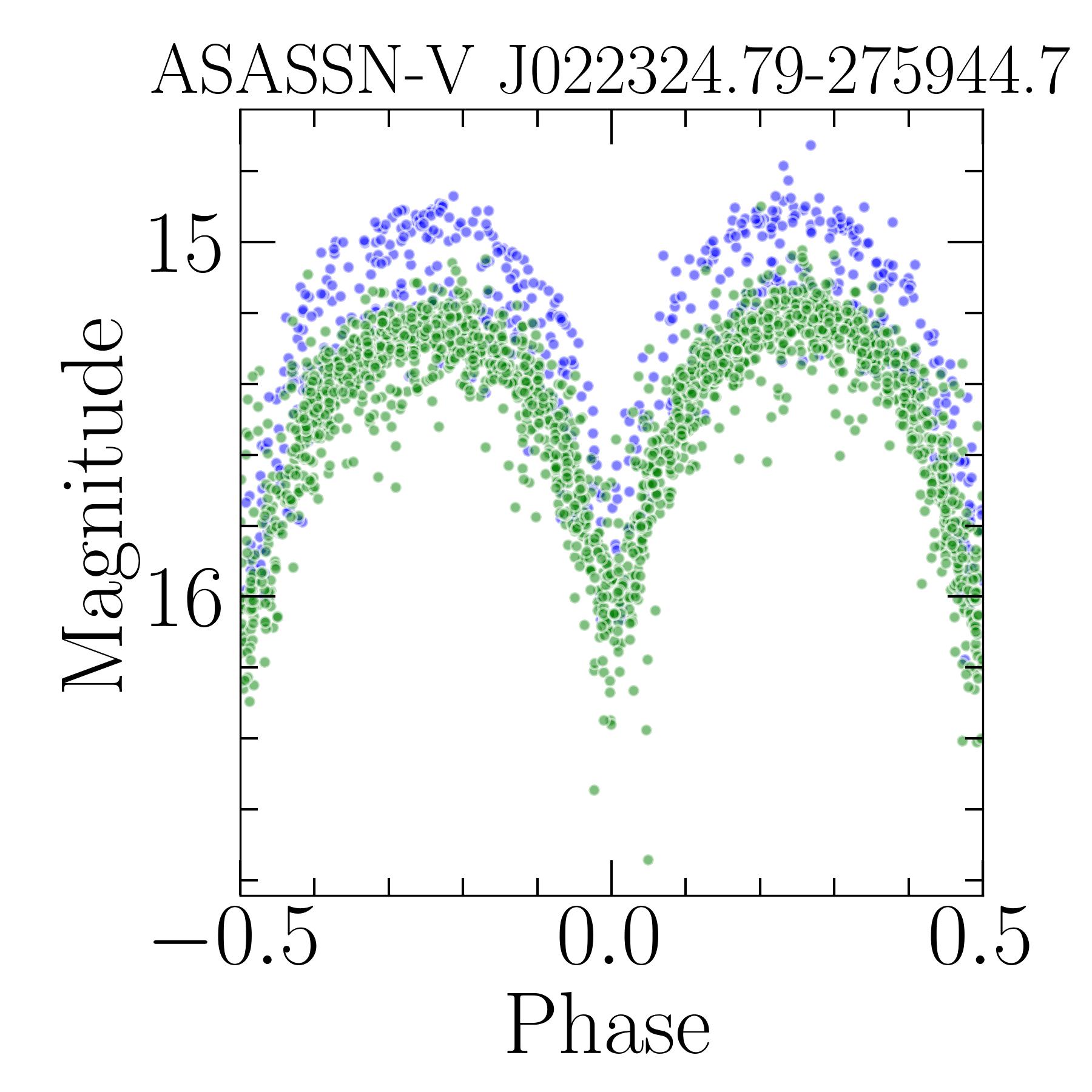}{0.245\textwidth}{}
    \fig{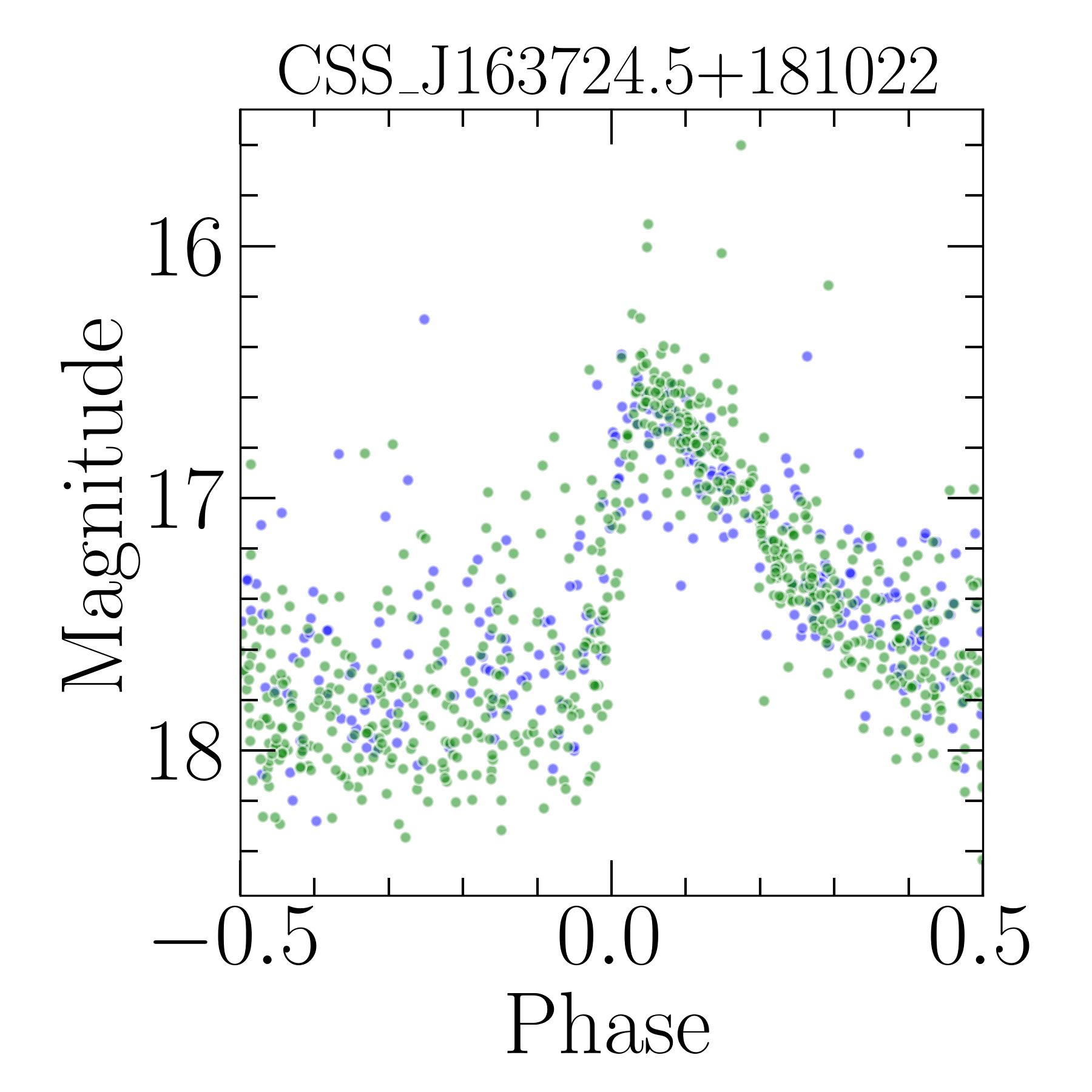}{0.245\textwidth}{}
    \fig{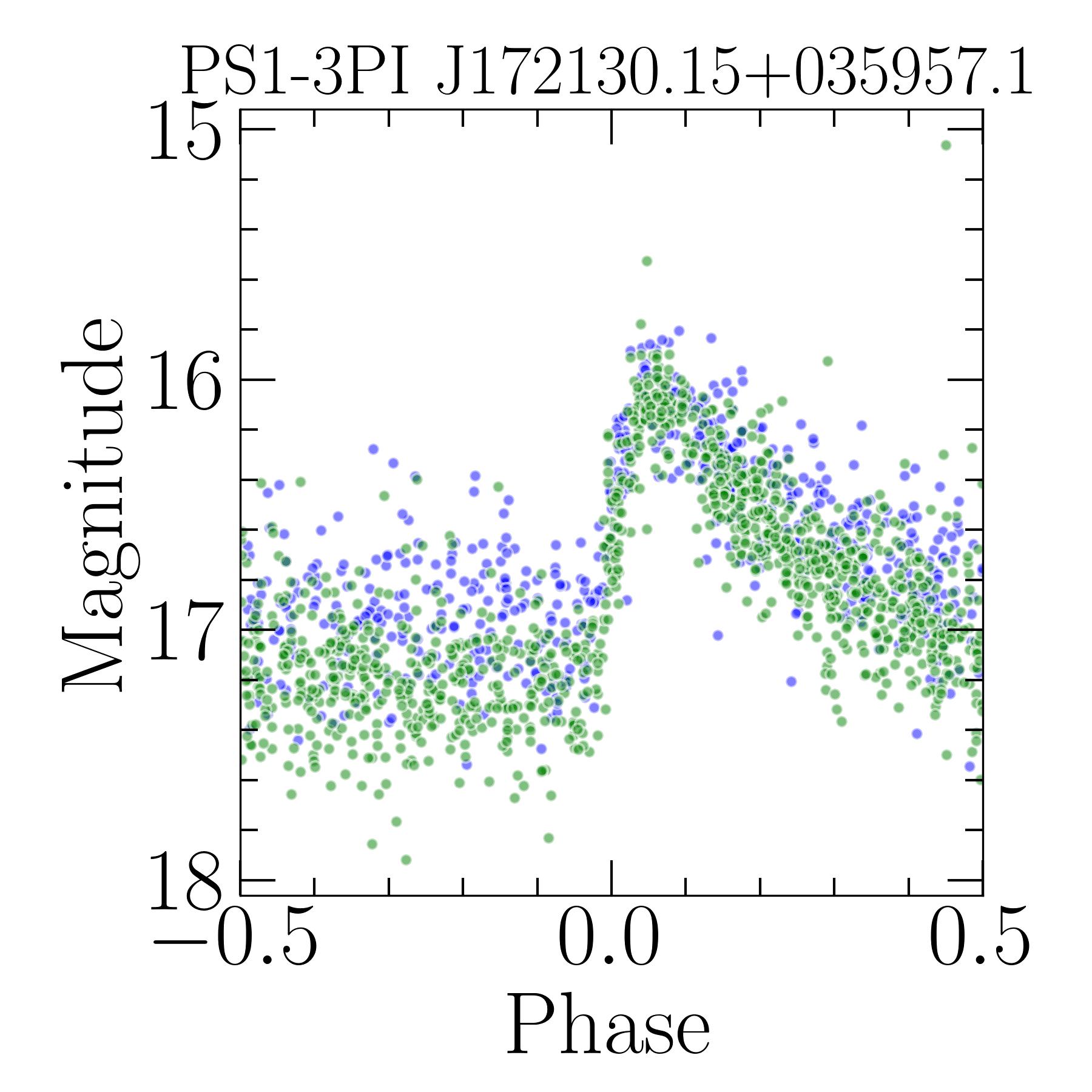}{0.245\textwidth}{}
  }
  \caption{Phase-folded light curves of the anomalies ranked $1$--$12$ in Table~\ref{tab:anomalies}, with each panel labeled by the source name. The first nine sources were unanimously voted anomalies by all five agents in the consensus review. The $V$-band is shown in blue, the $g$-band in green.}
  \label{fig:gallery_1}
\end{figure*}

\begin{figure*}
  \centering
  \gridline{
    \fig{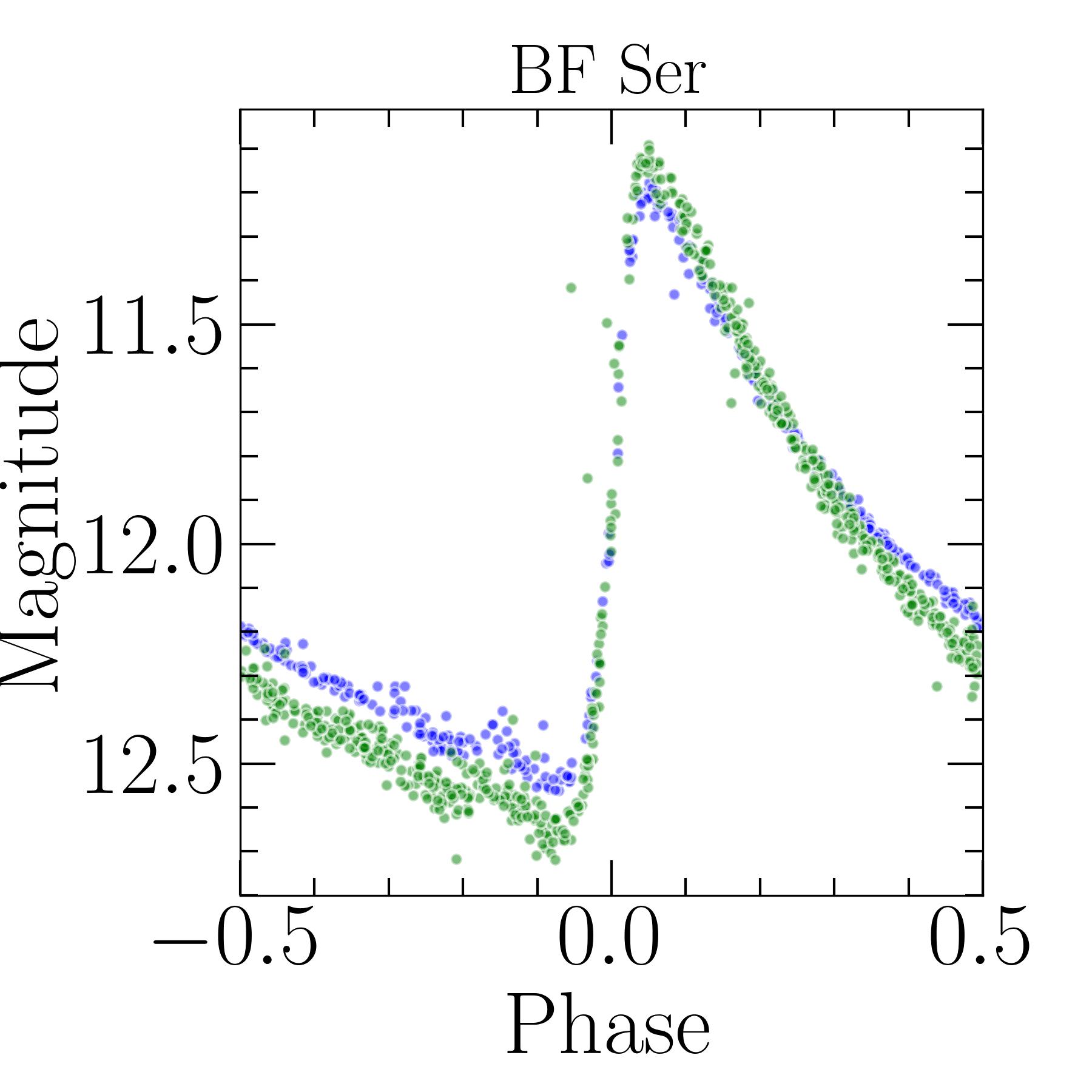}{0.245\textwidth}{}
    \fig{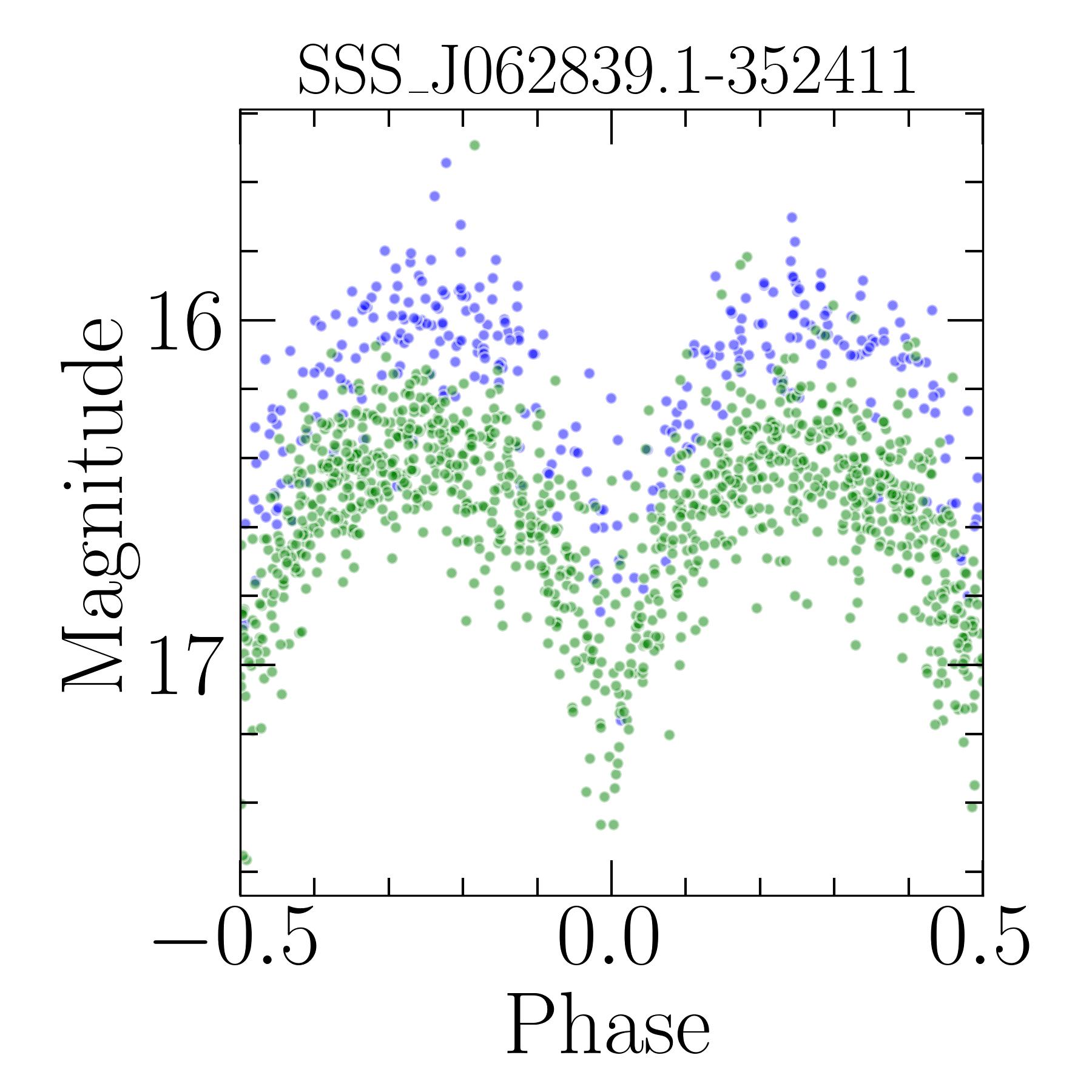}{0.245\textwidth}{}
    \fig{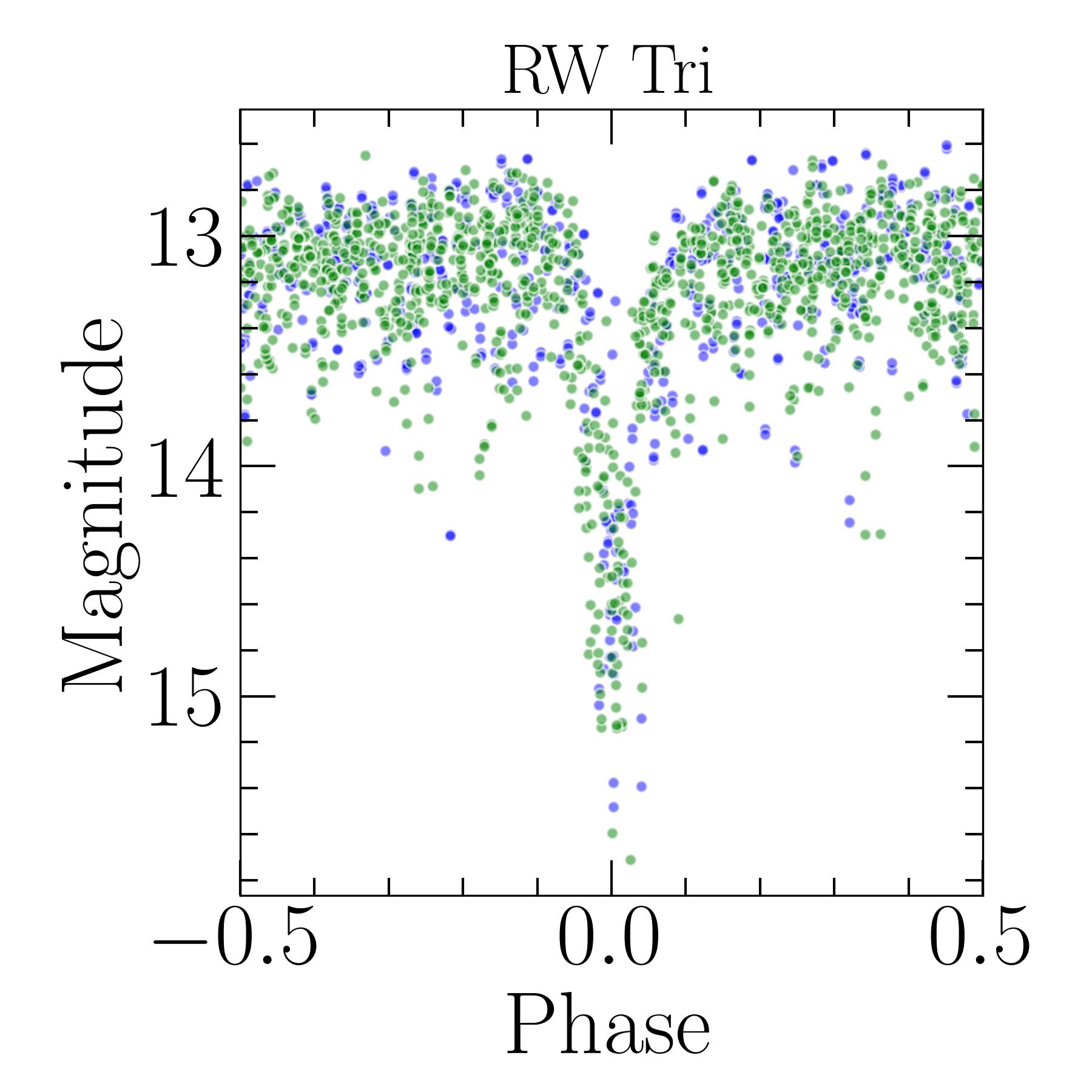}{0.245\textwidth}{}
    \fig{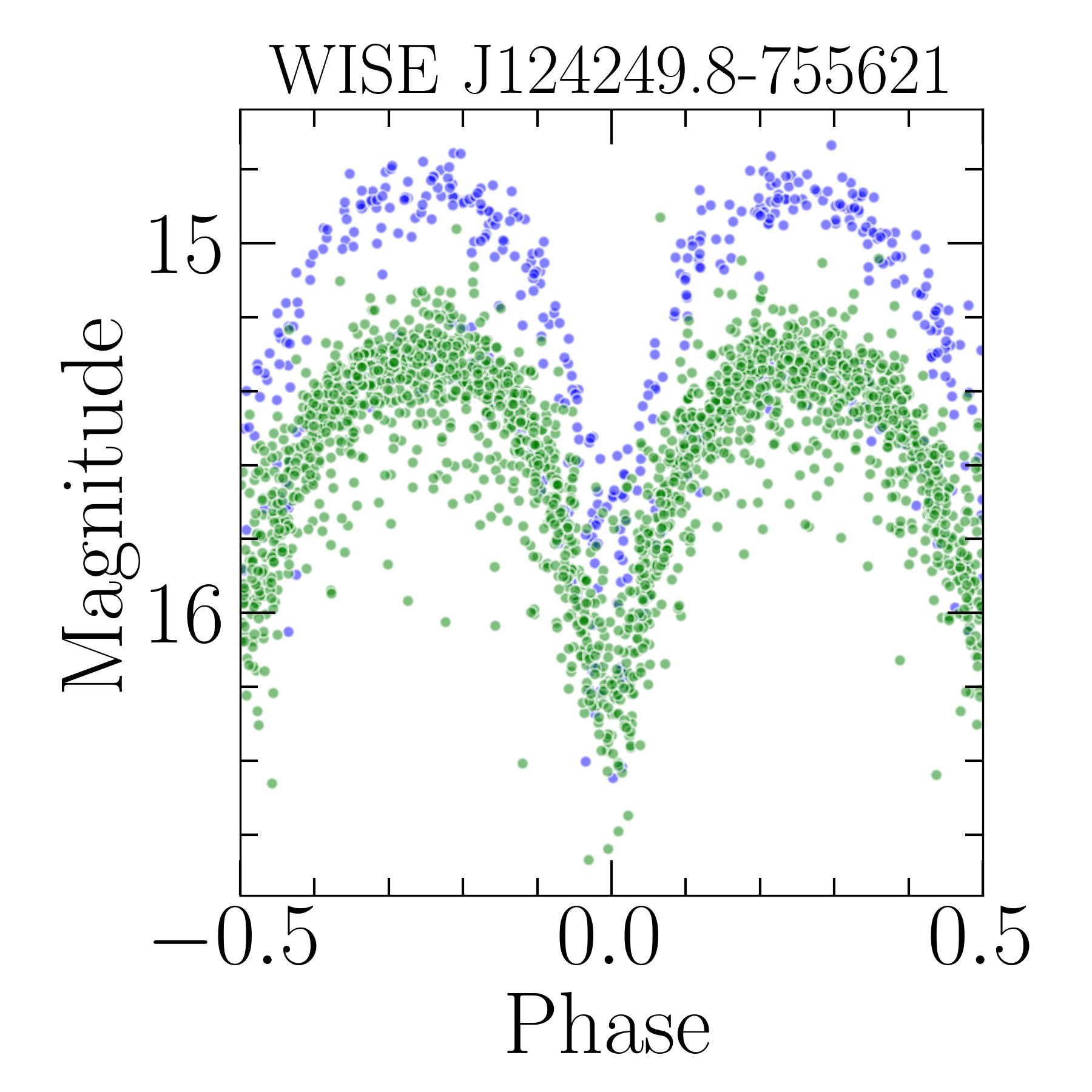}{0.245\textwidth}{}
  }
  \vspace{-2em}
  \gridline{
    \fig{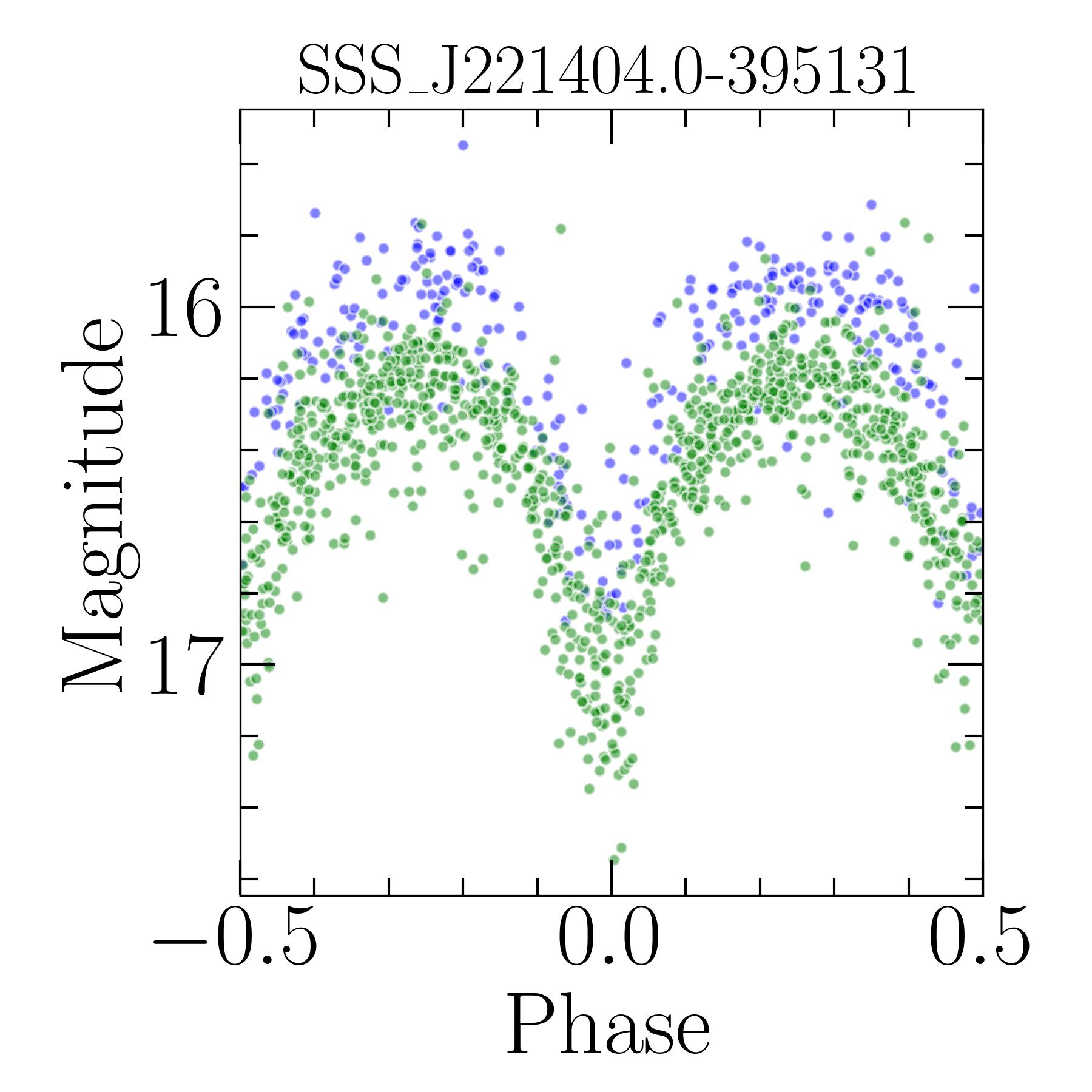}{0.245\textwidth}{}
    \fig{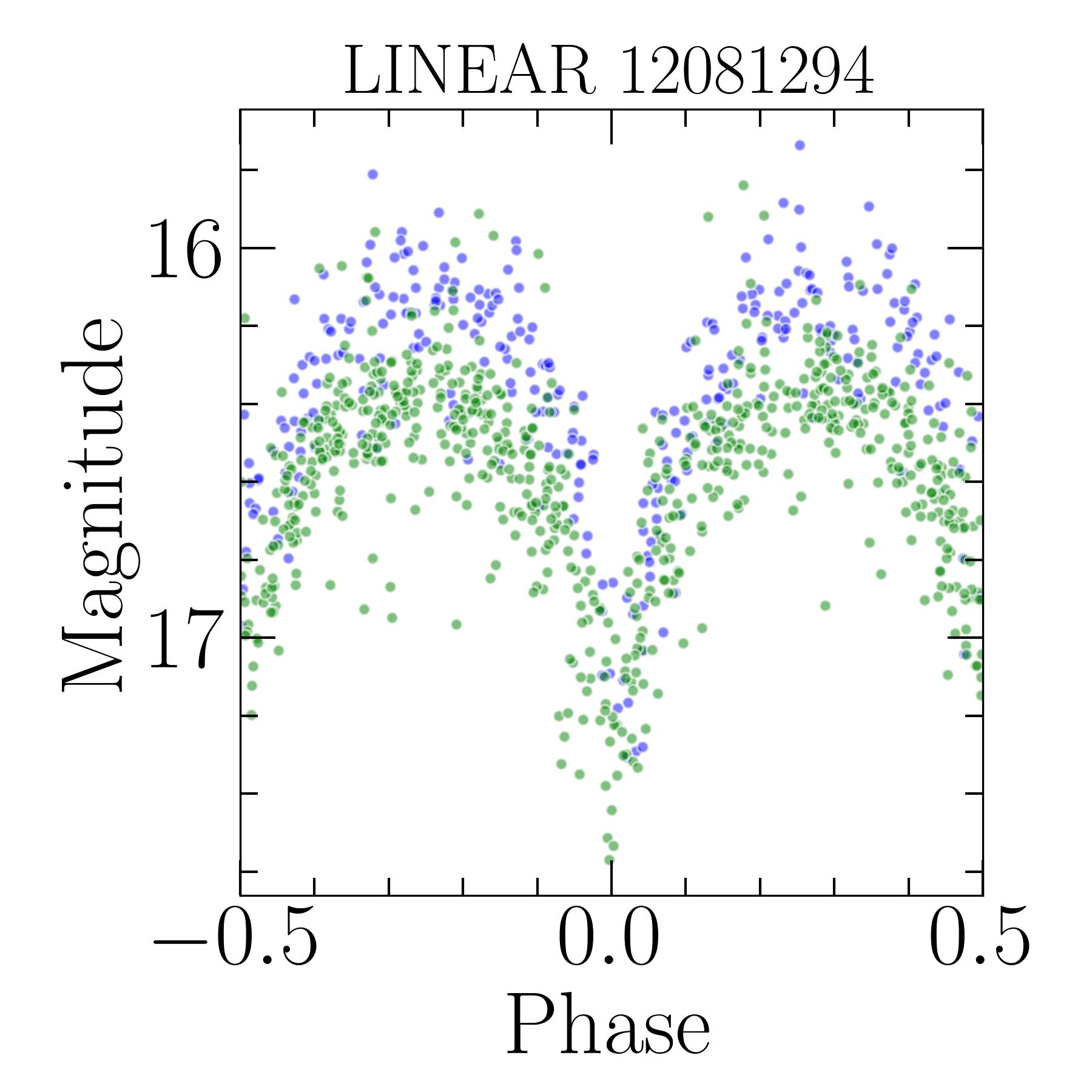}{0.245\textwidth}{}
    \fig{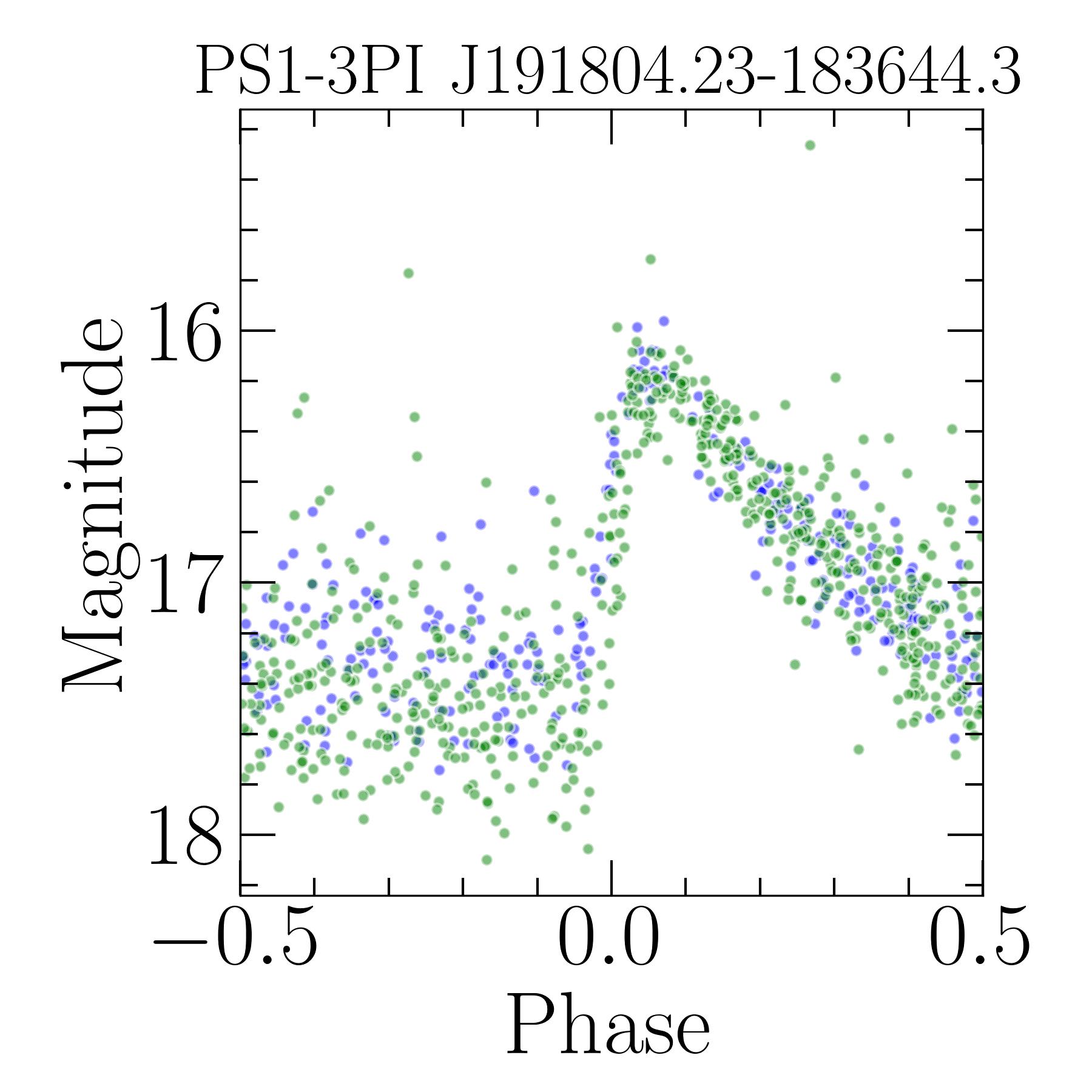}{0.245\textwidth}{}
    \fig{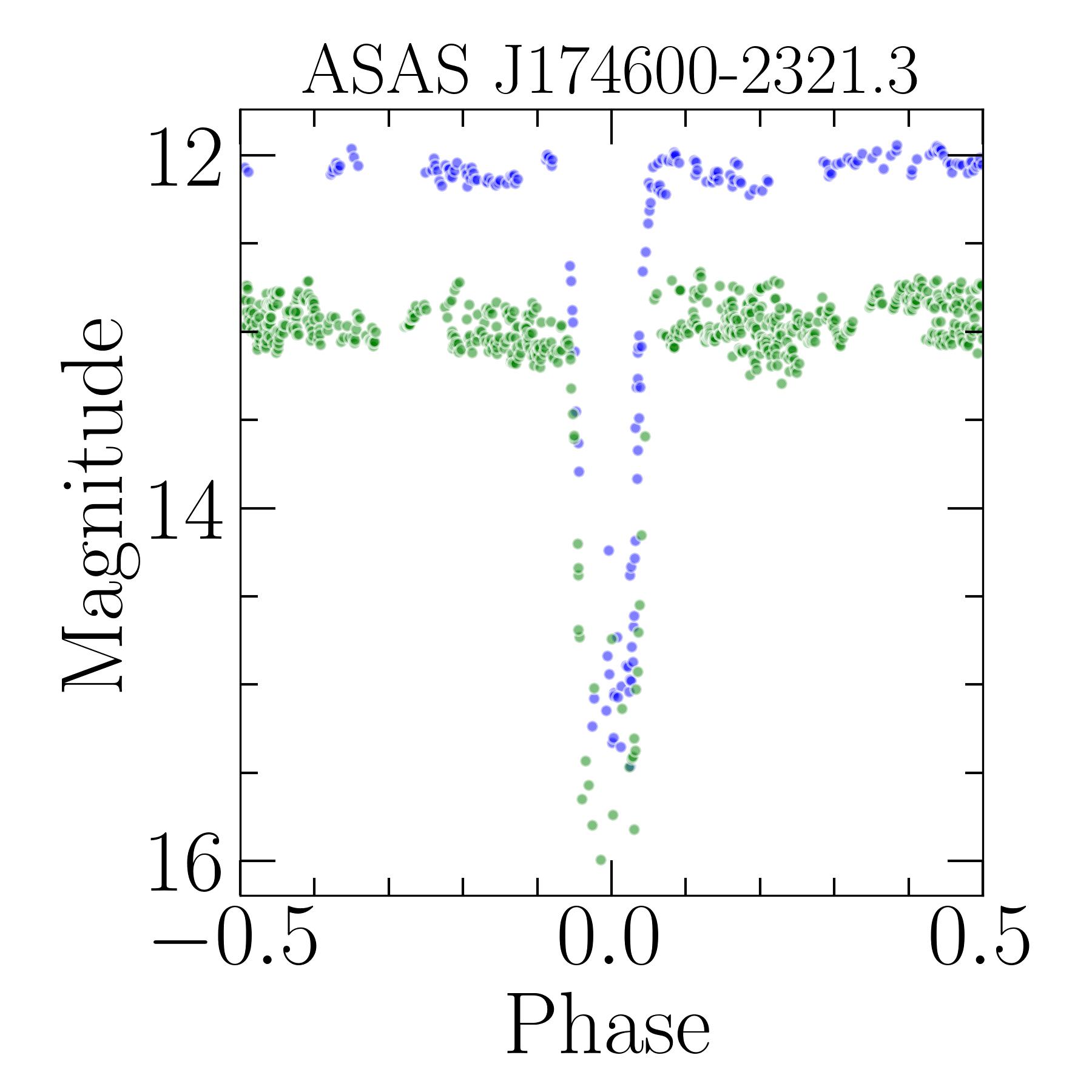}{0.245\textwidth}{}
  }
  \vspace{-2em}
  \gridline{
    \fig{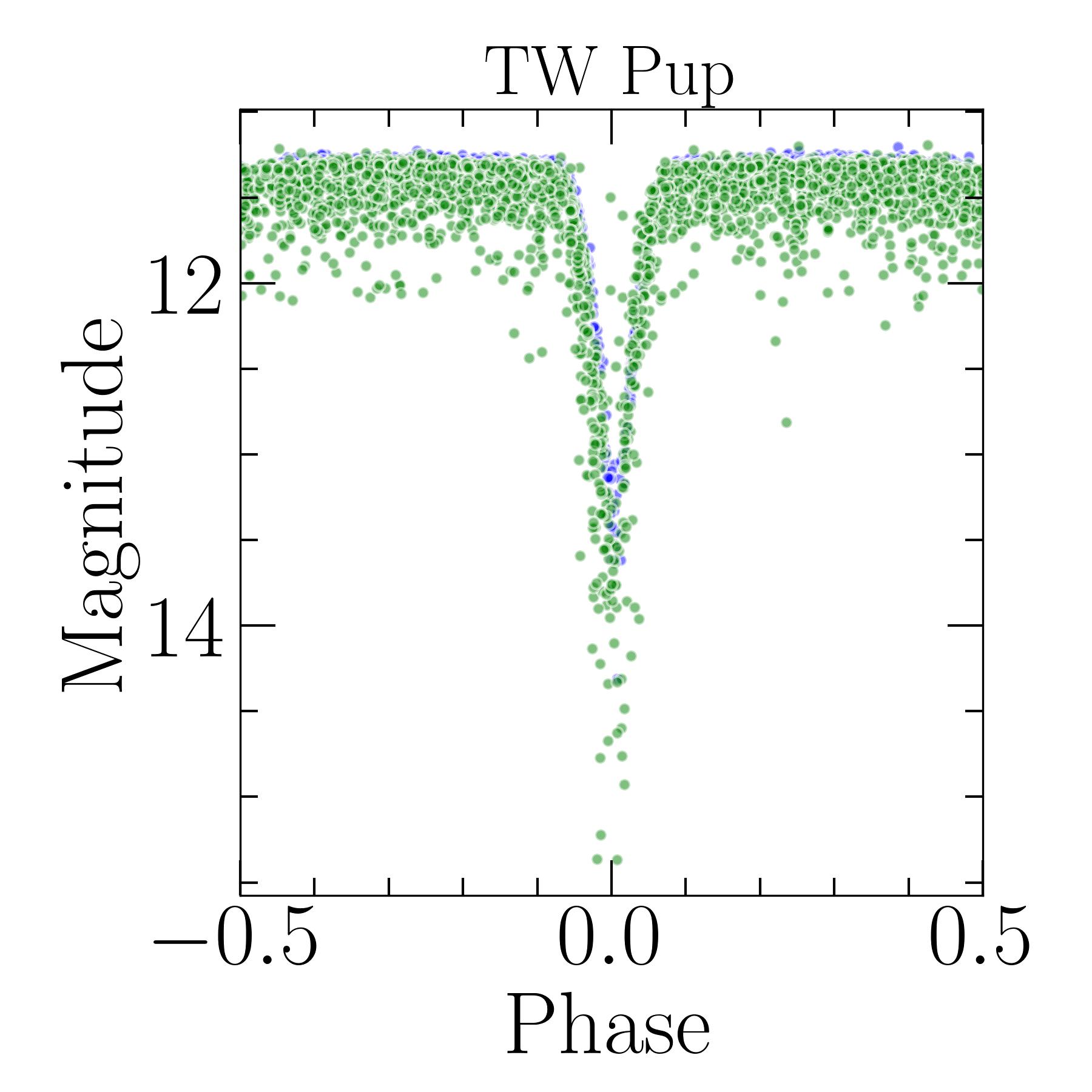}{0.245\textwidth}{}
    \fig{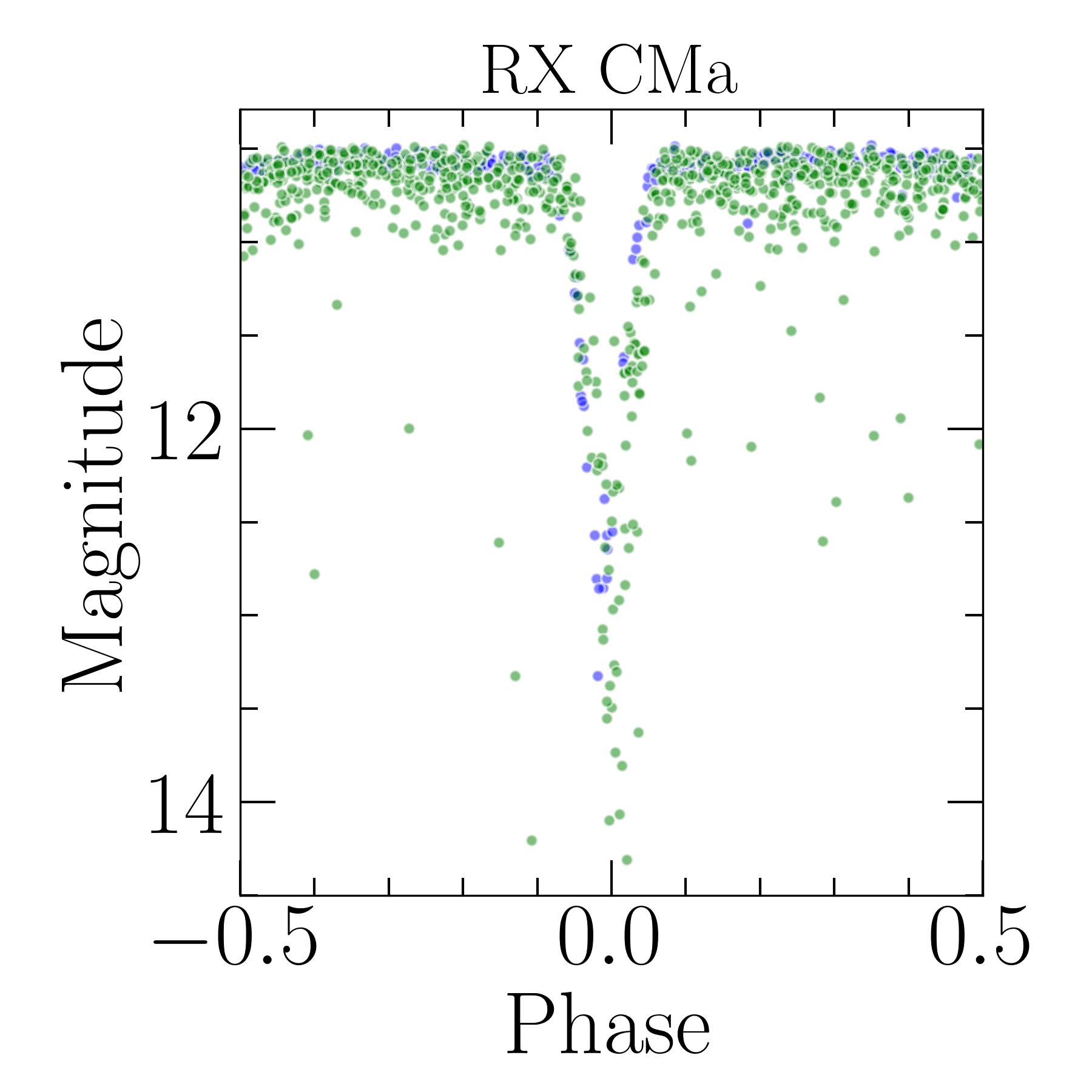}{0.245\textwidth}{}
    \fig{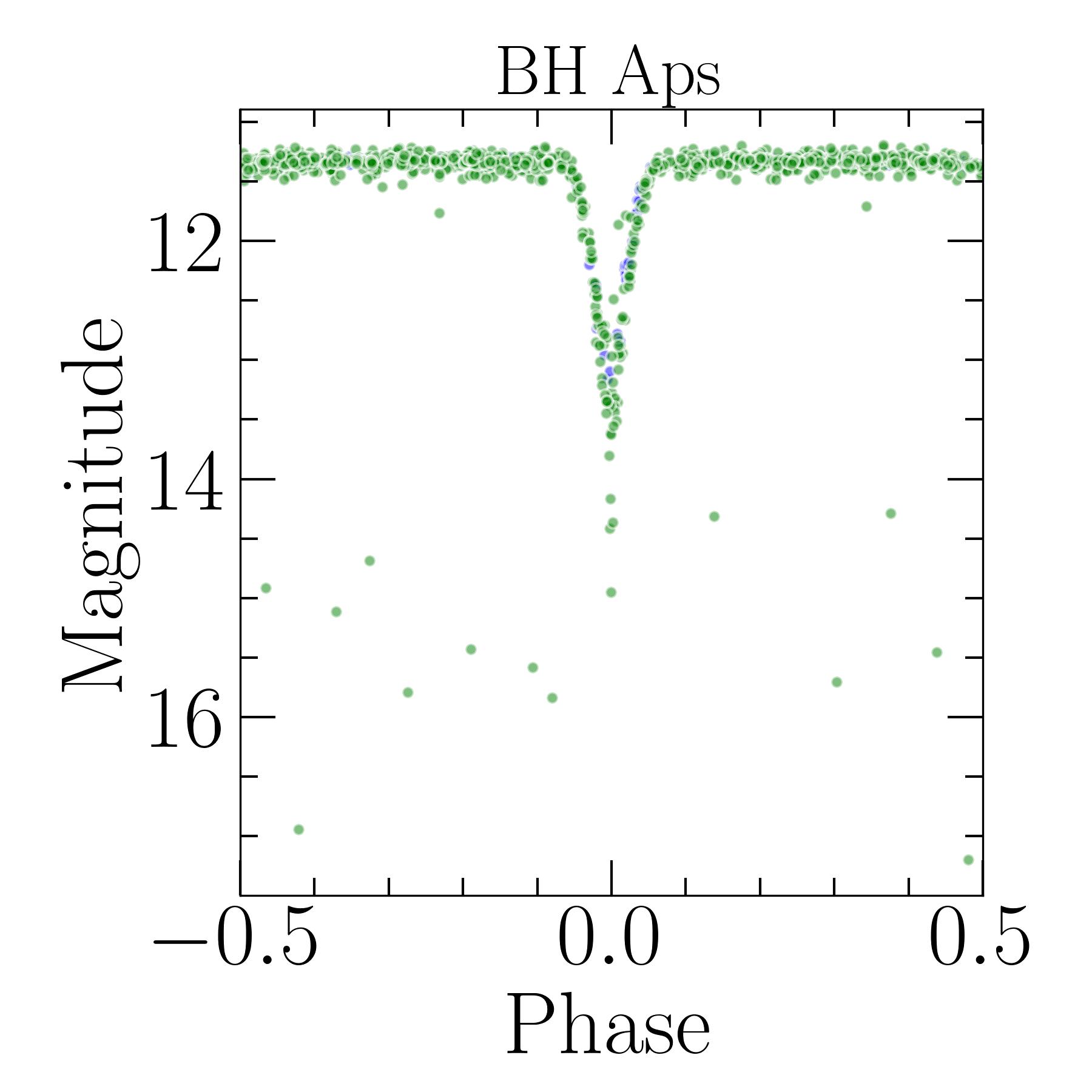}{0.245\textwidth}{}
    \fig{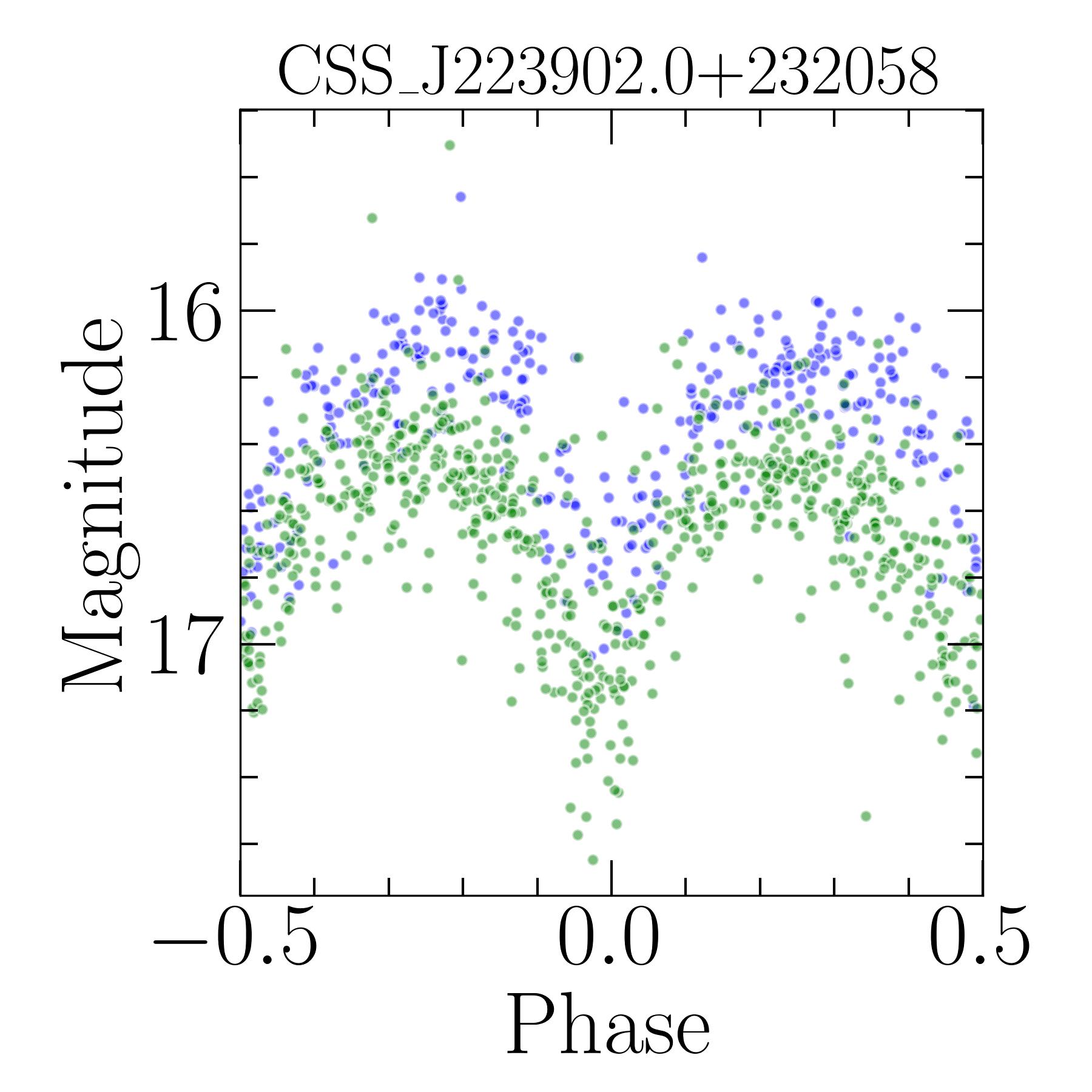}{0.245\textwidth}{}
  }
  \caption{Same as Figure~\ref{fig:gallery_1}, for the anomalies ranked $13$--$24$ in Table~\ref{tab:anomalies}.}
  \label{fig:gallery_2}
\end{figure*}

In Table~\ref{tab:anomalies}, we list all $24$ anomalies identified in the multi-agent review, together with their periods, VSX class labels, majority-vote classifications, mean per-filter amplitudes, and mean relevance scores from the consensus review. We show their phase-folded light curves in Figures~\ref{fig:gallery_1} and~\ref{fig:gallery_2}. By agent classification, the anomaly sample spans $11$ EA-type eclipsing binaries, $7$ contact binaries, $3$ RRab variables, $2$ Type~II Cepheids, and a single dwarf nova. Of these, $9$ received unanimous anomaly scores from all five agents, making them the highest-confidence detections in the sample: Z~Crt, SW~Phe, RU~CMa, NSV~1789, AG~Pav, Y~Leo, V0441~Oph, HM~Pup, and CSS\_J085817.6$-$075719.

\begin{deluxetable*}{clcccccc}
  \tablecaption{List of the $24$ anomalies identified in the multi-agent consensus review.\label{tab:anomalies}}
  \tablewidth{0pt}
  \tablehead{
  \colhead{Rank} &
  \colhead{Name} &
  \colhead{$P$ (d)} &
  \colhead{VSX Class} &
  \colhead{Agent Class} &
  \colhead{$A_V$ (mag)} &
  \colhead{$A_g$ (mag)} &
  \colhead{$\bar{y}$}
  }
  \startdata
  1  & Z~Crt                          & 3.05   & EA/SD       & Eclipsing binary (EA) & $3.96 \pm 0.14$ & $4.07 \pm 0.09$ & 2.0 \\
  2  & SW~Phe                         & 2.55   & EA/SD       & Eclipsing binary (EA) & $4.20 \pm 0.13$ & $4.38 \pm 0.06$ & 2.0 \\
  3  & RU~CMa                         & 1.98   & EA/SD       & Eclipsing binary (EA) & $3.55 \pm 0.46$ & $4.23 \pm 0.21$ & 2.0 \\
  4  & NSV~1789$^\dagger$             & 109.1  & CWA         & Type~II Cepheid       & $1.14 \pm 0.14$ & $1.67 \pm 0.09$ & 2.0 \\
  5  & AG~Pav                         & 1.96   & EA          & Eclipsing binary (EA) & $3.44 \pm 0.20$ & $4.16 \pm 0.09$ & 2.0 \\
  6  & Y~Leo$^\dagger$                & 1.69   & EA/SD+DSCT  & Eclipsing binary (EA) & $4.82 \pm 0.24$ & $4.82 \pm 0.21$ & 2.0 \\
  7  & V0441~Oph                      & 3.06   & EA/SD       & Eclipsing binary (EA) & $4.21 \pm 0.12$ & $4.33 \pm 0.12$ & 2.0 \\
  8  & HM~Pup$^\dagger$               & 2.59   & EA+DSCT     & Eclipsing binary (EA) & $4.18 \pm 0.10$ & $4.34 \pm 0.23$ & 2.0 \\
  9  & CSS\_J085817.6$-$075719        & 0.30   & EW          & Contact binary        & $1.30 \pm 0.16$ & $1.30 \pm 0.09$ & 2.0 \\
  10 & ASASSN-V~J022324.79$-$275944.7 & 0.27   & EW          & Contact binary        & $1.09 \pm 0.15$ & $1.13 \pm 0.12$ & 1.8 \\
  11 & CSS\_J163724.5$+$181022        & 0.47   & RRAB        & RR~Lyrae (RRab)       & $1.60 \pm 0.21$ & $1.72 \pm 0.20$ & 1.8 \\
  12 & PS1-3PI~J172130.15$+$035957.1  & 0.54   & RRAB        & RR~Lyrae (RRab)       & $1.43 \pm 0.16$ & $1.53 \pm 0.14$ & 1.8 \\
  13 & BF~Ser$^\dagger$               & 1.17   & AHB1        & Type~II Cepheid       & $1.14 \pm 0.10$ & $1.38 \pm 0.16$ & 1.6 \\
  14 & SSS\_J062839.1$-$352411         & 0.27   & EW          & Contact binary        & $0.94 \pm 0.08$ & $1.12 \pm 0.15$ & 1.6 \\
  15 & RW~Tri$^\dagger$               & 0.23   & EA/WD+NL    & Dwarf nova            & $2.78 \pm 0.14$ & $2.93 \pm 0.16$ & 1.6 \\
  16 & WISE~J124249.8$-$755621        & 0.29   & EW          & Contact binary        & $1.20 \pm 0.09$ & $1.13 \pm 0.09$ & 1.6 \\
  17 & SSS\_J221404.0$-$395131         & 0.30   & EW          & Contact binary        & $0.99 \pm 0.28$ & $1.13 \pm 0.19$ & 1.6 \\
  18 & LINEAR~12081294                & 0.46   & EW          & Contact binary        & $1.17 \pm 0.14$ & $1.11 \pm 0.18$ & 1.6 \\
  19 & PS1-3PI~J191804.23$-$183644.3  & 0.50   & RRAB        & RR~Lyrae (RRab)       & $1.43 \pm 0.28$ & $1.47 \pm 0.26$ & 1.6 \\
  20 & ASAS~J174600$-$2321.3$^\dagger$ & 1011.5 & NC+EA+SR   & Eclipsing binary (EA) & $3.76 \pm 0.15$ & $3.03 \pm 0.23$ & 1.6 \\
  21 & TW~Pup                         & 2.89   & EA          & Eclipsing binary (EA) & $4.27 \pm 0.09$ & $4.33 \pm 0.07$ & 1.6 \\
  22 & RX~CMa                         & 2.07   & EA/SD       & Eclipsing binary (EA) & $3.51 \pm 0.53$ & $4.02 \pm 0.04$ & 1.6 \\
  23 & BH~Aps                         & 3.65   & EA/SD       & Eclipsing binary (EA) & $4.69 \pm 1.04$ & $4.72 \pm 1.07$ & 1.4 \\
  24 & CSS\_J223902.0$+$232058        & 0.28   & EW          & Contact binary        & $0.85 \pm 0.11$ & $1.13 \pm 0.26$ & 1.4 \\
  \enddata
  \tablecomments{\textit{Rank} is set by decreasing mean relevance score $\bar{y}$, with ties broken by the minimum score across the five agents in the consensus review. \textit{VSX Class} is taken from the International Variable Star Index; slash (/) marks a subtype, plus (+) a combination of variability types, and colon (:) an uncertain classification. \textit{Agent Class} is the consensus variability type adopted in the multi-agent review. $P$ is the VSX-reported period in days. $A_V$ and $A_g$ are the per-filter amplitudes (in magnitudes), reported as mean $\pm$ standard deviation across the agents. Dagger ($\dagger$) marks objects previously studied in the literature, with references given in Section~\ref{subsec:new_discoveries}.}
  \end{deluxetable*}

\subsubsection{Data Quality and Amplitude Checks}\label{subsubsec:validation}

We manually inspected the light curves of all $24$ anomalies for data artifacts and compared the agent-estimated amplitudes with our own. Several light curves contain sparse outliers, but in most cases these remain isolated enough from the main signal that the agents did not absorb them into their amplitude estimates. The exception is BH~Aps, where three out of five agents confused the outliers with the eclipse floor, inflating the mean $g$-band amplitude to $4.72\pm1.07$\,mag compared to our estimate of $3.5$\,mag. In all remaining cases, the agent-estimated $g$-band amplitudes agree with ours to within $15\%$. The agreement is poorer in the $V$ band, where the agents overestimated the eclipse depths of several EA-type binaries with visually overlapping bands (e.g., Z~Crt, SW~Phe), apparently extrapolating the $V$-band points toward the deeper $g$-band minimum. The overestimation is absent when the two bands are well separated or only marginally overlap, supporting this interpretation.

\subsubsection{Eclipsing Binaries}
With $11$ sources, $7$ of which were unanimous detections, EA-type eclipsing binaries form the largest group in our anomaly sample. Most have periods of a few days and show deep, narrow, V-shaped eclipses with extreme $g$-band depths of $4.02$--$4.82$\,mag ($97$--$99\%$ flux decrease) and no well-defined secondary minima, pointing to high-inclination configurations in which a hot primary is occulted by a larger but much cooler secondary. \citet{liakos2022} confirm this configuration for HM~Pup, identifying it as a semidetached binary observed at inclination $89.7\pm0.3$\,deg with an A6V primary and a larger K8IV/III secondary. Similarly, \citet{giuricin1980} found Y~Leo to be a semidetached binary consisting of an A3 primary and a larger K-type secondary, observed at inclination $85.3\pm0.2$\,deg. In addition, Z~Crt, SW~Phe, and V0441~Oph all have primaries of spectral type A according to VSX, consistent with this picture.

The two sources that depart from this pattern are BH~Aps and ASAS~J174600$-$2321.3. The corrected eclipse depth of BH~Aps is about $3.5$\,mag, potentially placing it within the normal range for EA systems and casting doubt on its anomaly status. This uncertainty is also reflected in its mean relevance score of $1.4$, which is among lowest in the sample. ASAS~J174600$-$2321.3, on the other hand, has a long period of $1011.5$\,d and shows a wide eclipse profile with agent-estimated depths of $A_V$~$=$~$3.76\pm0.15$\,mag and $A_g$~$=$~$3.03\pm0.23$\,mag. \citet{hummerich2015} identified it as an eclipsing symbiotic binary, in which a white dwarf primary undergoing a slow nova outburst is periodically occulted by a late M giant companion. \citet{hambsch2015} analyzed the system further and found a slight decrease in the mean brightness, suggesting that the eruption had begun to subside. Nearly a decade later, our data show that the white dwarf is still in outburst.

Beyond the deep eclipses, most light curves show out-of-eclipse scatter at the level of ${\gtrsim}0.5$\,mag, far above the expected photometric noise. The scatter is one-sided, extending toward fainter magnitudes, which points to saturation-induced bias rather than intrinsic variability. Indeed, the out-of-eclipse baselines of these sources lie at $g$~$\approx$~$10$--$11.5$\,mag, well into the saturation regime of the survey, which sets in at $g$~$\approx$~$12.5$\,mag \citep{hart2023}. However, this does not undermine the extreme eclipse depths, since a baseline biased toward fainter magnitudes can only make the eclipses appear shallower, not deeper.

\subsubsection{Contact Binaries}
The seven contact binaries, including the unanimously voted CSS\_J085817.6$-$075719, all show the smooth double-wave morphology characteristic of the class. Their periods of $0.27$--$0.46$\,d fall well within the typical range, but $A_V$~$=$~$0.85$--$1.30$\,mag and $A_g$~$=$~$1.11$--$1.30$\,mag place them in the high-amplitude tail of the distribution. The amplitude dependence analysis in \citet{pesta2023} indicates that these systems are most likely high-inclination, near-equal-mass binaries in deep contact (fill-out factor $\geq0.5$). The combination of the short periods and large fill-out factors suggests that some of these systems may be close to filling their outer critical Roche lobes, at which point mass and angular momentum loss through the L2 Lagrange point would drive rapid merger, making them particularly interesting targets for follow up.

\subsubsection{Pulsators}
The five pulsators consist of three RRab variables and two Type~II Cepheids as classified by the agents. The RRab systems display the asymmetric sawtooth profile typical of the class, but their faint magnitudes ($16$--$18$\,mag) introduce substantial photometric scatter in the folded light curves, possibly compounded by Blazhko modulation. This could have inflated the agent-estimated amplitudes of $A_V$~$=$~$1.43$--$1.60$\,mag and $A_g$~$=$~$1.47$--$1.72$\,mag, but even the lower bounds on these estimates (Table~\ref{tab:anomalies}) still exceed or fall just below the $1.3$\,mag threshold we adopted for RRab variables, placing the systems at the high-amplitude end of the distribution.

The agents classified NSV~1789 as a Type~II Cepheid, matching its VSX label. However, the long period of $109.1$\,d is more consistent with that of a classical Cepheid, which is also supported by the literature \citep[e.g.][]{groenewegen2023,carson1984}. At this period, the high amplitude $A_g$~$=$~$1.67\pm0.09$\,mag falls in the high-end tail of the distribution for both classes. BF~Ser presents a similarly ambiguous case. The agent-assigned class of Type~II Cepheid agrees with \citet{yacob2022}, while VSX listed the star as AHB1 (above horizontal branch variable of subtype 1) at the time of our data retrieval. The VSX label has since been updated to ACEP, corresponding to an anomalous Cepheid pulsating in the fundamental mode. The distinction between Type~II and anomalous Cepheids is not important here, since the combination of the short period $P$~$=$~$1.17$\,d and high amplitude $A_g$~$=$~$1.38\pm0.16$\,mag makes BF~Ser unusual under either classification.

\subsubsection{Nova-like Variable}
The agents classified RW~Tri as a dwarf nova in quiescence, but the steady baseline near $13$\,mag is too bright for this class. The VSX catalog instead lists RW~Tri as EA/WD+NL with spectral type pec(cont+e)+M0V, corresponding to an eclipsing binary with a white dwarf primary, red dwarf secondary, and nova-like variability. Most likely, the agents conflated the two classes due to the absence of a nova-like class description in the prompt. Despite the physical differences, their typical amplitude ranges overlap, justifying the extreme amplitude flag even in the nova-like case. With an eclipse depth of about $2.8$\,mag, RW~Tri represents one of the deepest eclipsing nova-like systems \citep{kjurkchieva2015}, and as such it has been analyzed extensively in the literature \citep[e.g.,][]{boyd2023,subebekova2020,han2018}, making it a valid detection in the context of our search.

\subsubsection{Demoted Anomalies and Potentially Interesting Objects \label{subsec:demoted_and_pot_interesting}}

We also inspected the $23$ sources that were initially flagged as anomalies and then demoted during the consensus review. We found most demotions justified, with two exceptions and one ambiguous case. CSS\_J123713.5$-$114008 is an RRab variable misclassified as EW in VSX, which led one of the agents to flag it for aliasing, preventing a majority anomaly vote despite its high amplitude $A_g$~$=$~$1.58\pm0.17$\,mag. AQ~Ind is an eclipsing binary with an extremely deep primary eclipse ($A_g$~$=$~$4.34\pm0.22$\,mag) that, due to the arbitrary phase zero-point, appeared as two sparse streaks at the edges of the image submitted to the agents. Three agents interpreted the streaks as data quality issues, resulting in a majority-vote score of $0$. The ambiguous case is the eclipsing binary UU~Oph with an agent-estimated eclipse depth $A_g$~$=$~$2.59\pm1.24$\,mag and three sparse points extending it to ${\approx}4.8$\,mag that four agents interpreted as outliers. Without additional photometry, it is difficult to assess whether the demotion was justified.

\begin{figure*}
  \centering
  \gridline{
    \fig{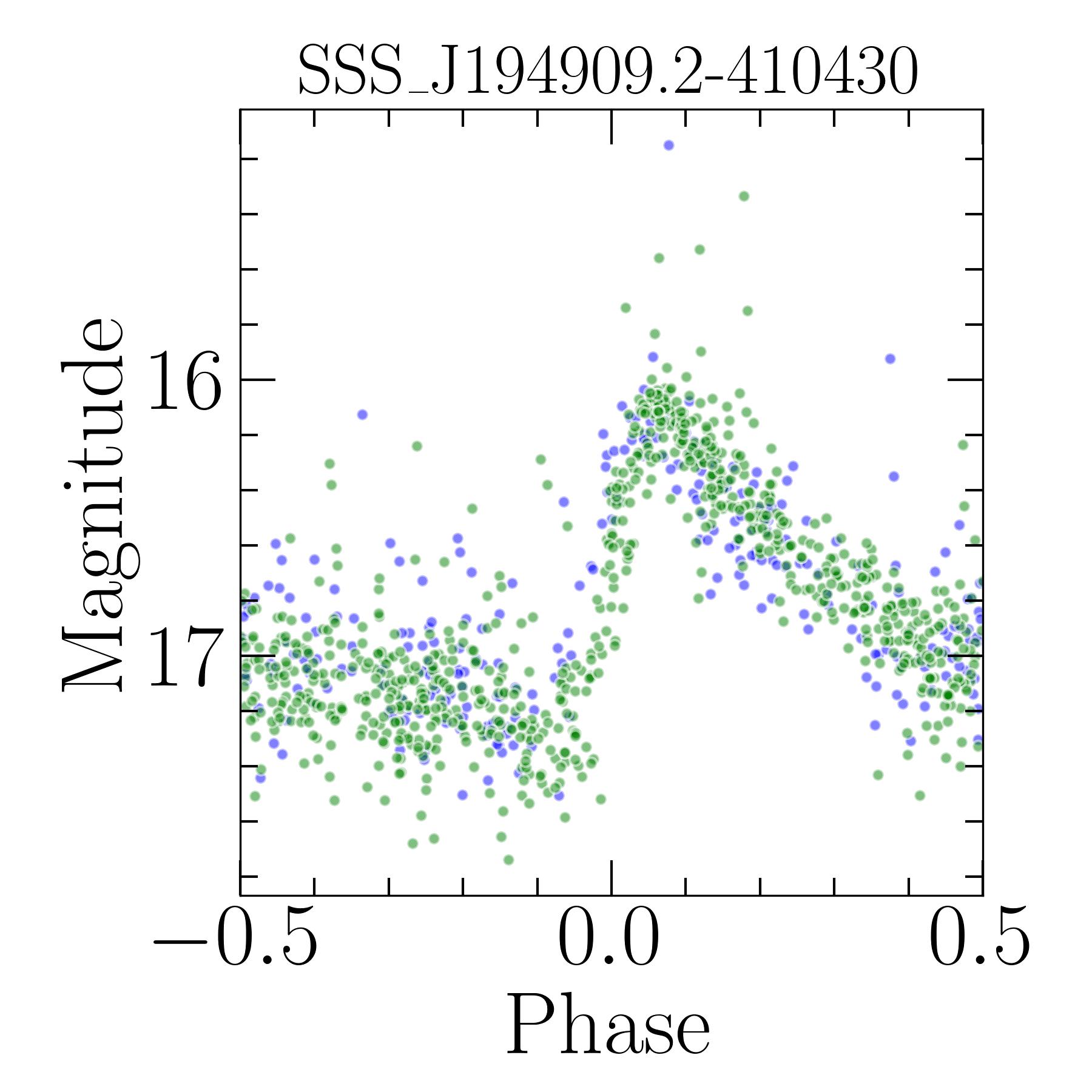}{0.245\textwidth}{}
    \fig{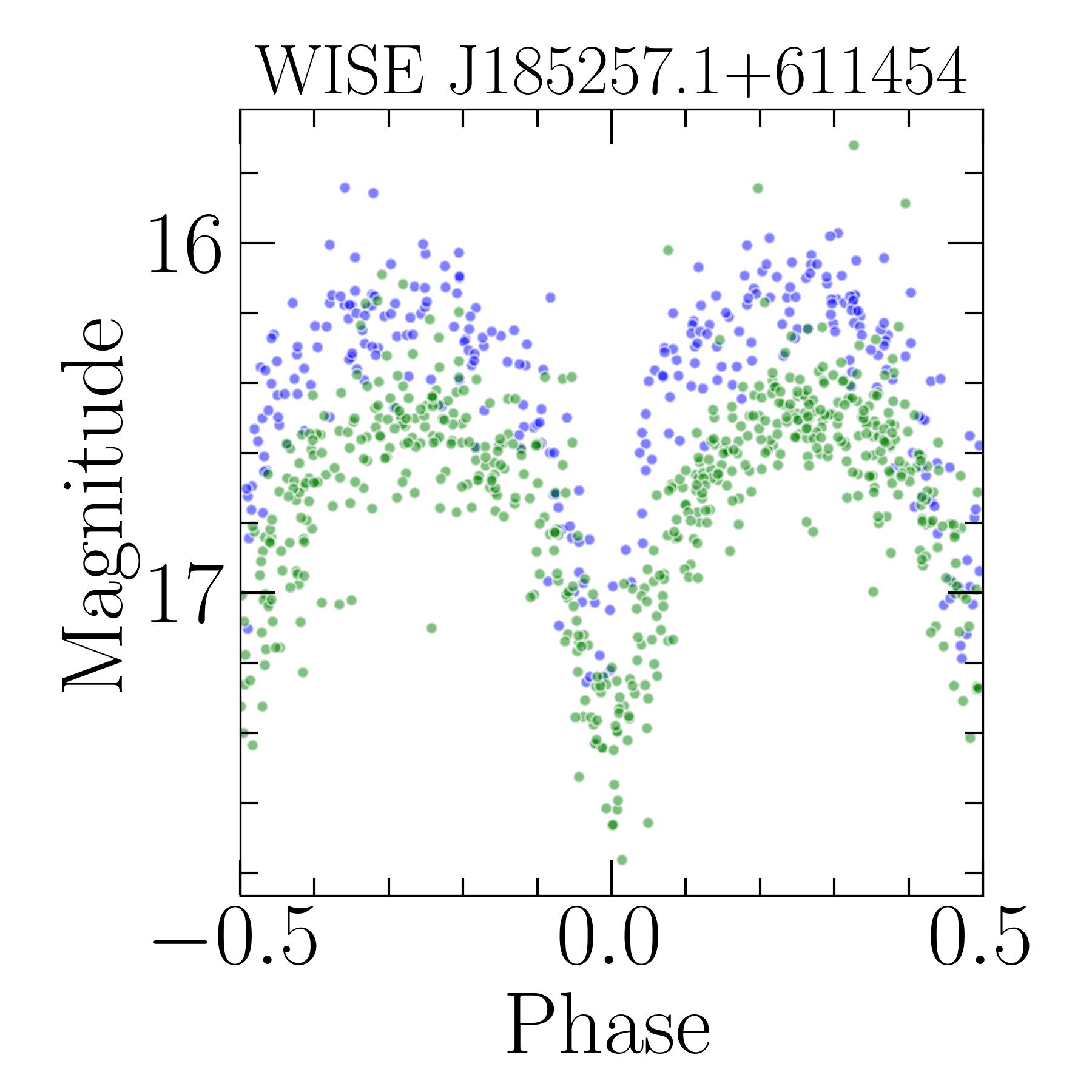}{0.245\textwidth}{}
    \fig{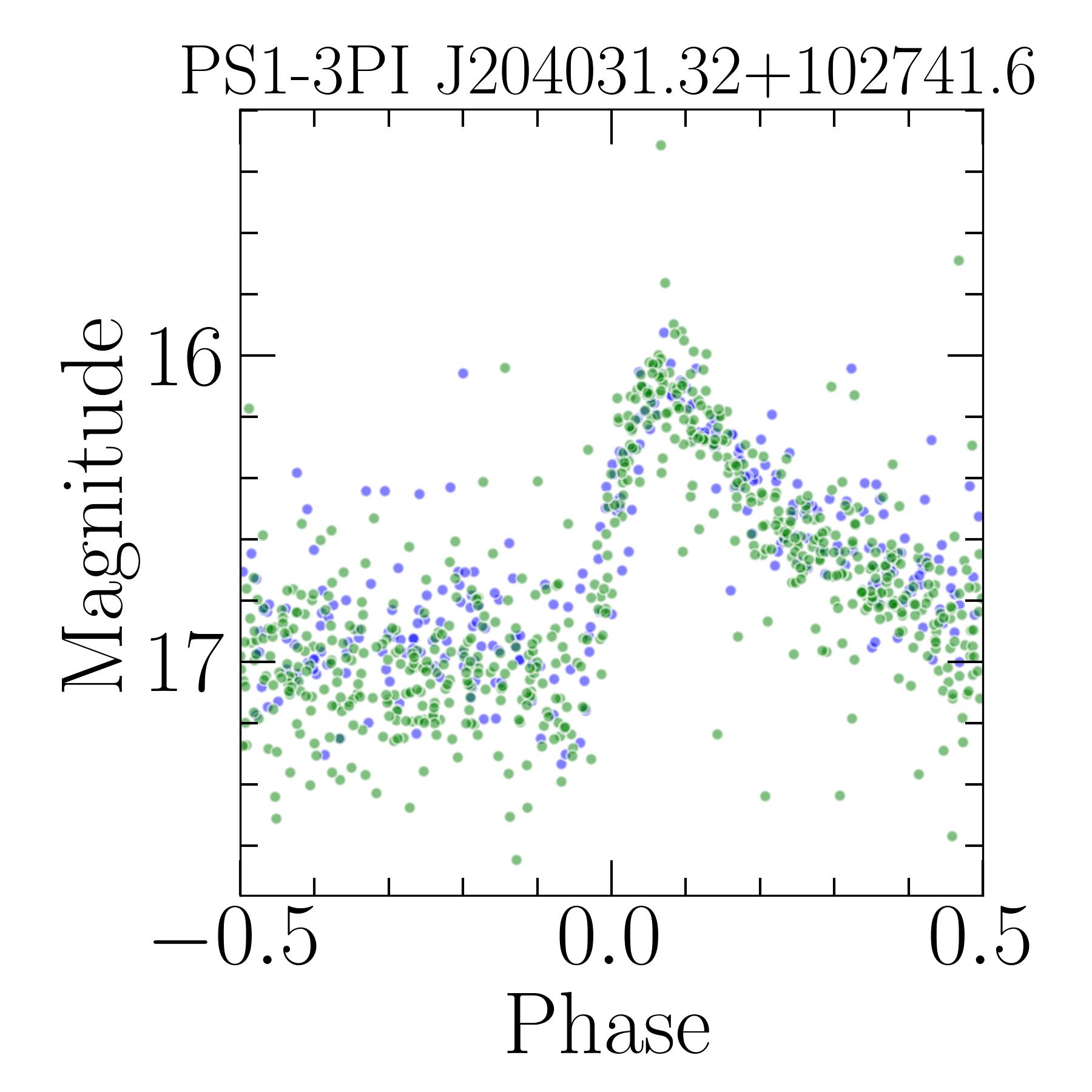}{0.245\textwidth}{}
    \fig{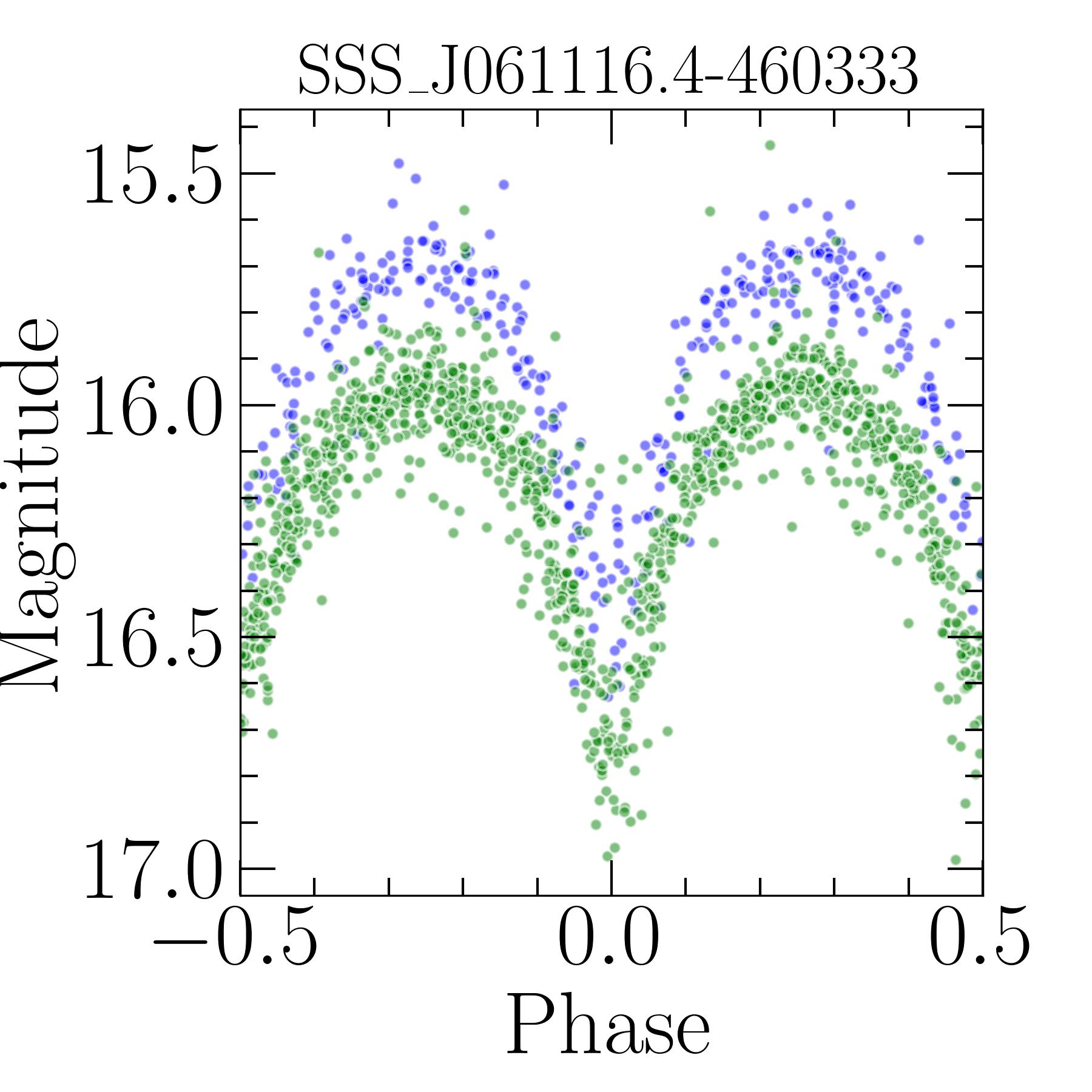}{0.245\textwidth}{}
  }
  \vspace{-2em}
  \gridline{
    \fig{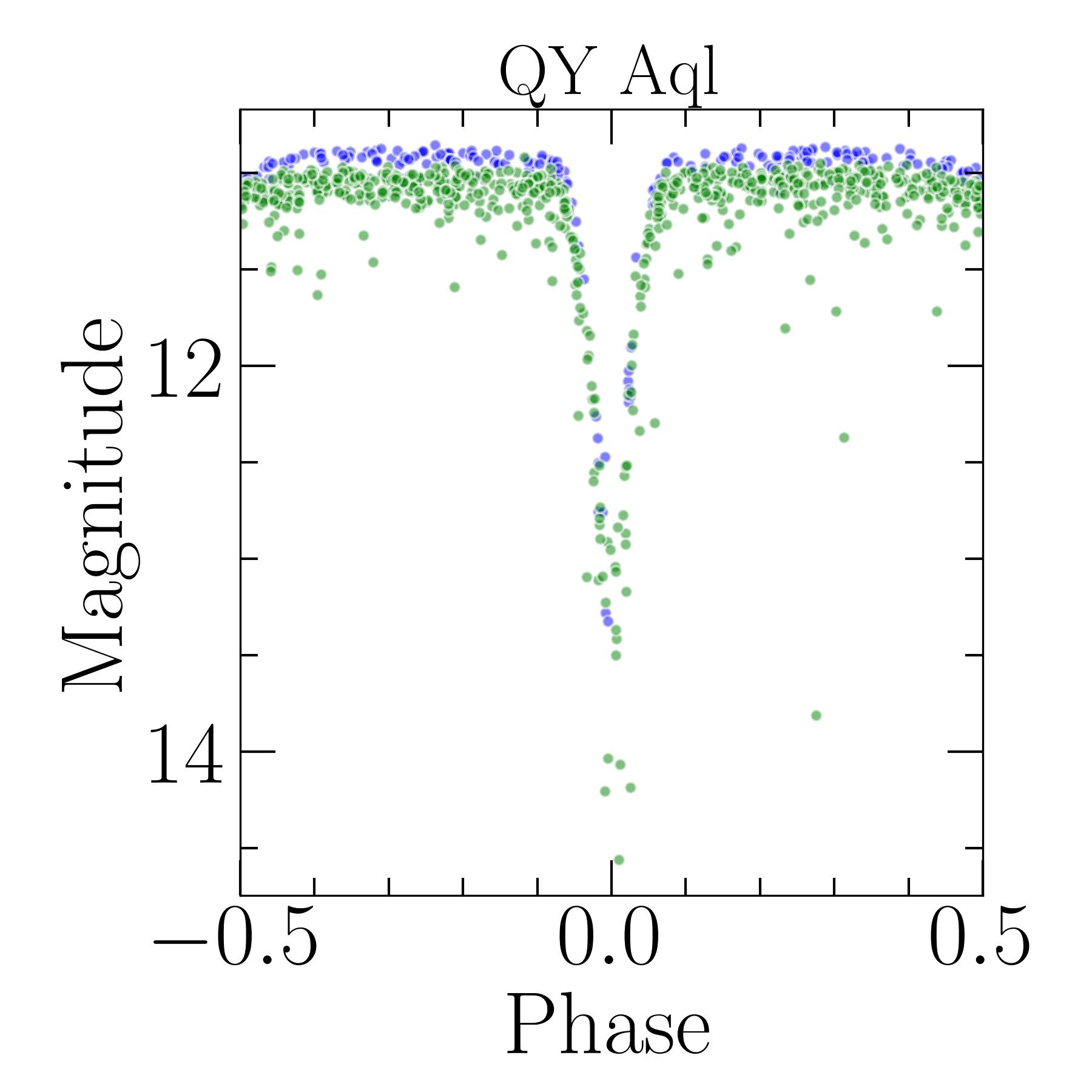}{0.245\textwidth}{}
    \fig{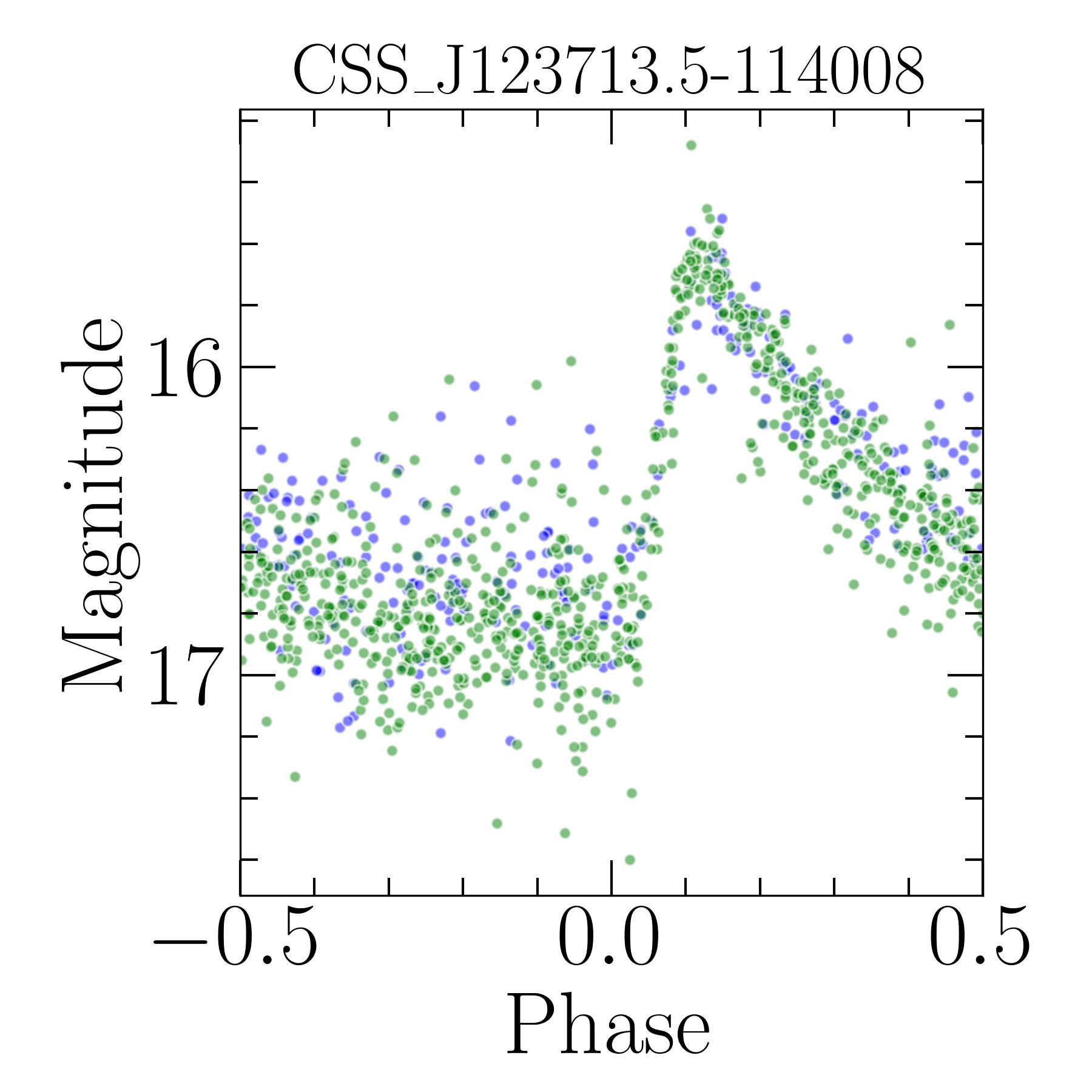}{0.245\textwidth}{}
    \fig{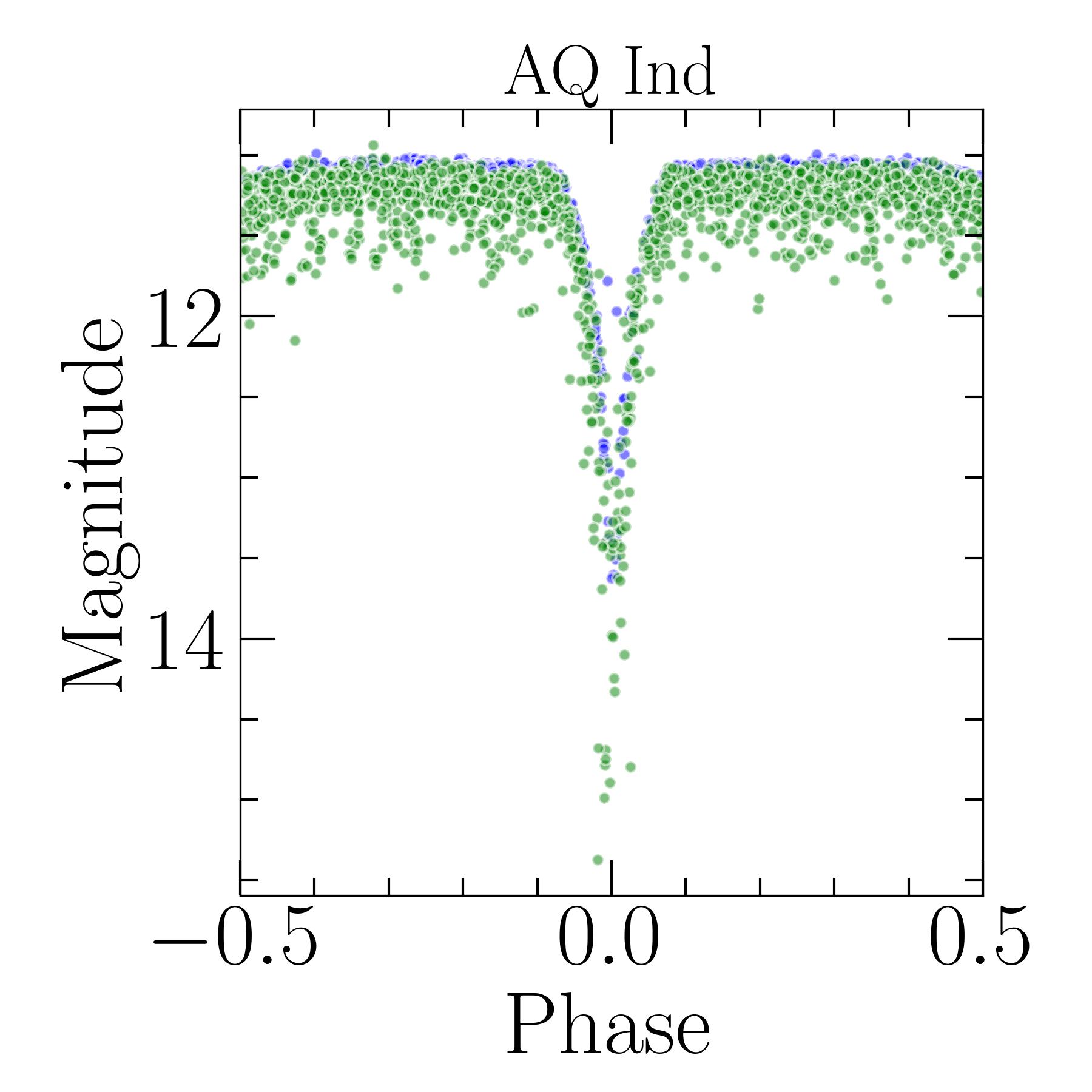}{0.245\textwidth}{}
    \fig{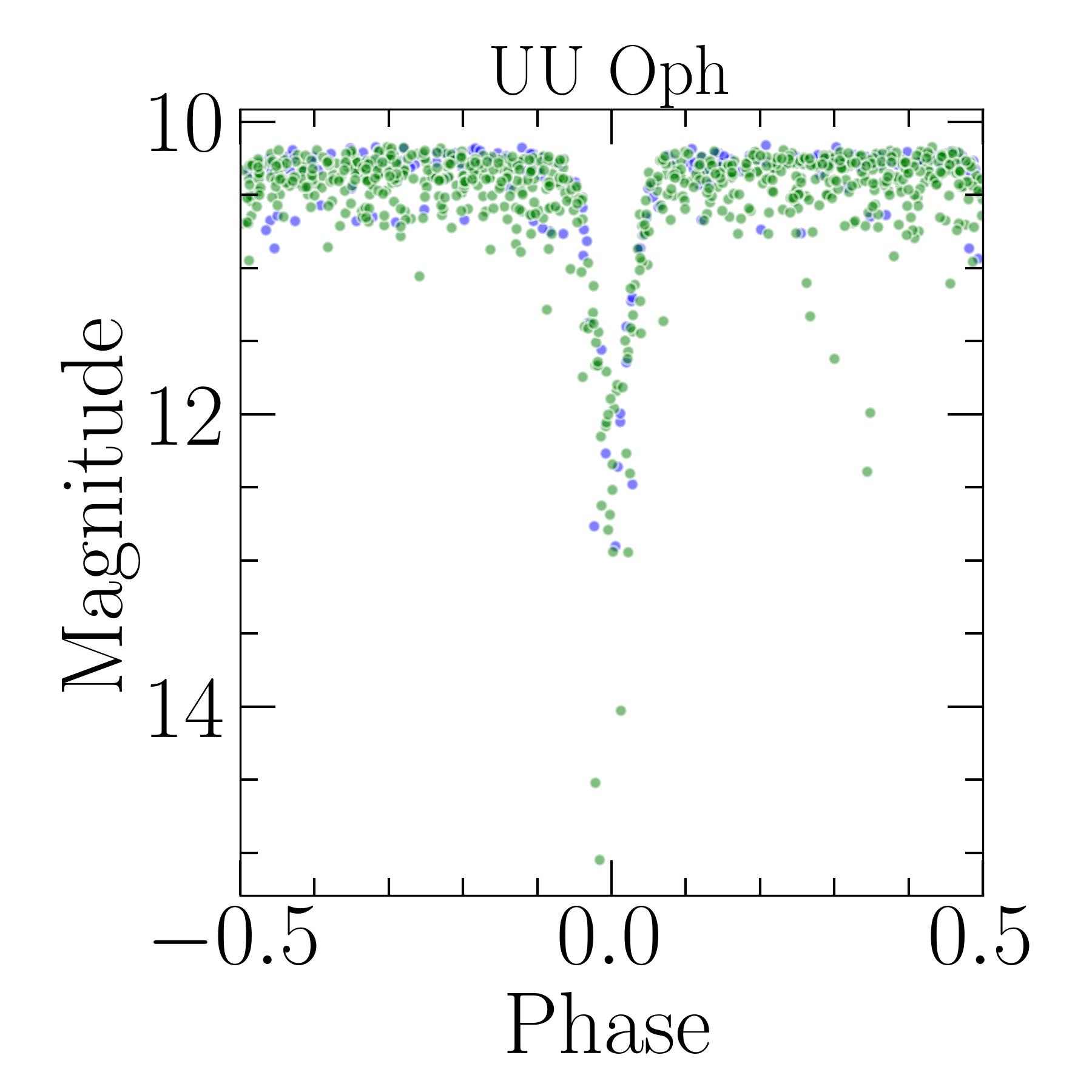}{0.245\textwidth}{}
  }
  \caption{Phase-folded light curves of the five highest-ranked potentially interesting sources from the multi-agent consensus review, followed by three demoted sources warranting further investigation: CSS\_J123713.5$-$114008, AQ~Ind, and UU~Oph. Each panel is labeled by source name. The $V$-band is shown in blue, the $g$-band in green.}
  \label{fig:gallery_3}
\end{figure*}

Beyond the anomalies, the consensus review also flagged $153$ potentially interesting sources, with mean relevance scores ranging from $0.6$ to $1.4$. These sources show borderline behavior such as near-boundary amplitudes or periods, mild modulation, or possible secondary periodicities, none of which were pronounced enough to secure a majority anomaly vote. Still, the five highest-ranked potentially interesting objects were flagged as anomalous by two out of five agents, making them worth revisiting. We list the five potentially interesting sources together with the three demoted anomalies in Table~\ref{tab:demoted_anomalies} and show all eight light curves in Figure~\ref{fig:gallery_3}.

\begin{deluxetable*}{clcccccc}
  \tablecaption{Five highest-ranked potentially interesting objects from the multi-agent consensus review and three demoted sources warranting further investigation. \label{tab:demoted_anomalies}}
  \tablewidth{0pt}
  \tablehead{
  \colhead{Rank} &
  \colhead{Name} &
  \colhead{$P$ (d)} &
  \colhead{VSX Class} &
  \colhead{Agent Class} &
  \colhead{$A_V$ (mag)} &
  \colhead{$A_g$ (mag)} &
  \colhead{$\bar{y}$}
  }
  \startdata
  \noalign{\vskip .8ex}
  \multicolumn{8}{c}{Potentially interesting} \\
  \noalign{\vskip .8ex}
  \hline
  25  & SSS\_J194909.2$-$410430         & 0.60 & RRAB   & RR~Lyrae (RRab)       & $1.39 \pm 0.14$ & $1.41 \pm 0.14$ & 1.4 \\
  26  & WISE~J185257.1$+$611454        & 0.34 & EW     & Contact binary        & $0.99 \pm 0.21$ & $1.16 \pm 0.21$ & 1.4 \\
  27  & PS1-3PI~J204031.32$+$102741.6  & 0.55 & RRAB   & RR~Lyrae (RRab)       & $1.37 \pm 0.11$ & $1.42 \pm 0.11$ & 1.4 \\
  28  & SSS\_J061116.4$-$460333         & 0.34 & EW     & Contact binary        & $0.87 \pm 0.11$ & $1.04 \pm 0.10$ & 1.4 \\
  29  & QY~Aql                         & 7.23 & EA/DS: & Eclipsing binary (EA) & $3.42 \pm 0.30$ & $3.73 \pm 0.19$ & 1.4 \\
  \hline
  \noalign{\vskip 0.8ex}
  \multicolumn{8}{c}{Demoted from anomaly status} \\
  \noalign{\vskip .8ex}
  \hline
  38  & CSS\_J123713.5$-$114008        & 0.52 & EW     & RR~Lyrae (RRab)       & $1.56 \pm 0.19$ & $1.58 \pm 0.17$ & 1.2 \\
  178 & AQ~Ind                         & 2.28 & EA     & Eclipsing binary (EA) & $4.33 \pm 0.22$ & $4.34 \pm 0.22$ & 0.8 \\
  195 & UU~Oph                         & 4.40 & EA/SD  & Eclipsing binary (EA) & $2.49 \pm 1.15$ & $2.59 \pm 1.24$ & 0.4 \\
  \enddata
  \tablecomments{Columns as in Table~\ref{tab:anomalies}. Ranks continue the same ordering, which spans all $276$ candidates entering the consensus review.}
\end{deluxetable*}

\subsubsection{New discoveries\label{subsec:new_discoveries}}

Of the $24$ anomalies identified by the pipeline, $6$ have been previously studied in the literature for their unusual properties: NSV~1789 \citep{groenewegen2023,bird2009}, Y~Leo \citep{turcu2011,pop2011}, HM~Pup \citep{liakos2022,moriarty2013}, BF~Ser \citep{ashbrook1950,schmidt2003}, RW~Tri \citep{vojkhanskaya1984,smak2019}, and ASAS~J174600$-$2321.3 \citep{hummerich2015,hambsch2015}, with NSV~1789, Y~Leo, and HM~Pup receiving unanimous anomaly scores from all five agents. For the remaining $18$ sources, prior work is limited to their identification as variable stars and the determination of their periods and ephemerides, making them new discoveries in the context of anomaly detection, though spectroscopic follow-up is needed to confirm their nature. We make the full catalog of all $24$ anomalies and $153$ potentially interesting objects available in our online data release.

\subsection{Adaptive Re-ranking}\label{subsec:adaptive_reranking}

\begin{figure*}
  \centering
  \includegraphics[width=0.9\textwidth]{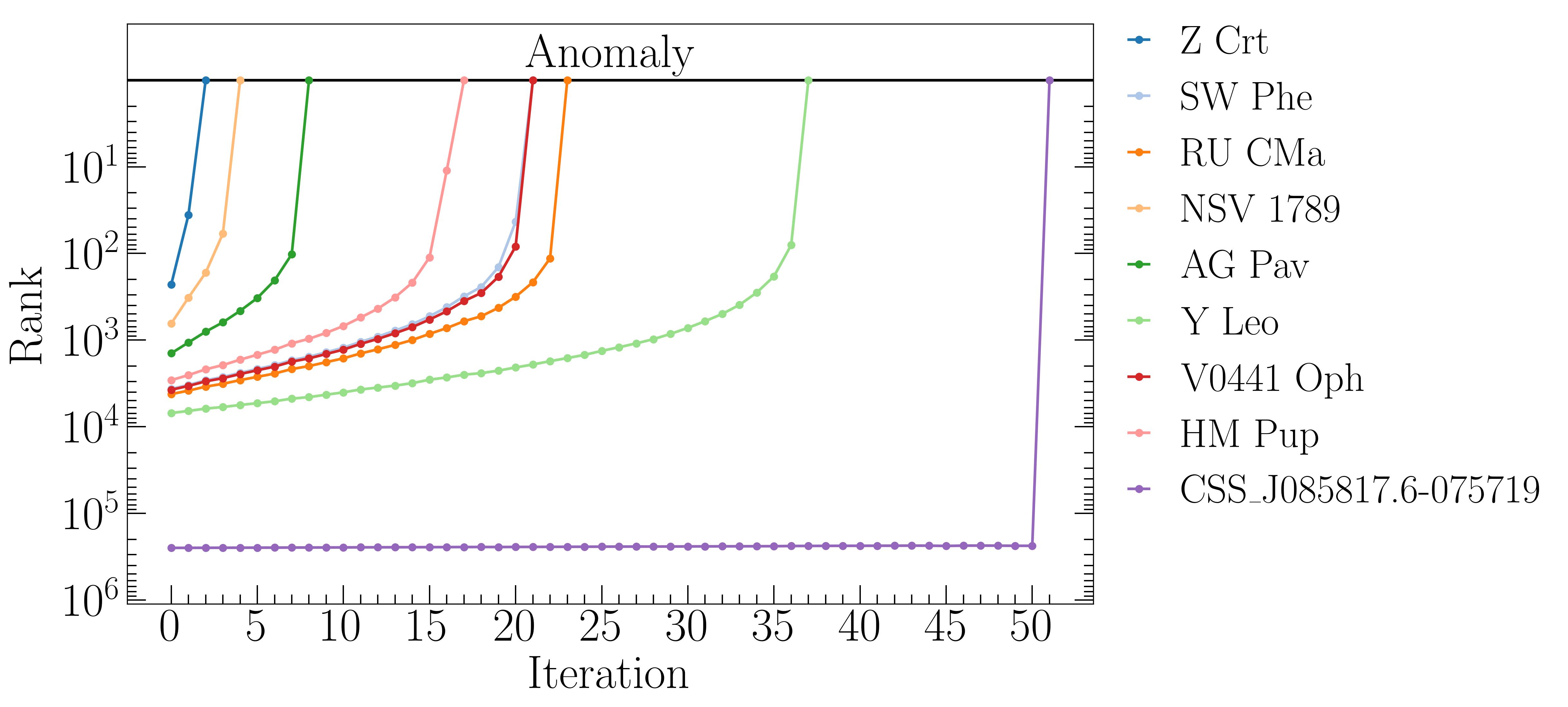}
  \caption{Rank trajectories of the nine unanimously voted anomalies (ranks $1$--$9$ in Table~\ref{tab:anomalies}) as a function of completed iterations, with the highest-ranked sources at the top. In each iteration, the agents labeled the $100$ top-ranked sources from the unlabeled pool. The intersection of each trajectory with the black horizontal line labeled ``Anomaly'' marks the iteration at which the source was flagged as an anomaly in the initial single-agent stages. Iteration~$51$ corresponds to the logistic regression step. Two of the nine anomalies have initial ranks (iteration~$0$) outside our labeling budget of $5\,100$ sources and were found only through neighbor-based label propagation (Y~Leo, rank $6\,942$) or the logistic regression step (CSS\_J085817.6$-$075719, rank $250\,222$).}
  \label{fig:rank_trajectory}
\end{figure*}

The rank trajectories in Figure~\ref{fig:rank_trajectory} provide direct evidence of the efficiency of our active learning loop, showing the $9$ unanimous anomalies climbing toward the top of the ranking as the iterations proceed. Under the initial isolation-forest ranking, only $7$ of the $9$ unanimous anomalies ($78\%$) and $10$ of the $24$ anomalies identified in total ($42\%$) would have fallen within our labeling budget of $5\,100$ sources. Across all $24$ anomalies, the most extreme case is CSS\_J163724.5$+$181022, which started near rank $3.5\times10^5$ and was surfaced only at iteration~$51$ by the logistic regression classifier trained on the labels accumulated throughout the main loop. The effect is even stronger for potentially interesting objects, where only $43$ of the $153$ sources from our final catalog ($28\%$) would have been labeled under the fixed initial ranking.

\section{Discussion}\label{sec:discussion}

Utilizing pre-trained visual embeddings and off-the-shelf LLM agents, our agentic active learning pipeline identified $24$ anomalies, $18$ of them newly reported, and $153$ potentially interesting objects. We discuss the effectiveness of the visual embeddings in Section~\ref{subsec:visual_embeddings}, assess the reliability of the LLM judges in Section~\ref{subsec:llm_judges}, examine the impact of the active learning loop in Section~\ref{subsec:agentic_discussion}, consider the scalability of the pipeline in Section~\ref{subsec:scalability}, and outline its limitations in Section~\ref{subsec:limitations}.

\subsection{Visual Embeddings \label{subsec:visual_embeddings}}

Our results show that vision foundation models trained on natural images can encode light curve images despite the profound domain shift from everyday scenes to scatter plots of magnitude versus phase. Without fine-tuning or catalog-label supervision, the DINOv2 embeddings organized the variable stars in our sample by morphology, with different classes occupying distinct regions (Figures~\ref{fig:umap} and~\ref{fig:heatmap}). While fine-tuned visual embeddings have previously been shown to rival time-series representations in variable star classification \citep{moreno_cartagena2025}, our results demonstrate a successful transfer of pre-trained embeddings in the unsupervised regime. The DINOv2 representation also proved robust to arbitrary phase zero-points and the central preprocessing crop that discards the outermost $6.25\%$ of the image (Appendix~\ref{app:robustness}), indicating that the success of the transfer does not hinge on rendering details.

\subsection{LLM Judges \label{subsec:llm_judges}}

Of the $43$ anomaly labels assigned by the Gemini~3 Flash agents in the main loop and the logistic regression step, the consensus review demoted $23$ of them, while elevating $4$ potentially interesting objects to anomaly status. We manually inspected all surviving anomalies and overturned sources and found the majority-vote relevance scores to be generally consistent with our own assessments (Section~\ref{subsec:anomalies}). Removing the demoted sources therefore increased the purity of the anomaly sample (fraction passing our inspection) by a factor of ${\sim}2$. Since the five-agent review targeted only candidates that received $y\geq1$ in the single-agent stages ($5.4\%$ of the labeled set), this increase came at a modest ${\approx}25\%$ additional cost.

Purity can alternatively be improved by employing a more capable LLM backend, but this would be considerably more expensive. For example, assuming the same token usage, Gemini~3.1 Pro would increase the labeling cost of the main loop and the logistic regression step by a factor of ${\sim}4$ compared to Gemini~3 Flash. Aggregating votes from multiple weaker agents for a small set of flagged sources is therefore more cost-efficient than having all sources evaluated by a single stronger agent. However, stronger agents might still be required when the priority is completeness (recovered fraction of all anomalies in the sample) rather than purity, since the consensus review cannot recover anomalies that were not flagged during the single-agent stages.

\subsection{Agentic Active Learning \label{subsec:agentic_discussion}}

Placing our Gemini~3 Flash agents within an active learning loop greatly reduced the number of sources they needed to review. The agents labeled only ${\sim}1\%$ of the sample but surfaced anomalies across the full range of the initial ranking, with some ranked as low as $3.5\times10^5$. Reaching these sources under the fixed unsupervised isolation-forest ranking would have required ${\sim}70$ times our labeling budget of $5\,100$ sources. At the same budget, only $30\%$ of our final catalog would have been recovered (Section~\ref{subsec:adaptive_reranking}), demonstrating the importance of active learning for both extending the reach and increasing the yield of anomaly searches.

\subsection{Scalability\label{subsec:scalability}}

Together, the active learning loop, logistic regression step, and consensus review finished in roughly $3$ hours. The agents labeled $5\,100$ sources and performed consensus review of $276$ anomaly candidates at a total cost of approximately $\$40$ in API calls. The low cost permits scaling to much larger data sets, but active learning rarely requires more than a few thousand labels. For example, \citet{etsebeth2024} found that the anomaly yield diminished rapidly beyond $6\,000$ labeled samples in their data set of ${\sim}4$ million galaxy images. The natural direction for scaling is therefore across multiple data sets rather than within a single large sample. Our pipeline is well aligned with this strategy, as pre-trained visual embeddings and off-the-shelf LLM agents make it readily applicable to other variable-star samples \citep[e.g.,][]{chen2020,heinze2018,soszynski2016}, requiring no fine-tuning or adaptation beyond prompt modification. With a labeling budget similar to ours, the cost of processing each additional data set should stay within a few tens of dollars.

More generally, agentic active learning has potential applications in settings that require expert judgment across large samples, such as source classification, artifact detection, or alert triage. A predictive model trained on a small agent-annotated subset can extend the labels to the entire data set at a fraction of the cost of exhaustive review. The approach could be especially valuable in applications where expertise is scarce or the task must be repeated regularly. For example, assuming a generous labeling budget of $10^4$ sources and per-call API costs comparable to those of our pipeline, an agentic active learning loop applied to the $10^6$--$10^7$ nightly alerts expected from LSST \citep{ivezic2019} could prioritize the most anomalous events for spectroscopic follow-up at ${\lesssim}\$100$ per night.

\subsection{Limitations \label{subsec:limitations}}

The most fundamental limitation of our pipeline is representational. Although visual embeddings proved effective at capturing the morphology of phase-folded periodic light curves, this representation does not readily extend to aperiodic sources or transients. One alternative is to embed raw photometry directly using a pre-trained time-series foundation model, but these models typically assume regular sampling, which is rarely the case for light curves. Despite this, \citet{li2025} found that \texttt{Chronos-Bolt-tiny}\footnote[4]{\url{https://huggingface.co/amazon/chronos-bolt-tiny}}, a transformer-based model trained on $100$ billion non-astronomical time-series observations, outperformed hand-crafted features in out-of-distribution detection on a sample of $40\,000$ ZTF variable-star light curves, even without any explicit treatment of the irregular sampling. Time-series foundation models therefore present a promising path for extending our pipeline to non-periodic sources at no additional training cost, allowing deployment on live alert streams and near-real-time identification of anomalous events.

A second major limitation is the lack of external tools available to the agents during evaluation. While we deliberately chose this minimal design to keep our proof-of-concept pipeline simple, equipping the agents with tools for light-curve modeling, outlier detection, or database queries would allow them to gather additional evidence beyond the image and help them make better-informed decisions. Additionally, access to a retrieval-augmented generation (RAG) knowledge base of extended class descriptions, example light curves, and relevant literature would provide the agents with targeted expert knowledge on each class without bloating the prompt. Used selectively when the image alone is inconclusive, these tools could improve the reliability of agent evaluations at modest additional cost.

Several additional design choices also constrain the pipeline. We used isolation forests for the initial ranking, but other methods, such as the local outlier factor \citep{breunig2000} or the sum of the distances to the $k$ nearest neighbors \citep{angiulli2002}, might have produced more informative rankings and increased the anomaly yield. We plotted the $V$- and $g$-band light curves in a single image, which biased the estimates of the $V$-band eclipse depths in EA-type binaries when the light curves overlapped (Section~\ref{subsec:anomalies}). Rendering the two bands separately would likely remove this bias. Finally, all evaluations relied on Gemini~3 Flash, a proprietary model that imposes strict API rate limits and ties per-call costs to commercial pricing. Migrating to a locally hosted open-weight multimodal LLM, such as Qwen3-VL \citep{bai2025} or Gemma 4 \citepalias{gemmateam2026}, would remove the API-imposed limits and reduce the cost to that of the compute needed to run the model.

Despite these constraints, the pipeline identified $24$ anomalies, including $6$ previously studied systems. The agents were not provided with the object names or coordinates and had access to no information beyond what Gemini~3 Flash was trained on, so they flagged these sources based on the light-curve images and metadata alone, making their recovery an independent validation of the pipeline. However, this does not let us assess the completeness of our final anomaly sample since we have no record of anomalies that were not recovered by the pipeline.

The most direct next step is to apply the pipeline to variable-star samples from other surveys, such as ZTF \citep{chen2020}, TESS \citep{qiang2025}, and OGLE \citep{soszynski2016}, which would require only rendering their light curves as images. In future work, we plan to address the limitations outlined above and deploy an enhanced version of the pipeline on live alert streams from ZTF and LSST, where it would act as an automated triage layer, surfacing the most anomalous events each night for spectroscopic follow-up.

\section{Conclusions}\label{sec:conclusions}

In this work, we presented an agentic active learning framework for anomaly detection in photometric samples of periodic variable stars (Figure~\ref{fig:pipeline}). The pipeline embeds phase-folded light-curve images with the self-supervised DINOv2 vision transformer (Section~\ref{subsec:embedding}), derives a baseline anomaly ranking from class-conditional isolation forests (Section~\ref{subsec:isolation}), refines it through an active learning loop in which Gemini~3 Flash agents label the top-ranked candidates, and propagates the labels to neighboring sources in the embedding space (Section~\ref{subsec:agentic}). A logistic regression step then surfaces anomalies outside these neighborhoods (Section~\ref{subsec:logreg}), and all flagged candidates are re-evaluated by five independent agents in a consensus review (Section~\ref{subsec:review}).

Applied to over $370\,000$ periodic variable stars from ASAS-SN Sky Patrol V2.0, the pipeline identified $24$ anomalies and $153$ potentially interesting objects among the $5\,100$ sources labeled by the agents (${\sim}1\%$ of the full sample) across $50$ active learning iterations and a single logistic regression step. Our main findings are as follows:

\begin{itemize}[leftmargin=*]

\item By our own classification, the anomalies comprise $10$ EA-type eclipsing binaries with extremely deep primary eclipses, $7$ contact binaries, $3$ RRab variables, and $2$ Cepheids in the high-amplitude tail of their classes, a deeply eclipsing nova-like system, and an eclipsing symbiotic nova that has been in outburst for over two decades (Section~\ref{subsec:anomalies}).
\item Six of the anomalies have previously been studied for their unusual properties, while the remaining $18$ are new discoveries requiring spectroscopic follow-up to determine their physical nature (Section~\ref{subsec:new_discoveries}). Among the latter, $6$ were unanimously voted anomalies in the consensus review, making them high-priority targets for follow-up (Section~\ref{subsec:anomalies}).
\item Without fine-tuning or label supervision, the DINOv2 embeddings organized our sample of variable stars by light-curve morphology, demonstrating successful zero-shot transfer from natural images (Sections~\ref{subsec:embedding} and~\ref{subsec:visual_embeddings}).
\item The multi-agent consensus review proved essential for correcting individual agent mistakes, demoting $36\%$ of the positive single-agent labels and roughly doubling the purity of our final anomaly sample (Sections~\ref{subsec:stats}, \ref{subsec:demoted_and_pot_interesting}, and \ref{subsec:llm_judges}).
\item The logistic regression step extended the reach of the pipeline beyond neighbor-based propagation, contributing $29\%$ of the anomalies and $19\%$ of the potentially interesting objects in our final catalog (Section~\ref{subsec:stats}).
\item Labeling only ${\sim}1\%$ of the sample, the agents found anomalies initially ranked as low as $3.5\times10^5$, which would have required ${\sim}70$ times our labeling budget to reach them under the fixed unsupervised ranking. At the same budget, only $42\%$ of the anomalies and $28\%$ of the potentially interesting objects from our final catalog would have been recovered (Sections~\ref{subsec:adaptive_reranking} and~\ref{subsec:agentic_discussion}).
\item The agentic stages of the pipeline finished in roughly $3$ hours at a total cost of ${\approx}\$40$ in API calls. The pipeline can be readily applied to other variable-star samples at a comparable cost, requiring no fine-tuning or adaptation beyond prompt modification (Section~\ref{subsec:scalability}).

\end{itemize}

These results establish our agentic active learning framework as a practical route to anomaly detection in large photometric surveys, although several limitations remain, namely the reliance on phase-folded images, the absence of external tools available to the agents, and the dependence on a proprietary LLM backend imposing strict API rate limits (Section~\ref{subsec:limitations}). In future work, we plan to address these limitations and extend the pipeline to unphased light curves, which will bring aperiodic sources and transients within its scope and allow deployment on live alert streams from ZTF and LSST.

\begin{acknowledgments}

The research of MP was supported by the Czech Science Foundation (GACR) Project No. 25-16846O. YST was supported by NSF Grant AST-2406729 and a Humboldt Research Award from the Alexander von Humboldt Foundation. We thank Chris Kochanek for detailed comments on the manuscript and Camber Inc. for helpful discussions on this project. We also thank the ASAS-SN team for making their light curves publicly available through Sky Patrol V2.0. This research has made use of the SIMBAD database, CDS, Strasbourg Astronomical Observatory, France \citep{wenger2000}, and of NASA's Astrophysics Data System Bibliographic Services. We acknowledge the use of various versions of GPT (OpenAI), Gemini (Google), and Claude (Anthropic) AI agents for literature review, code development, prompt design, and manuscript preparation. All content was reviewed and verified by the authors, who take full responsibility for the accuracy of this work.

\end{acknowledgments}

\facility{ASAS-SN}

\software{
\texttt{Matplotlib} \citep{hunter2007},
\texttt{NumPy} \citep{harris2020},
\texttt{pandas} \citep{pandas2026},
\texttt{PyTorch} \citep{paszke2019},
\texttt{scikit-learn} \citep{pedregosa2011},
\texttt{SciPy} \citep{virtanen2020},
\texttt{seaborn} \citep{waskom2021},
\texttt{SkyPatrol} \citep{hart2023},
\texttt{Transformers} \citep{wolf2019},
\texttt{UMAP} \citep{mcinnes2018}
}

\bibliography{references}

\appendix

\section{Prompt}\label{app:prompt}

All agent evaluations in the active learning loop (Section~\ref{subsec:agentic}), logistic regression step (Section~\ref{subsec:logreg}), and consensus review (Section~\ref{subsec:review}) relied on a single prompt implementing eight sequential steps: visual inspection, data quality checks, amplitude estimation, classification, extreme-parameter flagging, scoring, self-audit, and structured JSON response. Figure~\ref{fig:prompt} presents a condensed version of the full $1\,460$-word prompt, which is available in our GitHub repository (\url{https://github.com/milanpesta/aal_anomalies_asassn}).

\begin{figure*}[p]
\setlength{\fboxsep}{6pt}
\begin{lrbox}{\promptbox}
\begin{minipage}{\dimexpr\textwidth-2\fboxsep-2\fboxrule\relax}
\begin{lstlisting}[style=prompt]
You are an astrophysical time-series and variable star expert. Analyze a single phased light curve, assign discrete relevance score 0, 1, or 2 based on how anomalous AND astrophysically relevant the light curve appears. Return ONE JSON ONLY—no prose/markdown/fences.
(*\gap*)
INPUTS:id(str),period_days(days),y_scale_mag(mag),catalog_label(str),image(518×518px;filters:V=blue,g=green;gridlines=gray)
IMAGE GUIDANCE
axes:X=phase[−0.5,+0.5],Y=inverted mag(bright=TOP,faint=BOTTOM,span=y_scale_mag);blue may hide behind green
grid:20x20;1 col=0.05 phase,1 row=y_scale_mag/20
(*\gap*)
STEPS:0)Visual inspection 1)Data+folding+morphology checks 2)Amplitude 3)Class 4)Extreme flags 5)Score 6)Audit 7)JSON
PRIORITIES:Detect quality/systematic issues;keep scores conservative when quality concerns exist
(*\gap*)
STEP 0:VISUAL INSPECTION(≤200w,skeptically describe what you see)
1.Global:dense/sparse,clean/noisy,filters(V&g),outliers,repeating shapes
[...5 further inspection items and guidelines...]
Output JSON fields:raw_visual_description
(*\gap*)
STEP 1:CHECKS
SKEPTICAL DEFAULTS:Issues=True,Positive attributes=False;overturn only with clear visual evidence
AGGREGATION:"V OR g"(evaluate per filter,True if either);"MAX"(evaluate per filter,report highest);[...]
(*\gap*)
|Check|Type|Agg|Rule|
|has_V_filter|bool|—|≥1 data point in V|
[...21 further checks...]
Output JSON fields:all filters+data_quality+folding+morphology checks
(*\gap*)
STEP 2:AMPLITUDE(Vertical extent of V&g backbones;NOT absolute position)
Per filter:
1)ID backbone:find coherent phase-correlated structure;exclude vertically spread phase-independent scatter[...]
[...3 further amplitude steps and guidelines...]
Output JSON fields:inferred_V_amplitude_mag,inferred_g_amplitude_mag
(*\gap*)
STEP 3:CLASSIFICATION
Allowed:RR_Lyrae_ab|RR_Lyrae_c|[...24 further classes...]|unknown
(*\gap*)
CLASS TABLE(P=period days,A=amplitude mag):
|Class|P|A|Neg(exclude)|Pos(support)|
|RR_Lyrae_ab|0.3-1.2|0.5-1.3|2 max+2 min;flat top;double-wave;LPV-broad min|Sharp sawtooth;1 max+1 min;fast↑slow↓;asym~0.5-0.9;mod~0-0.5|
[...26 further classes...]
(*\gap*)
Critical rules:>2 min→LPV/irregular NOT eclipsing;[...]
(*\gap*)
Synonyms(non-exhaustive,case-insensitive):
E,E/RS→eclipsing_binary_EA,eclipsing_binary_EB,contact_binary
[...28 further synonym rows and label-interpretation notes...]
(*\gap*)
CATALOG_LABEL SYMBOL CONVENTIONS:
|(pipe)=logical OR;uncertain classification,all possible types indicated(e.g.,ELL|DSCT=ellipsoidal OR delta_Scuti)
[...4 further symbols and parsing rule...]
(*\gap*)
Process:
1)catalog_label_interpreted:parse symbols(|,+,/,!);map each component via synonyms(sep with commas);[...]
[...5 further process steps...]
Output JSON fields:catalog_label_interpreted,possible_classes,inferred_variability_type,type_match
(*\gap*)
STEP 4:EXTREME FLAGS
|Flag|Rule|
|extreme_amplitude|A>1.1×upper for class AND data_quality_issues=False;compare both V&g vs class A-range;[...]|
[...7 further flags...]
(*\gap*)
Process:
1)Get P&A ranges for inferred_variability_type from Step 3 CLASS TABLE
[...2 further process steps...]
Output JSON fields:calculations(document all arithmetic),all extreme+boundary+mixed flags
(*\gap*)
STEP 5:SCORING(Evaluate anchors in order;STOP at first match)
BINS:0=not_interesting,1=potentially_interesting,2=anomaly
[...2 definitions...]
(*\gap*)
|Priority|Condition|Score|
|1|folded_correctly=False|0|
[...7 further anchors...]
(*\gap*)
SCORING LOGIC:
Priority 1-2:Folding/quality issues→0
[...6 further anchor descriptions and 2 scoring notes...]
Output JSON fields:relevance_score,label,confidence,rationale
(*\gap*)
STEP 6:AUDIT(verify STEPS 0-5 again before generating JSON,check for mistakes)
0.Visual inspection:raw_visual_description complete? [...]
[...5 further audit items...]
(*\gap*)
STEP 7:JSON SCHEMA(output raw JSON only)
{
  "meta":{[...4 metadata fields...]},
  "analysis":{[...8 analysis field groups...]},
  "results":{
    "relevance_score":"<0|1|2>",
    "label":"<not_interesting|potentially_interesting|anomaly>",
    [...2 further fields...]
  }
}
(*\gap*)
VALIDATION:Numeric in range;relevance_score must be 0,1,or 2(int);strings match allowed;[...]    
\end{lstlisting}
\end{minipage}
\end{lrbox}
\noindent\fbox{\usebox{\promptbox}}
\caption{Condensed version of the prompt used by the agents when evaluating anomaly candidates. Bracketed ellipses mark omissions. Retained text is verbatim.\label{fig:prompt}}
\end{figure*}

\section{Agent Reasoning Consistency}\label{app:ngram}

To verify that the reasoning of the agents was consistent with the assigned relevance scores, we performed an $n$-gram frequency analysis of the rationales from the consensus review. For each source, we collected the rationales of the agents whose individual scores matched the majority relevance label. We converted all text to lowercase, removed non-alphabetic characters and stop words (e.g., ``the'', ``and'', ``with''), discarded words shorter than three characters, and compiled the set of unique $n$-grams for each source so that a term repeated across rationales counted only once. We also recorded the unique diagnostic flags that the agents raised during their analysis of each source, providing an independent reference against which to compare the vocabulary of the rationales. Table~\ref{tab:ngram} reports the five most frequent unigrams, bigrams, and trigrams along with the triggering rates of selected diagnostic flags for each of the three relevance scores.

The vocabulary changes considerably across the relevance labels. For the $24$ anomalies, the dominant words are ``exceeding'' ($91.7\%$) and ``extreme'' ($83.3\%$), which is consistent with the flag analysis, where all anomalies triggered \texttt{extreme\_amplitude} and no data quality flags were raised. For the $153$ potentially interesting objects, the vocabulary shifts to ``boundary'' ($81.0\%$) and ``near'' ($79.1\%$), in line with \texttt{amplitude\_near\_boundary} being raised for $89.5\%$ of them, while \texttt{extreme\_amplitude} was triggered for only $1.3\%$. The rationales for the $99$ demoted sources split between overlooked data quality issues and incorrectly flagged sources that are standard for their class, with ``quality'' appearing for $78.8\%$ of the demoted sources and ``standard'' for $72.7\%$. The flag analysis shows the same split, with $56.6\%$ of the sources triggering \texttt{data\_quality\_issues}. Overall, each relevance label carries a distinct vocabulary that aligns well with its flag signature, confirming that the agents were internally consistent in their decision process.

\begin{deluxetable*}{llll}
  \tablecaption{Top five unigrams, bigrams, and trigrams in the agent rationales and triggering rates of selected diagnostic flags for each consensus relevance label.\label{tab:ngram}}
  \tabletypesize{\small}
  \tablewidth{0.99\textwidth}
  \tablehead{
    \multicolumn{1}{l}{} &
    \multicolumn{1}{l}{Anomaly} &
    \multicolumn{1}{l}{Potentially interesting} &
    \multicolumn{1}{l}{Not interesting}
  }
  \startdata
  \noalign{\vskip 3pt}
  Unigrams &
  \makecell[l]{amplitude ($100\%$)\\ mag ($95.8\%$)\\ exceeding ($91.7\%$)\\ extreme ($83.3\%$)\\ high ($83.3\%$)} &
  \makecell[l]{class ($92.8\%$)\\ amplitude ($92.2\%$)\\ boundary ($81.0\%$)\\ near ($79.1\%$)\\ morphology ($74.5\%$)} &
  \makecell[l]{quality ($78.8\%$)\\ extreme ($72.7\%$)\\ standard ($72.7\%$)\\ binary ($68.7\%$)\\ features ($67.7\%$)} \\
  \noalign{\vskip 3pt}\hline\noalign{\vskip 3pt}
  Bigrams &
  \makecell[l]{exceeding standard ($70.8\%$)\\ exceeding typical ($66.7\%$)\\ high amplitude ($62.5\%$)\\ mag significantly ($62.5\%$)\\ extreme amplitude ($58.3\%$)} &
  \makecell[l]{amplitude near ($71.9\%$)\\ near upper ($52.3\%$)\\ boundary class ($51.0\%$)\\ morphology amplitude ($45.1\%$)\\ high amplitude ($43.8\%$)} &
  \makecell[l]{quality issues ($62.6\%$)\\ data quality ($61.6\%$)\\ anomalous features ($49.5\%$)\\ period amplitude ($49.5\%$)\\ eclipsing binary ($36.4\%$)} \\
  \noalign{\vskip 3pt}\hline\noalign{\vskip 3pt}
  Trigrams &
  \makecell[l]{mag exceeding standard ($41.7\%$)\\ mag significantly exceeding ($33.3\%$)\\ mag significantly exceeds ($33.3\%$)\\ significantly exceeding typical ($33.3\%$)\\ significantly exceeding standard ($29.2\%$)} &
  \makecell[l]{amplitude near upper ($46.4\%$)\\ physical limit class ($28.8\%$)\\ classic contact binary ($26.1\%$)\\ near upper physical ($25.5\%$)\\ upper physical limit ($24.8\%$)} &
  \makecell[l]{data quality issues ($49.5\%$)\\ typical period amplitude ($30.3\%$)\\ binary typical period ($22.2\%$)\\ parameters well within ($20.2\%$)\\ standard contact binary ($19.2\%$)} \\
  \noalign{\vskip 3pt}\hline\noalign{\vskip 3pt}
  Flags &
  {\footnotesize\makecell[l]{\texttt{extreme\_amplitude} ($100\%$)\\ \texttt{extreme\_period} ($4.2\%$)\\ \texttt{amplitude\_near\_boundary} ($4.2\%$)\\ \texttt{period\_near\_boundary} ($4.2\%$)\\ \texttt{outlier\_dominated} ($0.0\%$)\\ \texttt{data\_quality\_issues} ($0.0\%$)}} &
  {\footnotesize\makecell[l]{\texttt{extreme\_amplitude} ($1.3\%$)\\ \texttt{extreme\_period} ($0.0\%$)\\ \texttt{amplitude\_near\_boundary} ($89.5\%$)\\ \texttt{period\_near\_boundary} ($6.5\%$)\\ \texttt{outlier\_dominated} ($0.0\%$)\\ \texttt{data\_quality\_issues} ($0.0\%$)}} &
  {\footnotesize\makecell[l]{\texttt{extreme\_amplitude} ($1.0\%$)\\ \texttt{extreme\_period} ($0.0\%$)\\ \texttt{amplitude\_near\_boundary} ($2.0\%$)\\ \texttt{period\_near\_boundary} ($1.0\%$)\\ \texttt{outlier\_dominated} ($51.5\%$)\\ \texttt{data\_quality\_issues} ($56.6\%$)}} \\
  \noalign{\vskip 3pt}
  \enddata
  \tablecomments{The $n$-grams are ranked by the fraction of objects for which at least one rationale contained them, given in parentheses. Flag percentages give the fraction of objects for which at least one agent raised the flag. Only outputs of agents whose individual scores matched the consensus label are included.}
\end{deluxetable*}
\onecolumngrid

\section{Embedding Robustness}\label{app:robustness}

The phase of each light curve is measured from an arbitrary initial epoch, which introduces a relative phase offset between light curves of the same class. To quantify its impact on the DINOv2 embeddings, we rendered $100$ randomly selected light curves from each of the ten most common VSX classes at phase offsets spanning the full cycle in steps of $0.05$, embedded them as in Section~\ref{subsec:embedding}, and computed their cosine similarity to the unshifted version. Across the classes, the mean cosine similarity ranged from $0.81$ for the sharply peaked RRAB to $0.93$ for the irregular VAR (Table~\ref{tab:phase_shift}). By comparison, the cosine similarity for distinct sources averaged $0.25$ within the same class and $-0.02$ between different classes. The high similarity between offset copies of the same light curve therefore cannot be explained by shared class-level morphology but rather points to an approximate phase-shift invariance of the embeddings.

A second concern arises from the standard DINOv2 preprocessing, which applies a central crop to each image, discarding the outermost $6.25\%$ of the frame and potentially removing informative data points near its edges. To test this, we drew a new sample of $100$ light curves per class and embedded them once with the crop and once without it. The mean cosine similarity between the two versions exceeded $0.97$ for all ten most frequent VSX classes, confirming that the embeddings are largely insensitive to the crop. This is consistent with DINOv2's image-level training objective, which aligns the \texttt{[CLS]} token across global and local views and encourages approximate invariance to cropping \citep{oquab2023}.

\begin{deluxetable*}{ccccccccccc}
  \tabletypesize{\footnotesize}
  \setlength{\tabcolsep}{5pt}
  \tablecaption{Mean cosine similarity between the DINOv2 embeddings of shifted and unshifted versions of the same light curve, listed by phase offset for the ten most frequent VSX classes.\label{tab:phase_shift}}
  \tablewidth{0pt}
  \tablehead{
    \colhead{Phase offset} & \colhead{SR} & \colhead{EW} & \colhead{ROT} & \colhead{EA} & \colhead{RRAB} &
    \colhead{SRS} & \colhead{MISC} & \colhead{VAR} & \colhead{RRC} & \colhead{EB}
  }
  \startdata
  $0.05$ & $0.923$ & $0.939$ & $0.915$ & $0.909$ & $0.932$ & $0.918$ & $0.947$ & $0.948$ & $0.918$ & $0.926$ \\
  $0.10$ & $0.887$ & $0.923$ & $0.900$ & $0.865$ & $0.886$ & $0.894$ & $0.925$ & $0.933$ & $0.889$ & $0.875$ \\
  $0.15$ & $0.875$ & $0.925$ & $0.879$ & $0.843$ & $0.848$ & $0.875$ & $0.901$ & $0.930$ & $0.862$ & $0.855$ \\
  $0.20$ & $0.856$ & $0.906$ & $0.864$ & $0.814$ & $0.795$ & $0.848$ & $0.909$ & $0.923$ & $0.833$ & $0.829$ \\
  $0.25$ & $0.856$ & $0.898$ & $0.864$ & $0.793$ & $0.764$ & $0.841$ & $0.906$ & $0.918$ & $0.837$ & $0.825$ \\
  $0.30$ & $0.843$ & $0.899$ & $0.847$ & $0.784$ & $0.746$ & $0.849$ & $0.903$ & $0.918$ & $0.826$ & $0.824$ \\
  $0.35$ & $0.836$ & $0.901$ & $0.855$ & $0.791$ & $0.751$ & $0.851$ & $0.905$ & $0.923$ & $0.835$ & $0.830$ \\
  $0.40$ & $0.848$ & $0.912$ & $0.860$ & $0.795$ & $0.756$ & $0.871$ & $0.907$ & $0.929$ & $0.842$ & $0.832$ \\
  $0.45$ & $0.860$ & $0.918$ & $0.856$ & $0.792$ & $0.757$ & $0.862$ & $0.900$ & $0.926$ & $0.849$ & $0.836$ \\
  $0.50$ & $0.849$ & $0.921$ & $0.867$ & $0.828$ & $0.766$ & $0.865$ & $0.902$ & $0.931$ & $0.856$ & $0.838$ \\
  $0.55$ & $0.835$ & $0.922$ & $0.858$ & $0.831$ & $0.782$ & $0.864$ & $0.906$ & $0.928$ & $0.844$ & $0.835$ \\
  $0.60$ & $0.833$ & $0.911$ & $0.863$ & $0.822$ & $0.781$ & $0.857$ & $0.901$ & $0.931$ & $0.837$ & $0.828$ \\
  $0.65$ & $0.817$ & $0.907$ & $0.852$ & $0.809$ & $0.789$ & $0.848$ & $0.894$ & $0.924$ & $0.831$ & $0.813$ \\
  $0.70$ & $0.828$ & $0.893$ & $0.853$ & $0.800$ & $0.794$ & $0.846$ & $0.895$ & $0.910$ & $0.833$ & $0.813$ \\
  $0.75$ & $0.828$ & $0.908$ & $0.854$ & $0.786$ & $0.805$ & $0.849$ & $0.899$ & $0.913$ & $0.843$ & $0.833$ \\
  $0.80$ & $0.842$ & $0.906$ & $0.852$ & $0.792$ & $0.828$ & $0.863$ & $0.907$ & $0.909$ & $0.849$ & $0.830$ \\
  $0.85$ & $0.855$ & $0.913$ & $0.870$ & $0.825$ & $0.851$ & $0.884$ & $0.908$ & $0.921$ & $0.862$ & $0.867$ \\
  $0.90$ & $0.883$ & $0.923$ & $0.884$ & $0.853$ & $0.875$ & $0.897$ & $0.925$ & $0.925$ & $0.884$ & $0.886$ \\
  $0.95$ & $0.922$ & $0.945$ & $0.921$ & $0.918$ & $0.927$ & $0.927$ & $0.942$ & $0.946$ & $0.922$ & $0.919$ \\
  \hline
  Mean & $0.857$ & $0.914$ & $0.869$ & $0.824$ & $0.812$ & $0.869$ & $0.910$ & $0.926$ & $0.855$ & $0.847$ \\
  \enddata
  \tablecomments{Classes are ordered from left to right by decreasing frequency in our sample. Each value is the mean over $100$ randomly selected light curves of the corresponding class. All embeddings were centered as in Section~\ref{subsec:embedding} before computing the cosine similarity. The bottom row gives the mean over all offsets.}

\end{deluxetable*}  
  
\end{document}